\documentclass[10pt]{article}

\usepackage{amssymb,amsmath,amsthm} 
\usepackage[margin=1in]{geometry}
\usepackage{graphicx}
\usepackage{verbatim}
\usepackage{enumerate}
\usepackage{float}

\usepackage{epstopdf}

\newcommand{\R}{\mathbb{R}}

\newcommand{\p}{\partial}
\newcommand{\eps}{\epsilon}
\newcommand{\dt}{\Delta t}
\newcommand{\dx}{\Delta x}
\newcommand{\du}{\Delta u}
\newcommand{\dv}{\Delta v}
\newcommand{\ds}{\Delta s}

\newcommand{\cross}{\times}
\newcommand{\bb}{\mathbf}

\newcommand{\grad}{\nabla}

\usepackage{cases}
\usepackage{verbatim}
\usepackage{multirow}
\usepackage{array}
\usepackage[percent]{overpic}

\usepackage{mathtools} 

\usepackage{multicol}

\usepackage[hidelinks]{hyperref}
\hypersetup{
     allcolors=black
}

\usepackage[font=footnotesize]{caption}

\let\oldtabular\tabular
\renewcommand{\tabular}{\footnotesize\oldtabular}

\usepackage{xcolor}

\newcolumntype{L}[1]{>{\raggedright\let\newline\\\arraybackslash\hspace{0pt}}m{#1}}
\newcolumntype{C}[1]{>{\centering\let\newline\\\arraybackslash\hspace{0pt}}m{#1}}
\newcolumntype{R}[1]{>{\raggedleft\let\newline\\\arraybackslash\hspace{0pt}}m{#1}}

\begin{document}

\title{Modeling the Mitral Valve}
\author{Alexander D. Kaiser$^{1,2}$, David M. McQueen$^{1}$, Charles S. Peskin$^{1}$}
\date{
{\footnotesize $^{1}$ Department of Mathematics, Courant Institute of Mathematical Sciences, \\
New York University, 251 Mercer Street, New York NY 10012 \\ 
$^{2}$ Present address: Institute for Computational \& Mathematical Engineering and Department of Cardiothoracic Surgery, \\
Stanford University, Clark Center E100, 318 Campus Drive, Stanford, CA 94305 \\ }
\vspace{3mm} 
\today}

\maketitle

\begin{abstract}
This work is concerned with modeling and simulation of the mitral valve, one of the four valves in the human heart. 
The valve is composed of leaflets, the free edges of which are supported by a system of chordae, which themselves are anchored to the papillary muscles inside the left ventricle.  
First, we examine valve anatomy and present the results of original dissections. 
These display the gross anatomy and information on fiber structure of the mitral valve. 
Next, we build a model valve following a design-based methodology, meaning that we derive the model geometry and the forces that are needed to support a given load, and construct the model accordingly. 
We incorporate information from the dissections to specify the fiber topology of this model. 
We assume the valve achieves mechanical equilibrium while supporting a static pressure load. 
The solution to the resulting differential equations determines the pressurized configuration of the valve model. 
To complete the model we then specify a constitutive law based on a stress-strain relation consistent with experimental data that achieves the necessary forces computed in previous steps. 
Finally, using the immersed boundary method, we simulate the model valve in fluid in a computer test chamber. 
The model opens easily and closes without leak when driven by physiological pressures over multiple beats. 
Further, its closure is robust to driving pressures that lack atrial systole or are much lower or higher than normal. 
\end{abstract}

\section{Introduction }

The mitral valve is one of the four valves in the human heart. 
It lies between the left atrium, which serves as a staging chamber for oxygenated blood returning to the heart from the lungs, and the left ventricle, which is the main muscular pumping chamber that sends blood to all of the tissues and organs of the body. 
The valve is composed of leaflets, thin membranous flaps of tissue, attached to a ring. 
The free edges of the leaflets are supported, like a parachute, by a system of strings called chordae tendineae, which themselves are anchored to muscles called papillary muscles  that protrude from the left ventricular wall.  

This article, based on \cite{thesis}, concerns modeling and simulation of the mitral valve.  
The primary goal of this study is to build a mathematical model of the mitral valve that qualitatively matches the anatomy of a real mitral valve, and produces physiological flows when driven by physiological pressures over multiple cardiac cycles.

To achieve this, we use a design-based approach in which we compute the model geometry and tensions that are needed to support a pressure load when the valve is closed, and then we assign material properties such that these tensions can be generated by uniform strain.  
Note that in this approach we do not rely on measured geometry or material properties of an excised specimen, although we do use generic material properties of collagen-reinforced tissue.  
Our goal is to build the model valve as much as possible from first principles, formulated as differential equations for the closed, pressurized valve.  A previous study on the aortic valve using similar methods was successful and revealed much about the anatomy of the valve itself \cite{PeskinH319}.  
Subsequent simulation studies involving fluid-structure interaction showed that this model functions effectively throughout the cardiac cycle when driven by physiological pressures \cite{Griffith_aortic}.  
Another study used related methods to investigate the fiber structure of arteries and veins; they use cylindrical model geometry, which simplifies their analysis considerably \cite{doi:10.1177/1081286514551495}. 
The mitral valve, however, has a much more complicated architecture.  
In particular, the aortic valve lacks chordae, and the tension within its leaflets is supported primarily by circumferential fibers, whereas the mitral valve is supported by trees of chordae that connect to radial and circumferential fibers running within the valve leaflets.

Mitral valve tissue is highly heterogeneous and anisotropic. 
It includes fiber bundles large enough to be visible to the naked eye; these contribute to wide variation in thickness. 
Even if the material properties of the fibers that reinforce the mitral valve leaflet were fully known, to utilize these properties would require detailed knowledge of the local fiber orientations and the local thickness associated with each local orientation.
All of these data are highly variable from point to point on a given valve, and differ across individuals even within a given species.  
These facts make it extremely challenging to construct a realistic mechanical model of the mitral valve from experimental observations alone.

Mitral valve modeling has seen much progress in recent years, especially in methods to build models using an excised specimen. 
Two recent papers by Toma et al. \cite{Toma2016, CNM:CNM2815} build models by scanning an excised valve using micro-computed tomography, resulting in highly anatomical model geometry. 
They acknowledge that ``a challenge is how to determine fiber orientations.'' 
They address this challenge by specifying fiber orientation at certain locations as boundary conditions, and then fill in the rest of the fiber orientation by solving a modified Laplace equation. 
This equation is selected to ensure that the resulting fiber orientation is smooth. 
We also solve a partial differential equation to determine the fiber structure of our model, but our partial differential equation is determined by the requirement that the valve supports a uniform pressure difference when it is closed. 

Griffith et al.~have conducted fluid-structure-interaction simulations of a model prosthetic mitral valve \cite{griffith2009simulating}, of a model natural mitral valve built from MRI data \cite{MA2013417}, 
and another using CT data \cite{gao_chordae}.
They do not use the design-based approach that we take here. 
Other groups focus on solid mechanics only, without fluid-structure interaction. 
Kuhl and collaborators measure strains in ovine mitral valves in vivo, fit material parameters to experimental data using inverse finite element analysis, then use these models to estimate in vivo stresses \cite{inverse_fe_kuhl, KRISHNAMURTHY20091909}. 
Ratcliffe and collaborators use MRI-based model valve geometry, and compare simulation results to experimental findings from surgeries on living animal specimens \cite{wenk2010first}. 
Sacks and collaborators focus on multi-scale modeling methods \cite{Lee2015, CNM:CNM2921} and use Bayesian inference to fit parameters to experimental data \cite{sacks_collagen_constitutive}.
Reviews on mitral valve modeling can be found in \cite{morgan2016finite, CNM:CNM1280, doi:10.1002/cnm.2858}.

Finally, a broad and comprehensive four-part review of heart-valve engineering discusses major challenges in mitral-valve modeling \cite{kheradvar2015emergingIV}, and includes the comment that ``There appear to be relatively few FSI [fluid-structure interaction] valve models that can perform multiple cardiac cycles and also simulate the closed, loaded configuration of the valve... Closure seems to be especially challenging to simulate because it fundamentally involves a delicate balance between the fluid dynamics and elasticity of the valve's leaflets.''
In the present article, we achieve exactly this. 
We build a model that performs robustly during fluid-structure interaction over multiple cardiac cycles when driven by physiological pressures, or even by pressures substantially higher or lower than normal.

In Section \ref{anatomy}, we examine valve anatomy. 
We present an original dissection, and use the results directly in constructing the model valve. 

In Section \ref{static}, we build a model of the valve following a design-based methodology. 
We assume the valve achieves mechanical equilibrium while supporting a static pressure load. 
The solution of the resulting partial differential equations specifies the pressurized configuration of the valve model. 
This provides information about the tension throughout the model valve. 
Combining this with stress-strain relations that are known to govern collagen reinforced tissue, we generate a constitutive law for the model in Section \ref{general_model}. 
This creates a complete mechanical model that is suitable for simulations. 

Finally, in Section \ref{Fluid-Structure_Interaction}, using the immersed boundary method, we simulate the model valve in fluid. 
We agree with the authors of \cite{LAU20101057} that structural mechanics is only
 part of the heart valve problem, and that ``... in order to simulate
 full dynamic behaviour fluid-structure interaction models are required.''
The valve is placed in a model test chamber, and simulations are driven by prescribed waveforms of the pressures upstream and downstream of the valve.
When driven by physiological pressures over multiple beats, the model valve opens freely and seals reliably. 

Source code for the project is freely available at \url{github.com/alexkaiser/mitral_valve}.

\section{Mitral Valve Anatomy}
\label{anatomy}

\begin{figure}[ht]
\centering 
\includegraphics[width=.8\columnwidth]{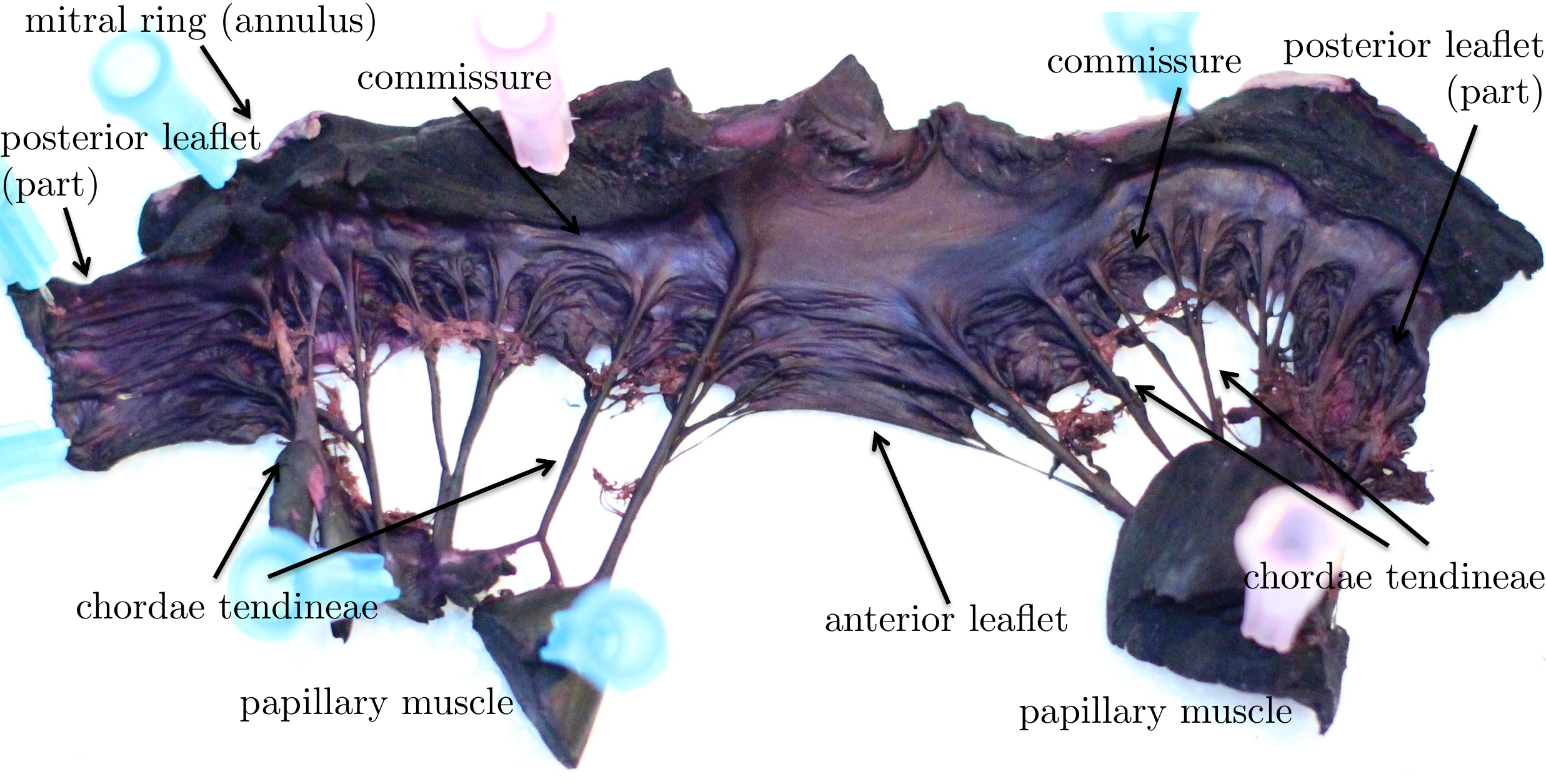}
\caption{Porcine mitral valve, ventricular side.}
\label{valve_labeled}
\end{figure}

The mitral valve apparatus is composed of leaflets, as well as a system of \emph{chordae tendineae} and two \emph{papillary muscles}. 
There are two main leaflets: the anterior, which covers a smaller fraction of the valve ring and is deeper, and the posterior, which takes up a larger fraction of the ring and is shallower. 
There are small flaps of tissue between the anterior and posterior leaflets, which are considered by some authors to be separate, commissural, leaflets, thus making the mitral valve into a four-leaflet structure. 
One side of each leaflet attaches to the \emph{mitral ring}, which separates the left atrium from the left ventricle. 
The separation between the leaflets at the commissures is not complete; there is a small annulus-like region of leaflet that runs without interruption below the valve ring. 
The free edges of the leaflets are attached to the chordae tendineae.
At the attachment to the leaflets, the chordae  branch and seamlessly blend into the surrounding leaflet tissue. 
Below the leaflets, the many branches of the chordae tendineae collect and form fewer, thicker chords, which in turn anchor into the papillary muscles, which protrude from the left-ventricular wall. 
A fully dissected valve is shown, stained for collagen and labeled in Figure \ref{valve_labeled} (original work, specimen from a local butcher); see also \cite{netter}. 
Additional dissection images and commentary are in Section \ref{anatomy_appendix}; see also \cite{thesis}.

\section{Construction of the Model Mitral Valve}
\label{static}

We assume that the closed valve supports a static pressure load, and in doing so achieves a state of mechanical equilibrium. 
That is, we specify how the valve has to function -- what forces it must support -- and determine its configuration by solving the associated differential equations.

\subsection{Assumptions}

The model geometry is built according to the following principles, which summarize the idealized anatomy and function of the mitral valve. 

\begin{enumerate}
  
\item The valve is composed of two leaflets, which are made up of fibers. 
These fibers exert tension only in the fiber directions. 

\item At any point internal to the leaflet, there are two families of fibers under tension. 
The first family of fibers is oriented radially. 
It connects chordae at the free edge to the valve ring. 
The second is circumferential. 
It runs approximately parallel to the valve ring. 
Each circumferential fiber either forms a closed ring or connects one point on the free edge of the valve to another such point. 
In the latter case, each end of the fiber makes contact with a chordal tree.

\item The leaflets are supported by a system of chordae tendineae, which anchor into two papillary muscles. 
Like the fibers in the leaflets, the chordae exert tensile forces only. 

\item Tension in the leaflets supports a static, uniform pressure load.
This is possible because the leaflets are curved.  
There is no pressure load acting directly on the chordae (since they are idealized as being one-dimensional), but the tension in the chordae indirectly supports the pressure load on the leaflets.
The whole structure, composed of leaflets and chordae, achieves a mechanical equilibrium in which all of these forces balance.

\end{enumerate}

\noindent 
Section \ref{models_literature} presents additional literature on mitral valve material properties and fiber structure, and existing models of the valve, along with additional commentary.

To justify the fourth assumption, note that the valve is closed for approximately 0.2s during each cardiac cycle. 
We estimate the inertio-elastic timescale of the pressurized valve to be of order $10^{-4}$ s (see \ref{Inertio-elastic_timescale}). 
Thus, analysis of a static, closed valve is relevant to the general dynamics of the valve. 

\subsection{Problem formulation}

First, we derive the continuous formulation of the equations of equilibrium in the leaflet. 
We represent the leaflet as an unknown parametric surface in $\R^{3}$, 
\begin{align}  
\bb X(u,v) : \Omega \subset \R^{2} \to \R^{3} . 
\end{align}
In this formulation, there are two families of fibers, one running along the curves $v$ = constant, and the other along the curves $u$ = constant.  
The fibers $v$ = constant, on which $u$ varies, will be called $u$-type fibers, and the fibers $u$ = constant, on which $v$ varies, will be called $v$-type fibers.  
In this paper, in contrast to \cite{thesis}, we take $u$ and $v$ to be dimensionless. 
Let subscripts denote partial derivatives and let single bars, $| \cdot |$, denote the Euclidean norm. 
The unit tangents to these two fiber families are 
\begin{align}
\frac{\bb X_{u}}{ | \bb X_{u} |} \quad \text{ and } \quad \frac{\bb X_{v}}{ | \bb X_{v} |}, 
\end{align}
respectively. 
Let $S(u,v)dv$ be the tension transmitted by the $u$-type fibers with $v$ in the interval $(v,v+dv)$, and similarly let $T(u,v)du$ be the tension transmitted by the $v$-type fibers with $u$ in the interval $(u,u+du)$.  
Note that $S$ and $T$ have units of force (since we assumed $u$ and $v$ are dimensionless), but they are best described as ``force per unit $v$'' and ``force per unit $u$'' respectively.  
In particular, the value of $S$ changes if $v$ is replaced by some function of $v$, and the value of $T$ changes if $u$ is replaced by some function of $u$.  
These are the most general changes of parameters that are allowed, since the parameterization is assumed to conform to the fibers.

Consider the static mechanical equilibrium of an arbitrary patch of leaflet $[u_{1}, u_{2}] \cross [ v_{1} , v_{2} ]$. 
Let $p$ denote the pressure, which acts in the normal direction to the patch. 
We assume that $p$ is constant; that is, the pressure load is spatially uniform. 
The total pressure force on the patch is given by 
\begin{align} 
\int_{v_{1}}^{v_{2}}  \int_{u_{1}}^{u_{2}}   p  (  \bb X_{u}(u,v) \cross \bb X_{v}(u,v) ) \;  du dv . 
\end{align}
The tension force due to $u$-type fibers acts on the edges of constant $u$, so the total force transmitted across the arc $u=u_{2}$, $v \in (v_{1}, v_{2} )$ is given by  
\begin{align} 
\int_{v_{1}}^{v_{2}}   S(u_{2}, v) \frac{ \bb X_{u} (u_{2}, v) }{|  \bb X_{u} (u_{2}, v) |}  \;  dv  . 
\end{align}
The total force due to $u$-type fibers on the patch is then given 
\begin{align} 
\int_{v_{1}}^{v_{2}}  \left( S(u_{2}, v) \frac{ \bb X_{u} (u_{2}, v) }{|  \bb X_{u} (u_{2}, v) |} - S(u_{1},v)   \frac{ \bb X_{u} (u_{1}, v) }{|  \bb X_{u} (u_{1}, v) |}    \right)  dv  . 
\end{align}
Similarly, the total force to to $v$-type fibers is given 
\begin{align} 
\int_{u_{1}}^{u_{2}}  \left( T(u, v_{2})  \frac{\bb X_{v} (u, v_{2})}{|\bb X_{v} (u, v_{2})|} - T(u,v_{1})  \frac{\bb X_{v} (u, v_{1}) }{|\bb X_{v} (u, v_{1})|} \right)  du . 
\end{align}

\begin{figure}[t]
\centering
\includegraphics[width=.6\columnwidth]{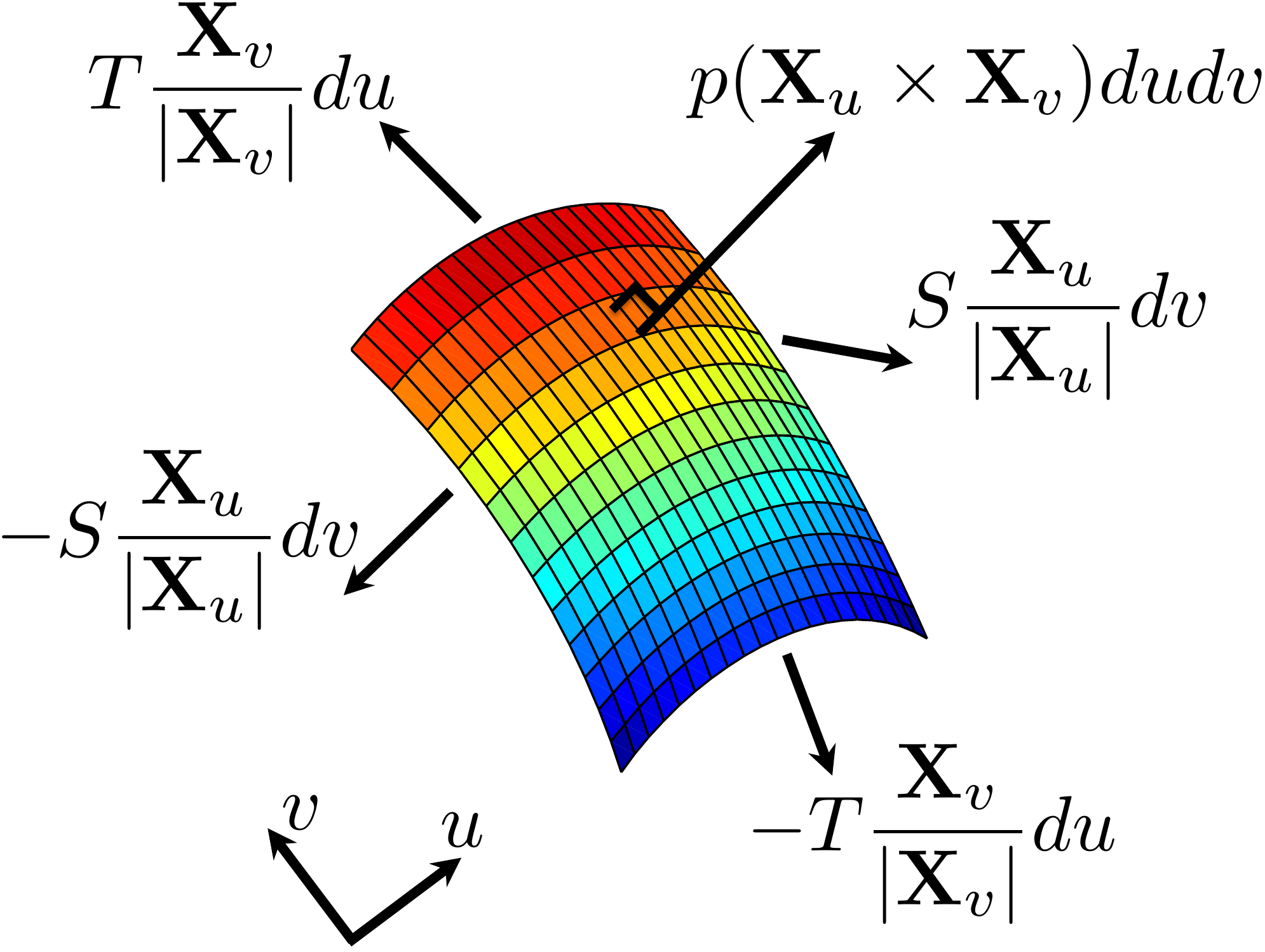}
\caption{Free body diagram for mechanical equilibrium on the leaflet. 
Note that the tension variables are evaluated on opposing boundaries of the patch, and thus do not cancel, but we drop arguments to all functions for visual clarity.}
\label{free_body_diagram_leaflet.pdf}
\end{figure}

A free body diagram of these forces applied to a patch of leaflet is shown in Figure \ref{free_body_diagram_leaflet.pdf}. 

The condition of static mechanical equilibrium dictates that the forces on the patch $[u_{1}, u_{2}] \cross [v_{1}, v_{2}]$ must sum to zero, so the integral form of the equations of equilibrium is
\begin{align} 
0 &= \int_{v_{1}}^{v_{2}}   \int_{u_{1}}^{u_{2}}    p  \left(  \bb X_{u}(u,v) \cross \bb X_{v}(u,v) \right)    du dv  \\ 
     &+ \int_{v_{1}}^{v_{2}}  \left(  S(u_{2}, v) \frac{ \bb X_{u} (u_{2}, v) }{|  \bb X_{u} (u_{2}, v) |} - S(u_{1},v)   \frac{ \bb X_{u} (u_{1}, v) }{|  \bb X_{u} (u_{1}, v) |}   \right)    dv \nonumber  \\ 
     &+ \int_{u_{1}}^{u_{2}} \left(  T(u, v_{2})  \frac{\bb X_{v} (u, v_{2})}{|\bb X_{v} (u, v_{2})|} - T(u,v_{1})  \frac{\bb X_{v} (u, v_{1}) }{|\bb X_{v} (u, v_{1})|} \right)  du . \nonumber
\end{align}
Apply the fundamental theorem of calculus to convert all the integrals into double integrals over the patch. 
Swap the order of integration formally as needed. 
This gives 
\begin{align} 
0 &= \int_{v_{1}}^{v_{2}}  \int_{u_{1}}^{u_{2}} \bigg(   p  (  \bb X_{u} \cross \bb X_{v} )   +   \frac{\p}{\p u}  \left( S \frac{ \bb X_{u} }{ |\bb X_{u}| } \right)  +  \frac{\p}{\p v}  \left( T \frac{ \bb X_{v} }{|\bb X_{v}|} \right)    \bigg)  \;  du dv , \nonumber
\end{align}
in which all variables are evaluated at $u,v$. 
Since the patch is arbitrary, the integrand must be zero. 
This gives the partial differential equation form of the equations of equilibrium as 
\begin{align} 
0 = p  (  \bb X_{u} \cross \bb X_{v} )  +   \frac{\p}{\p u}  \left( S \frac{ \bb X_{u} }{ |\bb X_{u}| } \right)  +  \frac{\p}{\p v}  \left( T \frac{ \bb X_{v} }{|\bb X_{v}|} \right).    
\label{eq_eqns}
\end{align}
Since the parameters $u$ and $v$ are dimensionless, dimensional consistency implies that the tensions $S$ and $T$ have units of force, and thus each term in equation \eqref{eq_eqns} has units of force.

\subsection{Closing the equilibrium equations}

Equation \eqref{eq_eqns} is not closed. 
It is three equations, one for each component of force, and it involves five unknowns, three for $\bb X$ and two for the tensions $S,T$. 
To close it, we need equations for $S$ and $T$. 
Here, we aim to find a tension law that does not require a reference configuration, since we do not have access to such a configuration. 

The simplest example of a tension law without a reference configuration would a be prescribed, constant tension for each fiber family. 
Although this is appealing conceptually and has the interesting consequence that fibers of both families are geodesics on the surface that they form, it does not produce a good model of the mitral valve. 
The difficulty is that when iterating to solve the discretized versions of these equations, the fibers bunch together near the free edges during iterations of the nonlinear system solver. 
Finite differences are evaluated between fibers, and when the points nearly coincide on the bunched-together fibers, the associated linear systems become ill-conditioned. Solvers then fail to find a solution to the equilibrium equations. (See Sections \ref{Discretization} and \ref{Numerical-method} for discussion of discretization and numerical methods.)

We propose the following tension law as an alternative, which we call the \emph{decreasing-tension} model.  
In this model, we control the maximum tension. 
Suppose that the maximum tension in $u$-type fibers is limited by $\alpha$, but goes smoothly to zero as $|\bb X_{u}|$ goes to zero. 
Take 
\begin{align}
S(u,v)  = \alpha \left( 1 - \frac{1}{1 + |\bb X_{u}|^{2} / a^{2} } \right), \label{tension_S} 
\end{align}
where $a$ is a tunable parameter. 
Similarly, let 
\begin{align}
T(u,v)  = \beta \left( 1 - \frac{1}{1 + |\bb X_{v}|^{2} / b^{2} } \right). \label{tension_T}
\end{align}
We find that the decreasing tension law prevents the bunching of fibers that would otherwise occur, and that the parameters $a$ and $b$ can be tuned to control fiber spacing. 
Equations \eqref{tension_S} and \eqref{tension_T} are not proposed as physical tension-strain relations, but rather as a means of arriving at a fiber-architecture in which prescribed maximum tensions $\alpha$ and $\beta$ are not exceeded in the closed, pressure-loaded configuration of the valve.  
Once that configuration has been found, we replace \eqref{tension_S} and \eqref{tension_T} by physical tension-strain relations in such a way that the pressure-loaded configuration of the valve, including its tension distribution, is unchanged. 
How this is achieved will be described below.

Substituting \eqref{tension_S} and \eqref{tension_T} into \eqref{eq_eqns}, we get the closed equation 
\begin{align} 
0 &= p  (  \bb X_{u} \cross \bb X_{v} )  \label{eq_eqn_dec_tension}  
+  \frac{\p}{\p u}  \left( \alpha \left( 1  -  \frac{1}{1 +  |\bb X_{u}|^{2} / a^{2}  } \right) \frac{ \bb X_{u} }{ |\bb X_{u}| } \right)  
+  \frac{\p}{\p v}  \left(  \beta \left( 1  -  \frac{1}{1 +  |\bb X_{v}|^{2} / b^{2} }     \right) \frac{ \bb X_{v} }{|\bb X_{v}|} \right) .   
\end{align}
Note that the coefficients $\alpha, \beta$, as well as the decreasing-tension parameters $a, b$ need not be constants.
Their values will be tuned to specific regions of the valve, and are shown in Section \ref{Values_coefficients}. 
The pressure is $p = 100$ mmHg, slightly less than the nominal human systolic blood pressure of 120 mmHg.

The domain on which equation \eqref{eq_eqn_dec_tension} is to be solved is defined by $u \in [0,1]$ and $v \in [v_{min}(u),1]$.  
The function $v_{min}(u)$ is continuous, piecewise linear with slopes $\pm 1$ or 0, and periodic with period 1, shown at the bottom of the Cartesian mesh in Figure \ref{mesh_schematic_II.eps}.  
It is selected to approximate the shape of the leaflets in the real valve, so that after solving the equation, the leaflets take on an anatomical shape. 
Each leaflet has slope -1 on either side of its approximate plane of symmetry, and slope 1 on the opposing side so that the preimage of the leaflet forms a triangle. 
A small patch of tissue is added between the two leaflets, and accounts for the region in which $v_{min}(u)$ has slope zero. 
See \ref{further_discretization} for further details.
The function $\bb X(u,v)$ is also periodic in $u$ with period 1. 
This means that we have selected the circumference of the valve ring as the length scale for this problem.
The curve $\bb x = \bb X(u,1)$ is the valve ring, and the curve $\bb x = \bb X(u,v_{min}(u))$ is the free edge of the valve.  
Both of these space curves are closed because of the assumed periodicity of $\bb X$ and $v_{min}$.  
The valve ring $\bb X(u,1)$ is prescribed as a boundary condition (see Section \ref{Boundary-conditions-mitral-valve-ring}), but the free edge curve $\bb X(u,v_{min}(u))$ is not. 
Instead, it is determined as part of the solution of equation \eqref{eq_eqn_dec_tension}, coupled to the equations of the chordae.  
The chordae are modeled as trees of fibers. 
Since the chordae are inherently discrete, they are best described along with the discretization of the leaflet equations, which is the subject to which we now turn.

\subsection{Fiber mesh and chordae tendineae}
\label{fiber_spec}

\begin{figure}[th]
\centering
\includegraphics[width=.8\columnwidth]{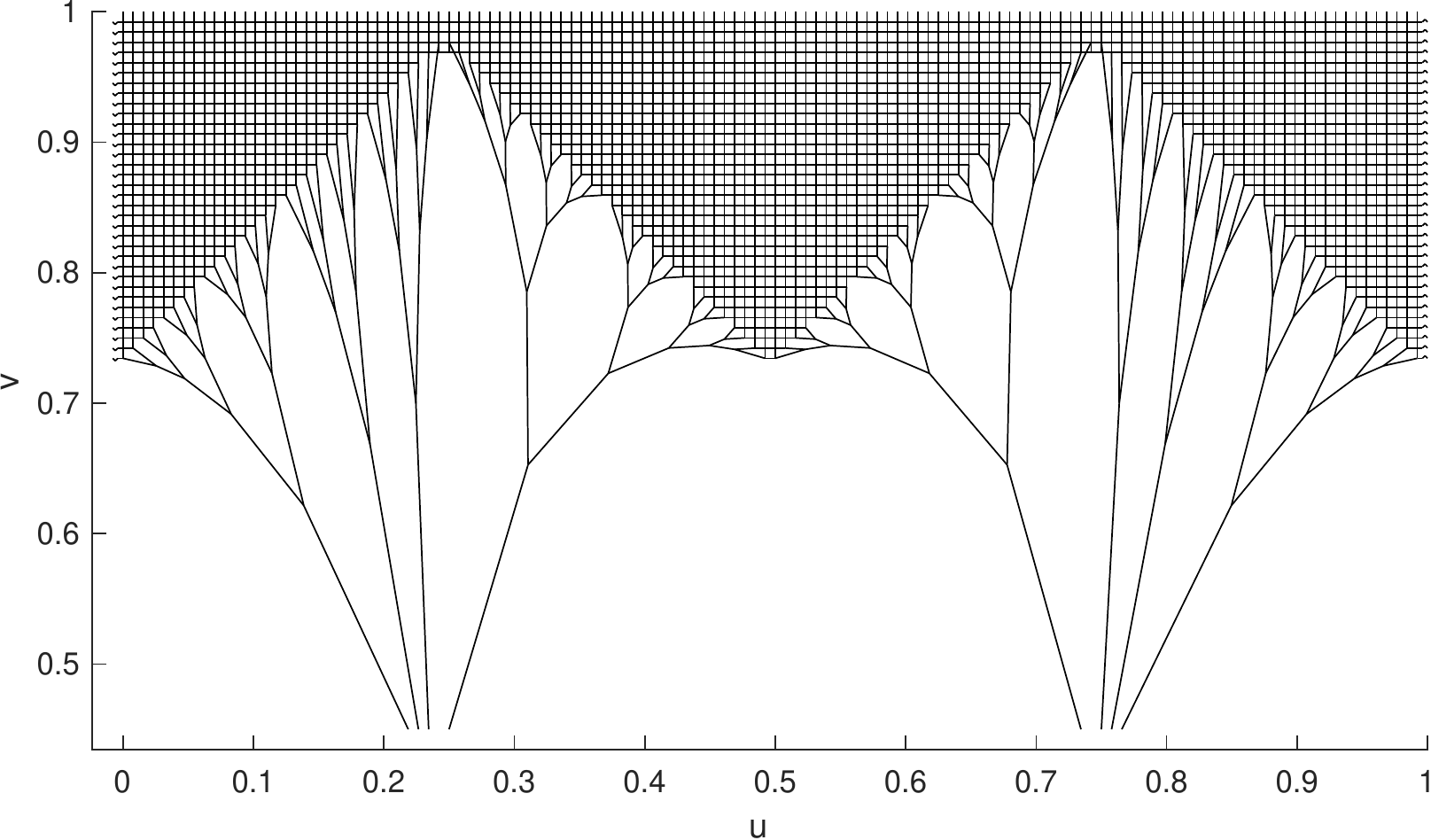}
\caption{Schematic of valve mesh, including trees of chordae. 
The anterior leaflet is centered, and the posterior leaflet is split in two, as in the dissection in Figure \ref{valve_labeled}.
The top of the Cartesian mesh shown here represents the valve ring in parameter space. 
The curve $v_{min}(u)$ represents the free edge of the valve in parameter space; it is the bottom of the Cartesian mesh shown here. 
For visual clarity, the mesh shown is four times more coarse here, $N = 128$, than the final mesh that is later used in fluid, $N = 512$. 
The positions of the trees of chordae within the $(u,v)$ plane have no meaning within our modeling framework, but their topological structures and their connections to the leaflets are important. }
\label{mesh_schematic_II.eps}
\end{figure}

The layout of fibers in our model is shown in Figure \ref{mesh_schematic_II.eps}, which depicts the discretized leaflets and the chordae tendineae in the $(u,v)$ parameter plane.  
Note, however, that the geometric positions of the chordae as shown in the figure have no significance, since nodes of the chordal trees are not assigned to specific values of $u$ and $v$, except for the terminal nodes, which coincide with points of the leaflets. 
Fibers within the leaflets form a Cartesian grid in the $(u,v)$ parameter plane, with meshwidths $\Delta u = \Delta v = 1/N$, where $N$ is a power of 2.  
Each time $N$ is doubled, one more generation is added to each of the chordal trees.

\subsection{Discretization}
\label{Discretization}

Equation \eqref{eq_eqn_dec_tension} is discretized as follows. 
Let $(j,k)$ denote the leaflet point whose coordinates in the $(u,v)$ plane are $(j\Delta u, k\Delta v)$, and let $\bb X^{j,k}$ be the position of that point in physical space.  
If $(j,k)$ denotes an interior point of the leaflet, then we use the discretization

\begin{align}
&0 =   p \left(  \frac{(\bb X^{j+1,k} - \bb X^{j-1,k} )}{2\du}  \cross \frac{ ( \bb X^{j,k+1} - \bb X^{j,k-1} ) }{2\dv} \right)  \label{equilbrium_eqn_discrete}  \\ 
            &+  
             \frac{\alpha}{\du} 
	     \left(  1 - 
	     1 \left/ \left( 1 + \dfrac{ | \bb X^{j+1,k}  -  \bb X^{j,k} |^{2}}{  a^{2} (\du)^{2} } \right) \right) \right.            
             \frac{ \bb X^{j+1,k}  -  \bb X^{j,k}  }{ | \bb X^{j+1,k}  -  \bb X^{j,k} |  }  \nonumber  \\  
            &  -    
             \frac{\alpha}{\du} 
	     \left( 1 - 
	     1 \left/ \left(1 + \dfrac{ | \bb X^{j,k}  -  \bb X^{j-1,k} |^{2}}{  a^{2} (\du)^{2}  } \right) \right) \right.
              \frac{ \bb X^{j,k}  -  \bb X^{j-1,k}  }{ | \bb X^{j,k}  -  \bb X^{j-1,k} |  }     \nonumber   \\   
             &  + 
             \frac{\beta}{\dv} 
	     \left( 1 -
	     1 \left/ \left( 1 + \dfrac{ | \bb X^{j,k+1}  -  \bb X^{j,k} |^{2}}{  b^{2} (\dv)^{2}  }  \right) \right) \right. 
             \frac{ \bb X^{j,k+1}  -  \bb X^{j,k}  }{ | \bb X^{j,k+1}  -  \bb X^{j,k} |  } \nonumber  \\ 
             &  -               
              \frac{\beta}{\dv} 
	     \left( 1 -
	     1 \left/ \left( 1 + \dfrac{ | \bb X^{j,k}  -  \bb X^{j,k-1} |^{2}}{  b^{2} (\dv)^{2}  } \right) \right) \right. 
             \frac{ \bb X^{j,k}  -  \bb X^{j,k-1}  }{ | \bb X^{j,k}  -  \bb X^{j,k-1} |  }    .  \nonumber 
\end{align}

\noindent
If the point $(j,k)$ is missing one or more of the four neighbors referenced in the above equation, then the term corresponding to each missing neighbor is simply deleted from \eqref{equilbrium_eqn_discrete}, and the corresponding difference formula in the pressure term is replaced by a one-sided difference.  
As written above, each  term in \eqref{equilbrium_eqn_discrete} has units of force, but this is actually force per unit area in the $(u,v)$ parameter plane (recall that $u$ and $v$ are dimensionless).  
In practice, we multiply both sides of \eqref{equilbrium_eqn_discrete} by $\Delta u \Delta v$, and then each term represents an actual force on the node $(j,k)$. 
If a node has a branch of a chordal tree connected to it, the force from that branch is added to the forces that are already accounted for by \eqref{equilbrium_eqn_discrete}.

\subsection{Model trees of chordae tendineae}

Internal to the chordae, there is no pressure directly applied, and the tensions applied to a given chordal junction must then sum to zero. 
In the trees of chordae, there is no notion of a continuum limit in the tension equations and we use the discrete equations alone. 
To make the finest level in the trees blend seamlessly into the valve mesh, we specify that there are $N$ total leaves in the trees.
For each point on the discretization of the valve ring, there is exactly one point on or near the free edge to which some leaf of a tree attaches. 
Note that if $N$ doubles, then the number of leaves in the trees must double to maintain this relationship. 
Thus, when $N$ is doubled, we double the number of leaves in each tree by adding another generation, and this also doubles the total number of leaves across all of the trees.
All trees are taken to be binary, in which each edge has precisely two descendants. 
A recent study concluded that a binary branching structure is anatomical in real mitral valves \cite{Khalighi2017}. 
The chordae in the model generally attach to the free edge, and as such are referred to as primary chordae. 
Real mitral valves also have secondary chordae, which attach to the ventricular surface of the leaflet, and tertiary, which connect chordae to other locations on the chordae \cite{Toma2016}. 
One recent numerical study explored three models of chordae with fluid structure interaction, and observed only minor differences in flow and in leaflet position \cite{gao_chordae}.
Another numerical study that focuses on solid mechanics only compared a very simple model of chordal forces (uniform force orthogonal to the annulus), attached all over the leaflets, at the free edge only and on a large distributed region of the leaflets. 
They report qualitatively similar strains in the leaflets with all three models, see Figure 8 in \cite{doi:10.1002/cnm.3142}.
We therefore consider the simplification of only including primary chordae sufficient for the current work, and leave including secondary and tertiary chordae for future work.

The tension in the trees takes the same form as the tension terms in equation \eqref{equilbrium_eqn_discrete}. 
Suppose that $\bb C$ denotes a particular junction in the tree, and $\bb C'$ denotes one of its neighbors. 
The force on $\bb C$ from its connection to $\bb C'$ is defined to be  
\begin{align}
\gamma 
\left( 1 -
	     1 \left/ \left( 1 + \dfrac{  |\bb C' - \bb C|^{2} } {c^{2}}  \right) \right) \right.     \frac{ \bb C'  -  \bb C }{ | \bb C'  -  \bb C| } . \label{chordae_tension}
\end{align}
in which the coefficient $\gamma$ has units of force and $c$ has units of length. 
A rule to determine how $\gamma$ and $c$ scale in the trees of chordae is described Section \ref{scaling_laws}.

\subsection{Results -- the model mitral valve}

\begin{figure*}[thb]
$\begin{array}{c c c}
\hspace{-0.15cm} \includegraphics[width=.5\textwidth]{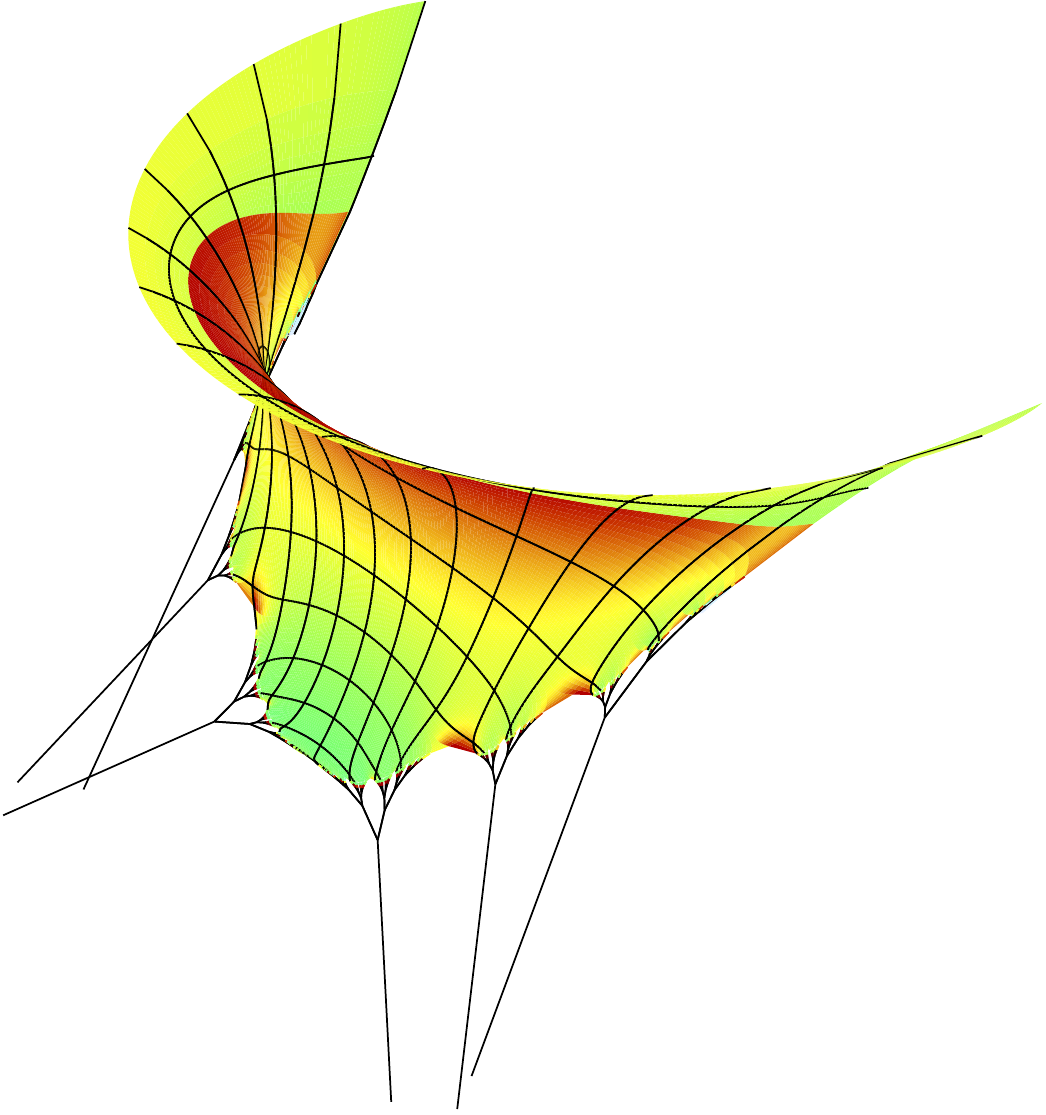}  & 
\hspace{-1.2cm}  \includegraphics[width=.5\textwidth]{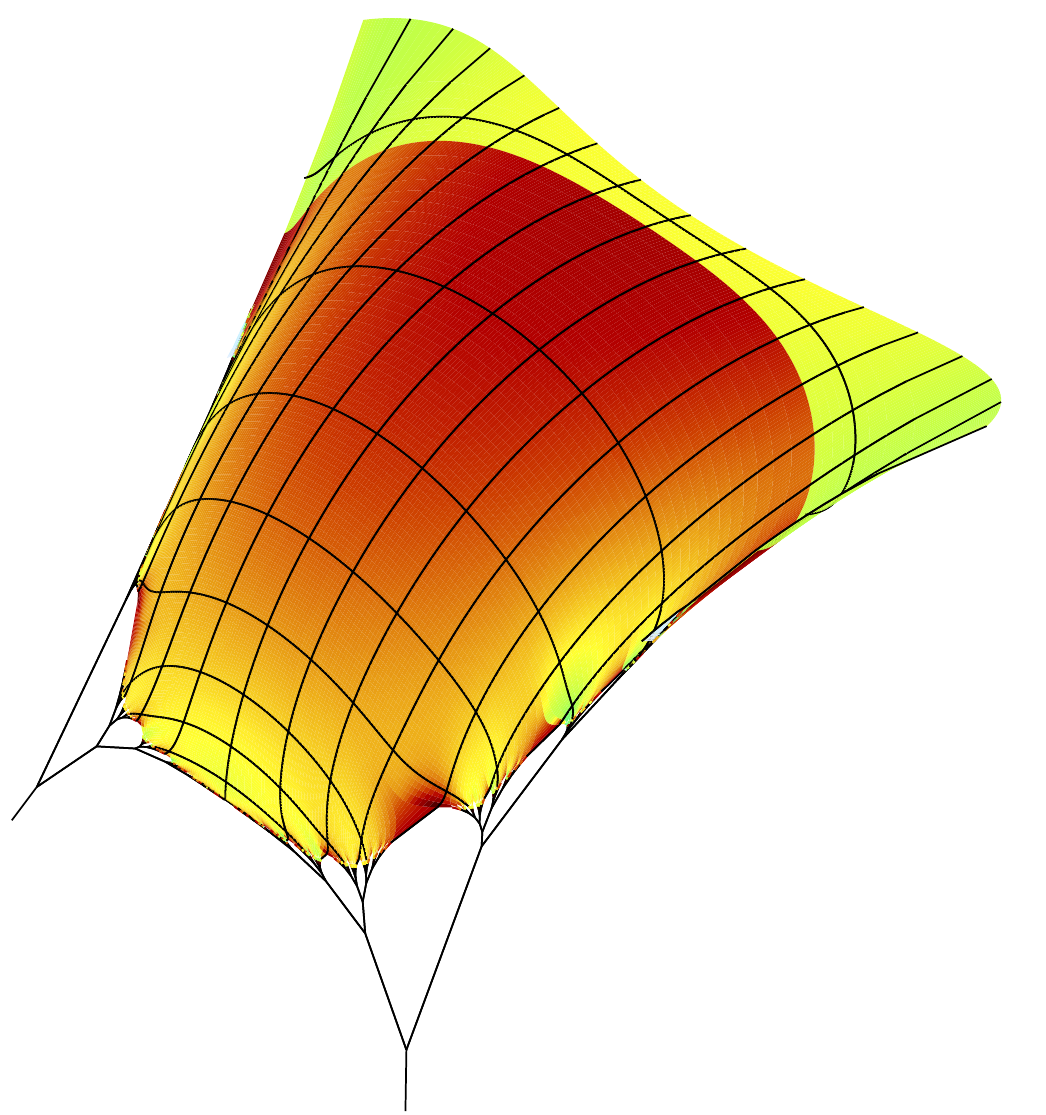} &
\hspace{-1.0cm}  \raisebox{2.4cm}{ \includegraphics[width=0.08\textwidth]{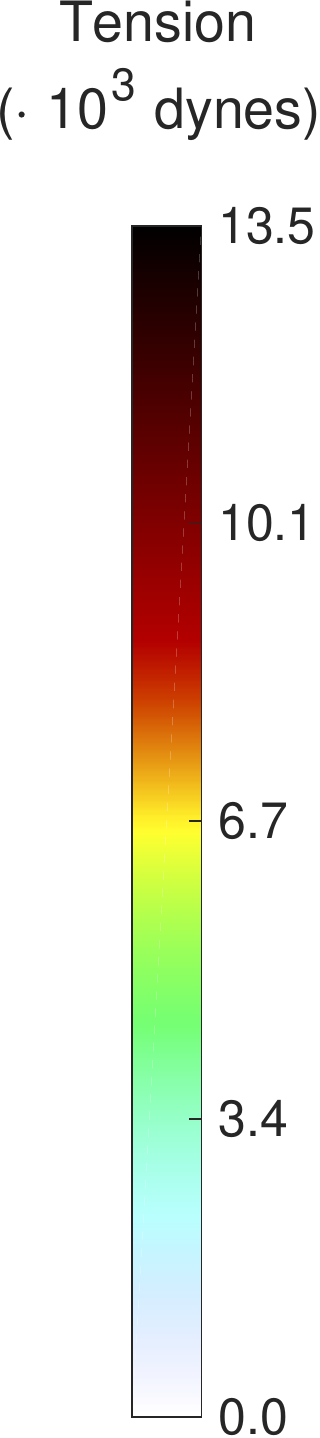} } 
\end{array}$
\caption{Closed geometry of model valve showing the total tension, the sum of the radial and circumferential tensions, in the posterior leaflet (left) and in the anterior leaflet (right). 
The color scale runs from zero to the maximum possible sum of the design tensions, $\max(\alpha\dv + \beta\du)$.
The leaflets are colored with respect to discrete fiber tension.
Tension in the chordae tendineae mostly exceeds the maximum allowed leaflet fiber tension; these chordae plotted in black. 
Some chordae at finer scales near the free edge have tension that is less than the maximum allowed leaflet fiber tension; these chordae are colored according to the leaflet color scale (zoom in to view). 
Every sixteenth fiber in the leaflets is also plotted in black as a visual cue. 
The pressure, $p = 100$ mmHg, is slightly less than the nominal human systolic blood pressure of 120 mmHg. 
$N=512$. }
\label{static_valves}
\end{figure*}

\begin{figure}[bth!]
\centering
\includegraphics[width=.6\columnwidth]{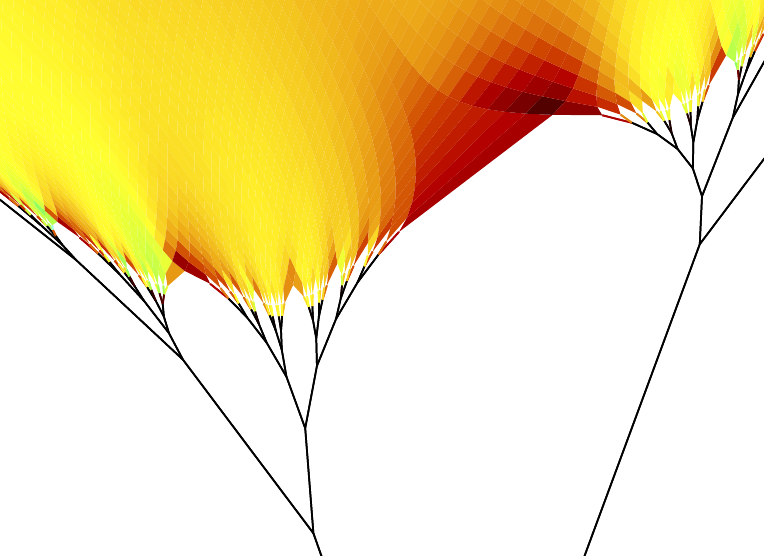} 
\caption{Detail on trees and free edge of the anterior leaflet. Total tension and tension in the chordae are shown as in Figure \ref{static_valves} with the same color scale. $N=512$.}
\label{tree_detail}
\end{figure}

\begin{figure*}[ht]
\centering
\begin{tabular}{c c c}
 \hspace{-0.6cm} 
 \text{Circumferential fibers}  \hspace{-0.9cm} 
& \text{Radial fibers}  
&        \\ 
 \hspace{-0.6cm} \begin{overpic}[width=.477\textwidth]{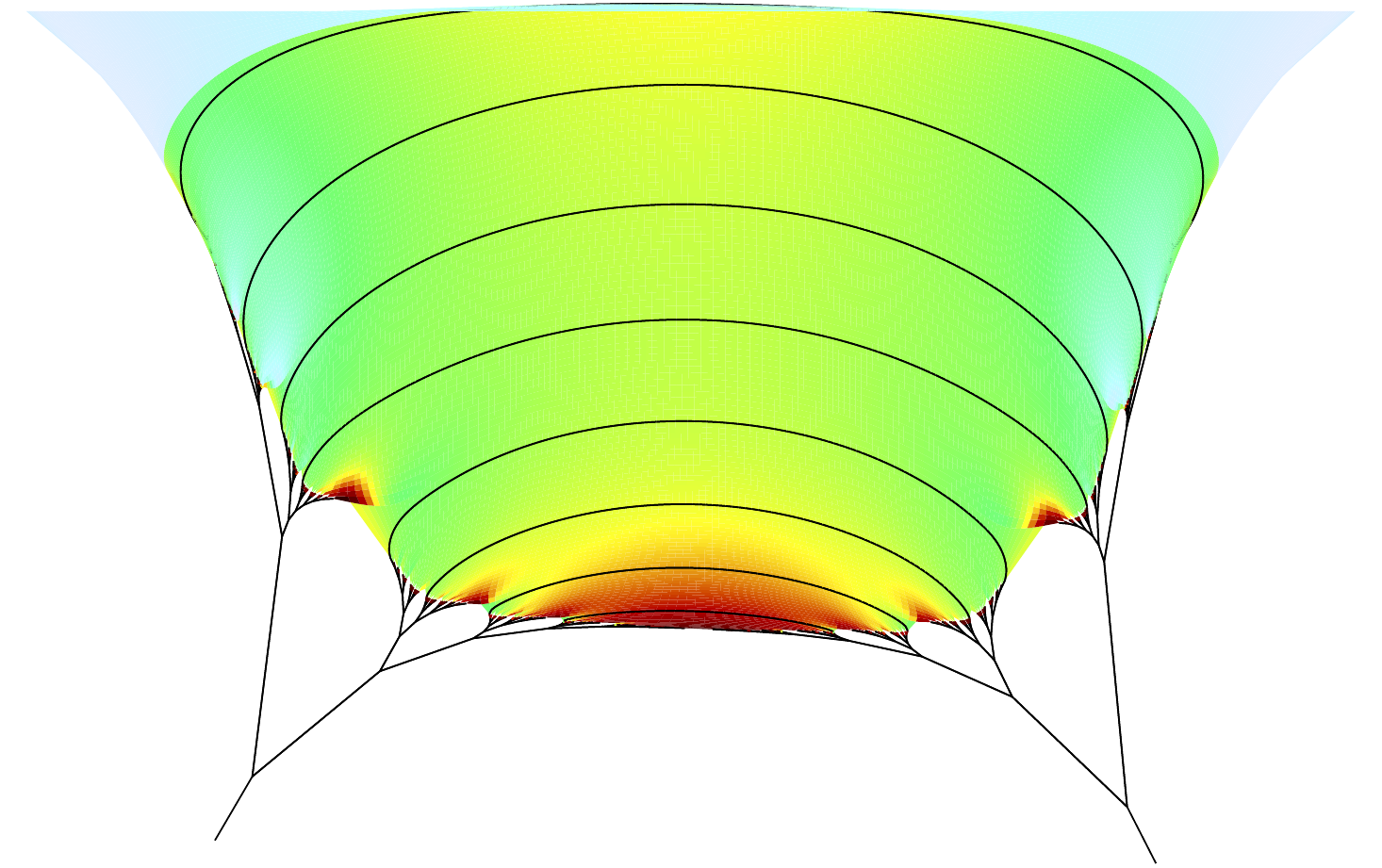} \put (3,20) {\rotatebox[origin=l]{90}{Anterior leaflet}}\end{overpic} \hspace{-0.4cm}    
&  \includegraphics[width=.477\textwidth]{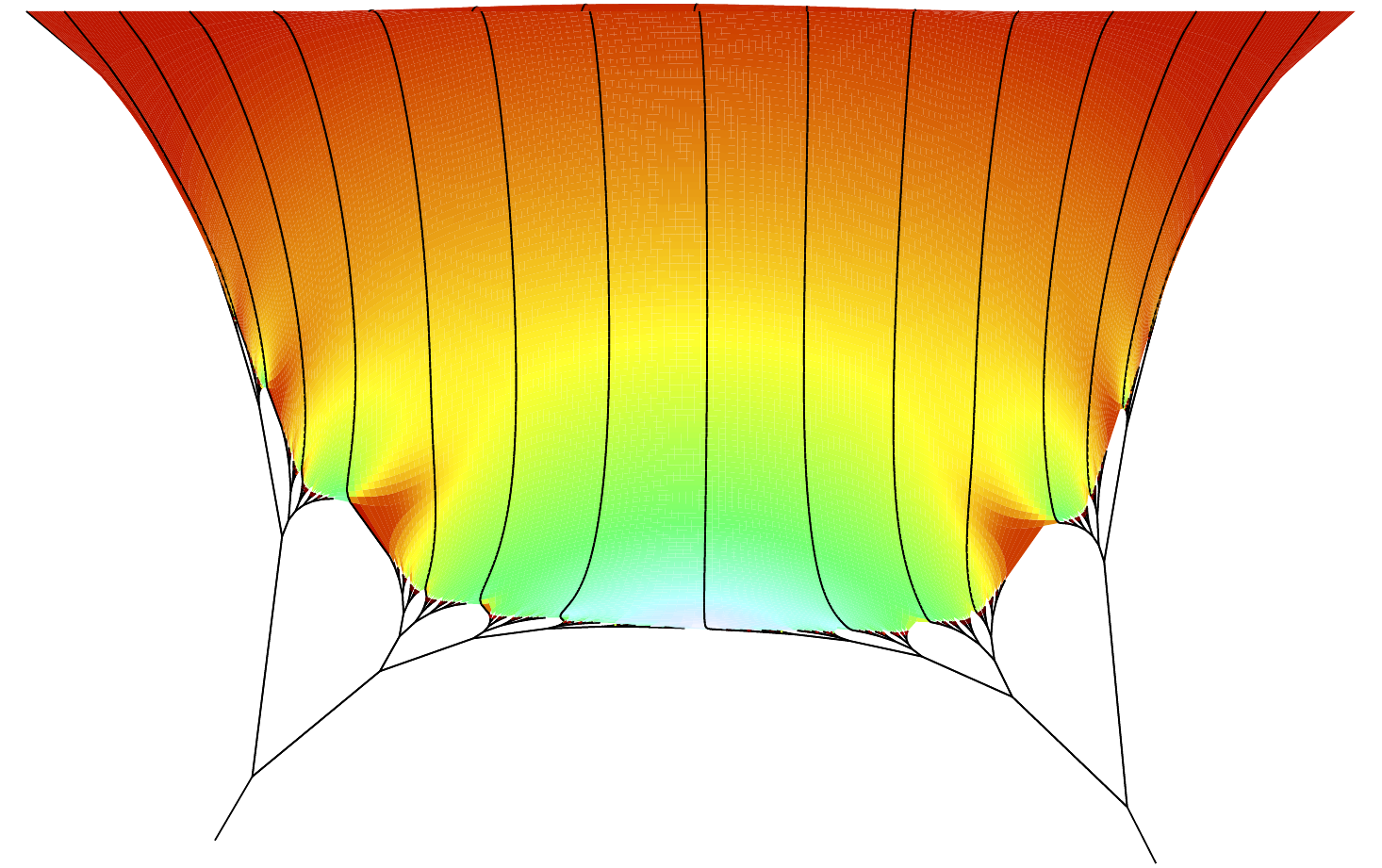} 
&  \hspace{-1.1cm} \raisebox{3.6cm}{ \multirow{2}{*}{ \includegraphics[width=.08\textwidth]{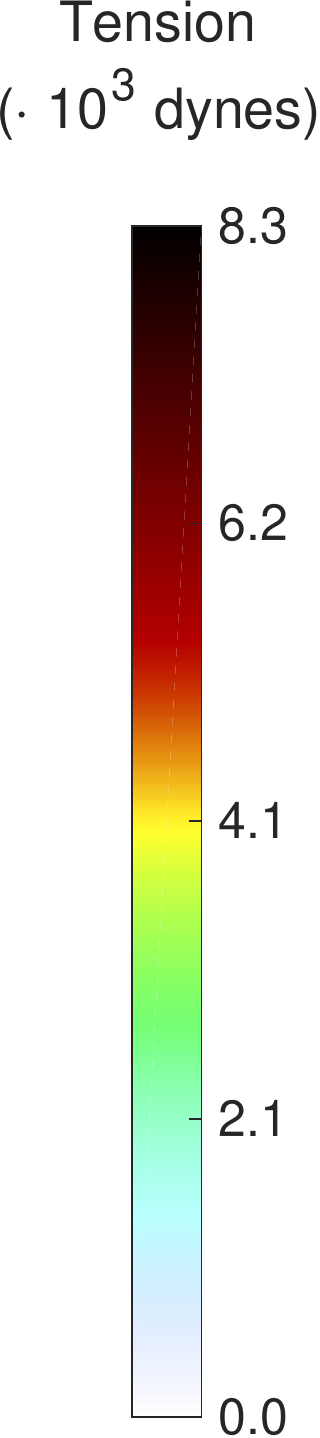} } } \hspace{-.4cm}  \\ 
 \hspace{-0.6cm} \begin{overpic}[width=.477\textwidth]{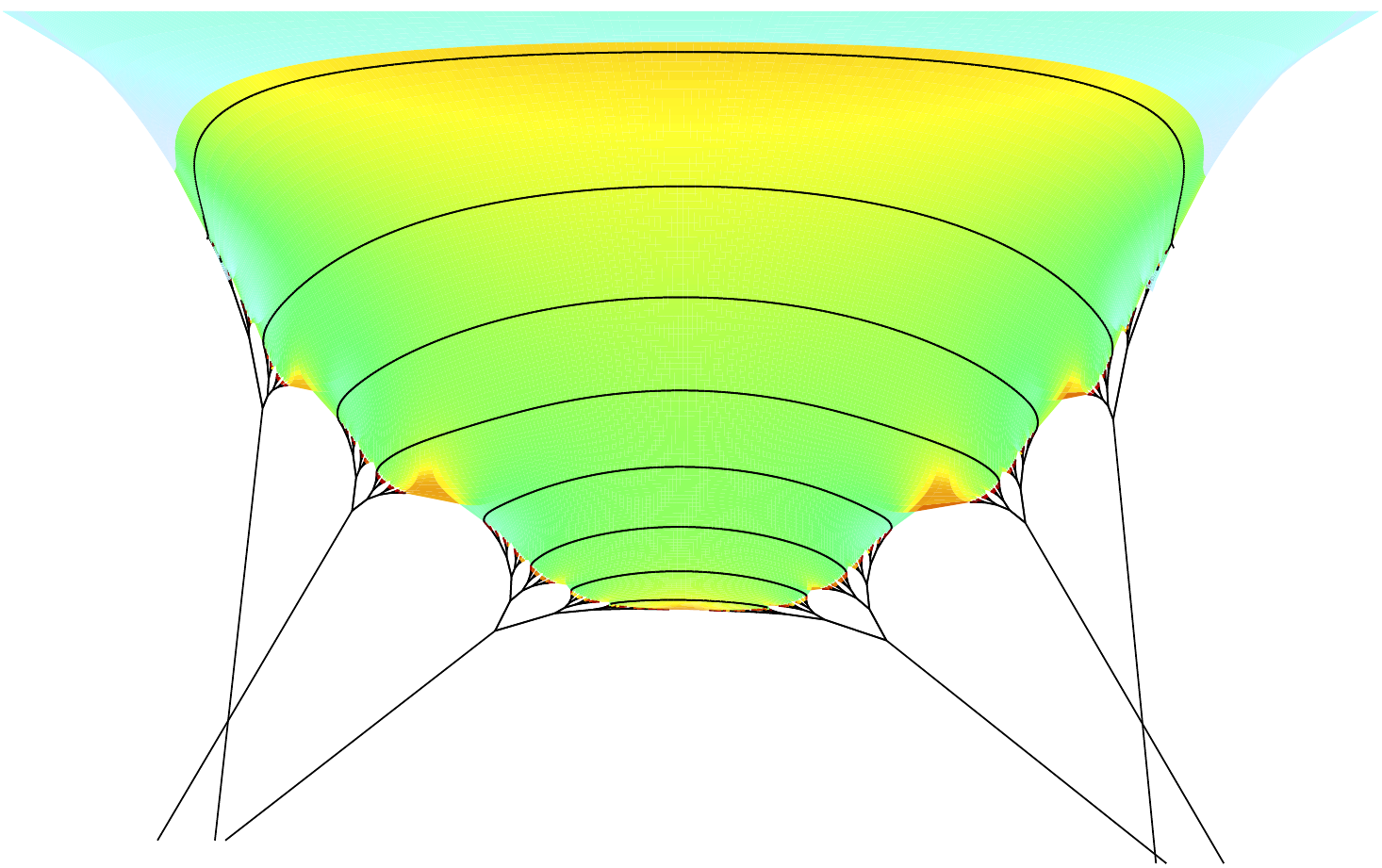}  \put (3,20) {\rotatebox[origin=l]{90}{Posterior leaflet}}\end{overpic} \hspace{-0.4cm}    
&  \includegraphics[width=.477\textwidth]{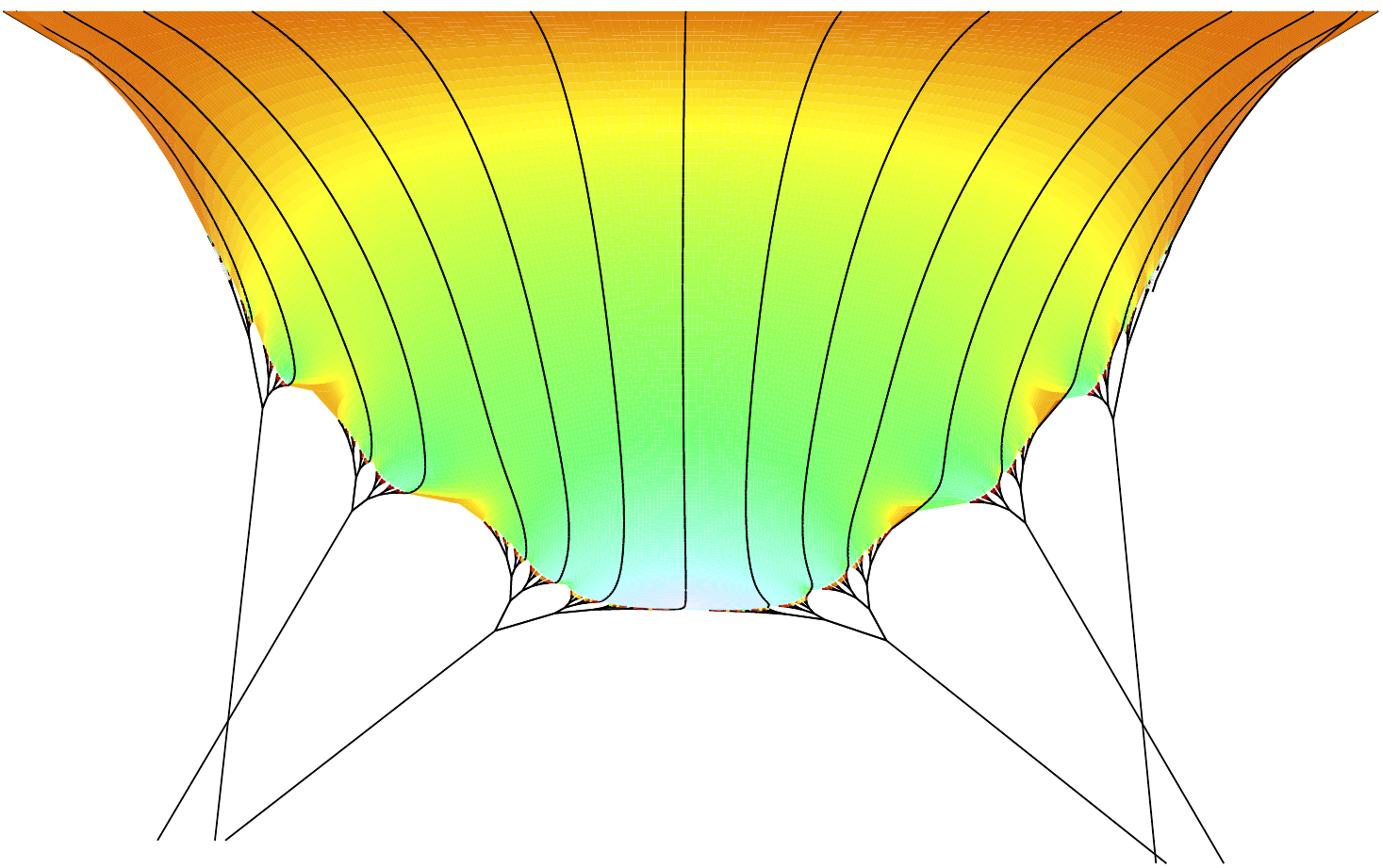} 
&   \\ 
\end{tabular}
\caption{Closed geometry of model valve showing tensions in each fiber family showing the circumferential fiber family (left) and radial fiber family (right) of the anterior leaflet (top row) and posterior leaflet (bottom row). 
The tension scale runs from zero to the maximum of the possible design tension in each family, $\max(\alpha\dv, \beta\du)$.
Fibers in the chordae tendineae that exceed the maximum of the tension scale are colored black. 
Every sixteenth fiber in the corresponding family in the leaflets is plotted in black as a visual cue. 
$p = 100$ mmHg; 
$N=512$. 
}
\label{one_family_tensions}
\end{figure*}

The equilibrium equations for the leaflets and chordae form a nonlinear system of difference equations. 
We solve this system using Newton's method with line search; see Section \ref{Numerical-method}.

The geometry of the leaflets and chordae in the closed, pressurized configuration of the valve emerges from the solution to the equilibrium equations. 
Figure \ref{static_valves} shows the anterior and posterior leaflets colored by the sum of the tensions in the two fiber families, $S\dv + T\du$. 
The leaflets are solved for simultaneously, but plotted separately so that both are visible. 

Note that the equilibrium equations, \eqref{equilbrium_eqn_discrete}, do not include any notion of contact (i.e., we allow the leaflets to intersect). 
We have found that tuning for a solution in which the leaflets actually intersect near their free edges is helpful in arriving at a design that does not leak.
Then, when the valve is placed in fluid, there is some extra model tissue to let the leaflets coapt. 
Before placing the model valve in fluid, we let it relax to a partially open configuration with no leaflet overlap, as described in Section  \ref{general_model}.  
Subsequent leaflet overlap during fluid-structure interaction is prevented by the immersed boundary method, see \cite{doi:10.1137/070699780}. 

Figure \ref{tree_detail} shows detail at the free edge where chordae insert into the leaflet. 
The vertices of the free edge, at which the mesh has a staircase shape, form approximately a smooth curve. 
The chordae form nested arches, inheriting a recursive structure from the trees.

\begin{figure}[ht]
\centering
\includegraphics[width=.6\columnwidth]{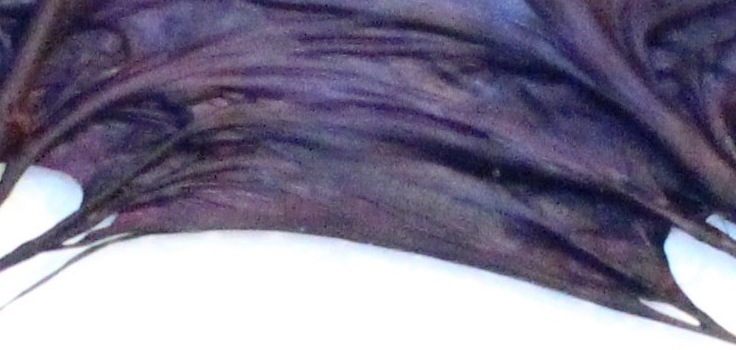} 
\includegraphics[width=.6\columnwidth]{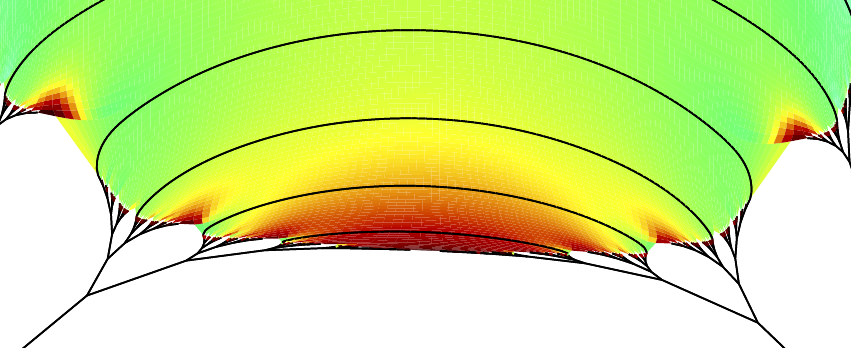} 
\caption{Detail on free edge of the anterior leaflet in a real valve from Figure \ref{valve_labeled}, (top) and detail of circumferential tension in the model valve as in Figure \ref{one_family_tensions} (bottom) with the same color scale. 
Every sixteenth fiber in the circumferential family in the leaflets is plotted in black as a visual cue. 
$N=512$.}
\label{free_edge_comparison}
\end{figure}

In Figure \ref{one_family_tensions}, each fiber family is plotted separately and color coded with respect to its tension. 
Tensions in the chordae typically exceed the maximum allowed leaflet fiber tension, and wherever this happens the chordae are colored black.
The emergent tension is highly heterogeneous. 
At the center of the free edge of the anterior leaflet, there is high circumferential tension and low radial tension. 
On real valves, this region has thick, circumferential fibers that are visible to the eye. 
A comparison of this region in a real valve to the corresponding region of the model valve is shown in Figure \ref{free_edge_comparison}.
Radial tension is generally lower near the free edge, and increases towards the valve ring in both leaflets. 
Circumferential tension is highest at the anterior free edge, and generally lower moving towards the valve ring.
The rings, which are topologically circles, closest to the valve ring and not directly supported by chordae, have generally lower circumferential tension. 
Both radial and circumferential families in both leaflets have isolated regions of higher tension at and around the chordal attachments. 
In the real valve, this region is rough and thick, and could feasibly support tensions that are heterogeneous in magnitude and direction. 
A model that prescribes uniform material properties could not capture this behavior.

\begin{figure}[bht]
\centering 
\includegraphics[width=.6\columnwidth]{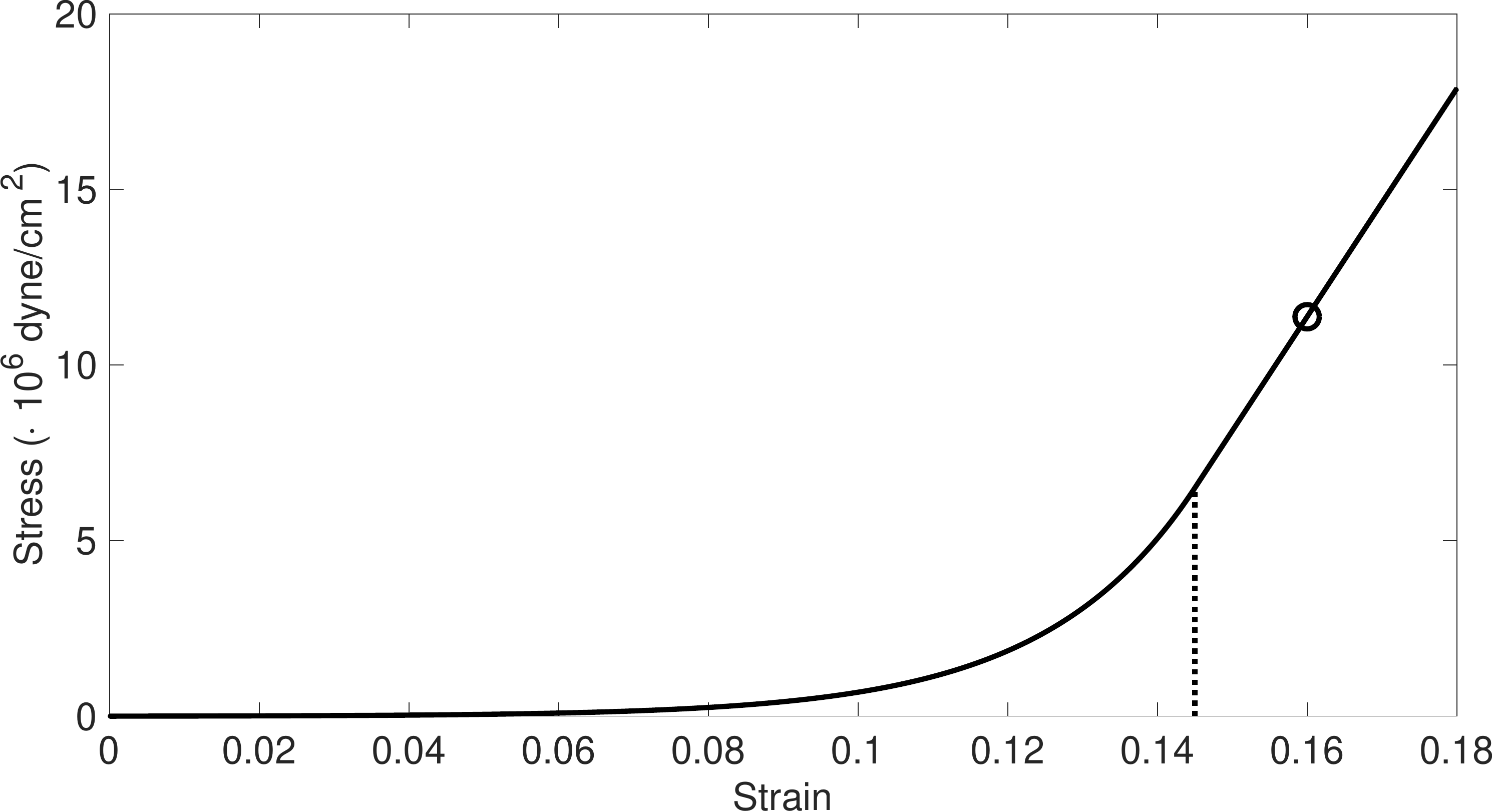}
\caption{Experimental stress-strain relation, based on \cite{sacks_collagen_constitutive}, Figure 3a. 
Fibers achieve \emph{full recruitment} strains above the dashed vertical line at $E = E^{*} = .145$. 
Uniform strains of $E_{0} = .16$ are used to compute material parameters for the constitutive law; this point on the curve is indicated with a circle.}
\label{collagen_curve.eps}
\end{figure} 
 
\section{An Elasticity Model for Use in Fluid}
\label{general_model}

We now use the tensions and geometry of the pressurized model to assign a physical constitutive law to the model valve.

The experiments reported in \cite{sacks_collagen_constitutive} motivate the following strain-tension relation for the elastic links of the model valve:
\begin{numcases}{ \tau(E) = }
0                           & $:  E < 0$  \label{constitutive_summary} \\
 \kappa (e^{\lambda E} - 1) & $:  0 \leq E < E^{*}$ \nonumber \\ 
 \kappa \big( ( e^{ \lambda E^{*}} - 1)     + \lambda e^{ \lambda E^{*}} (E - E^{*}) \big) & $:  E^{*} \leq E$ , \nonumber 
\end{numcases}
in which $\tau$ is the tension and $E$ is the strain, i.e., $E = L/R - 1$, where $L$ is the length of the link and $R$ is its rest length. 
The function $\tau(E)$ is plotted in Figure \ref{collagen_curve.eps}.  
This functional form has been widely used for modeling soft tissue, see for example \cite{FAN20142043, quapp1998material}.   
The tension is zero for negative strains, so the links of our model valve do not resist compression.
For strains between 0 and $E^{*}$ = 0.145, the tension is proportional to $(e^{\lambda E} - 1)$, so that the tension is zero at zero strain, and has an exponentially increasing slope that has been attributed to the gradual recruitment of wavy collagen fibers that straighten as strain increases \cite{sacks_collagen_constitutive}.  
For $E > E^{*}$, collagen is fully recruited (i.e., straight), and the tension becomes a linear function of the strain.  
Equation \eqref{constitutive_summary} is written such that the slope of the strain-tension relation is automatically continuous at $E=E^{*}$.
The parameter $\kappa$ simply scales the tension; we choose it and the rest length differently for every link in the manner described below.
The parameter $\lambda$ is dimensionless and affects the shape of the strain-tension curve.  
We use $\lambda =  49.96$, which fits well to the data in \cite{sacks_collagen_constitutive}, in the sense that for all $\kappa$ the curve is approximately a scalar multiple of their experimental shape.  
This constitutive law is phenomenological, simple and effective. 
It is applied to all links in all fibers in the model valve.

Having solved the design problem as described above (Section \ref{static}), we have both the length $L$ and the tension $\tau$ for every link in our model valve, both within the leaflets and also within the trees of chordae, in the fully loaded, closed configuration of the model valve.
To complete the model, we need only choose the rest length $R$ and also the tension parameter $\kappa$ for each link. 
To do so, we make the assumption that all of the links of the model have the same strain $E_0 = 0.16$ in this configuration.
This value was chosen arbitrarily to be slightly larger than $E^{*}$, representing strain at which collagen fibers are fully recruited. 
Then we solve $E_0 = L/R -1$ for $R$, and we choose $\kappa$ so that the link has the tension $\tau$ when $E=E_0$, see equation \eqref{constitutive_summary}.  
Note that we use the last line of equation \eqref{constitutive_summary} for this purpose, since $E_0 > E^{*}$.

Finally, having chosen a constitutive law and assigned its parameters, we solve a second equilibrium problem for the unloaded ($p$ = 0) configuration of the model valve with its constitutive law governed by equation \eqref{constitutive_summary}.  
In this unloaded configuration, the leaflets do not intersect, and we use this configuration as an initial condition in our fluid-structure interaction studies.
The model mitral valve turns out to have residual stresses and strains
 in this unloaded configuration, which is used as an initial configuration
 in our fluid-structure interaction studies.  This is consistent with
 experimental evidence that the in situ mitral valve has residual stresses
 and strains throughout the cardiac cycle, relative to a reference
 configuration provided by an excised specimen \cite{Amini2012,Rausch2013}.

In the model, the fully loaded strain of $E_0$ = 0.16 is defined in relation to the point of zero stress. 
This would be hard to identify experimentally, since the stress-strain curve is so flat where the stress is zero, see Figure \ref{collagen_curve.eps}.  
Thus, for comparison with experiment it may be better to extrapolate the linear part of the stress-strain curve to zero stress to get a reference point.  
In our case, this reference strain is 0.125.  
Using this as a reference gives the following recomputed strain for the fully loaded state: (0.16 - 0.125)/(1.125) = 0.031.
In some experimental studies, values of strain are reported using the valve configuration at minimum left ventricular pressure as the reference configuration, but in others an excised valve is used as the reference configuration. 
This difference dramatically affects the values of strain reported, since there is residual strain in vivo, but not in the excised specimen.
These issues are explored further in \cite{KRISHNAMURTHY20091909,Amini2012,Rausch2013}.

\section{Fluid-Structure Interaction}
\label{Fluid-Structure_Interaction}

The immersed boundary formulation \cite{ib_acta_numerica} of fluid-structure interaction is briefly described as follows. 
Fixed Cartesian coordinates $\bb x$ and time $t$ are used as independent variables for the fluid velocity field $\bb u(\bb x,t)$ and the pressure field $p(\bb x,t)$. 
The structure is described in terms of material coordinates previously called $u,v$, but which we now describe collectively by $\bb s$ to avoid confusion with fluid velocity. 
Thus $\bb x = \bb X(\bb s,t)$ at some fixed time $t$ gives the spatial configuration of the structure at that time. 
A patch $d\bb s = dudv$ of the structure applies a force $\bb F(\bb s,t)d\bb s$ to the surrounding fluid. 
This produces on the fluid a force per unit volume that we denote by $\bb f(\bb x,t)$. 
The force density $\bb f (\bb x, t)$ is singular: it is zero everywhere except at the location of the structure and infinite there, but in such a way that its integral over any finite volume of fluid is finite. 
The Dirac delta function (see below) provides a mathematical tool for the representation of such a singular force field.

The equations are:
\begin{align}
\rho \left( \frac{ \p \bb u}{\p t}  + \bb u \cdot \grad \bb u \right) &= - \grad p + \mu \Delta \bb u  + \bb f   \label{momentum} \\
\grad \cdot \bb u  &= 0   		\label{mass}		\\
\bb F( \, \cdot \, , t) &=  \mathcal S(\bb X( \, \cdot \, ,t))  \label{nonlinear_force}  \\    
\frac{ \p \bb X(\bb s,t)}{\p t}&=  \bb u(\bb X(\bb s,t), t) 	 \label{interpolate} 	 \\
		&= \int \bb u(\bb x, t)   \delta (  \bb x  - \bb X(\bb s,t) )  \;  d  \bb x   \nonumber  \\ 
\bb f(\bb x, t)   &= \int  \bb F(\bb s,t)  \delta(  \bb x - \bb X(\bb s,t)  )   \; d\bb s          \label{spreading} 
\end{align}

In equations \eqref{momentum}-\eqref{mass}, the variables $\bb u,p, \bb f$ are functions of the fixed Cartesian coordinates $\bb x$ and the time $t$.  
This is called the Eulerian description of a viscous incompressible fluid.  
Equation \eqref{momentum} expresses momentum conservation, and equation \eqref{mass} expresses volume conservation.
In equation \eqref{nonlinear_force}, $\bb F$ and $\bb X$ are functions of the material coordinates $\bb s$ and the time $t$; this is called a Lagrangian description of the immersed boundary.  
The mapping $\mathcal S$ from $\bb X$ to $\bb F$ gives the force density applied to the fluid as a function of the configuration of the model valve; it includes the constitutive law described previously.  
The blank arguments indicate that $\mathcal S$ takes the whole function $\bb X$ at time $t$ as input and produces the function $\bb F$ at time $t$ as output.  
Equations \eqref{interpolate} and \eqref{spreading} are interaction equations; they couple the Eulerian and Lagrangian descriptions to each other through convolutions with the Dirac delta function.  
The regularized delta function for all simulations is the 5-point delta function derived in \cite{IB5_arxiv}.

\begin{figure}[t!]
\centering
\includegraphics[width=.6\columnwidth]{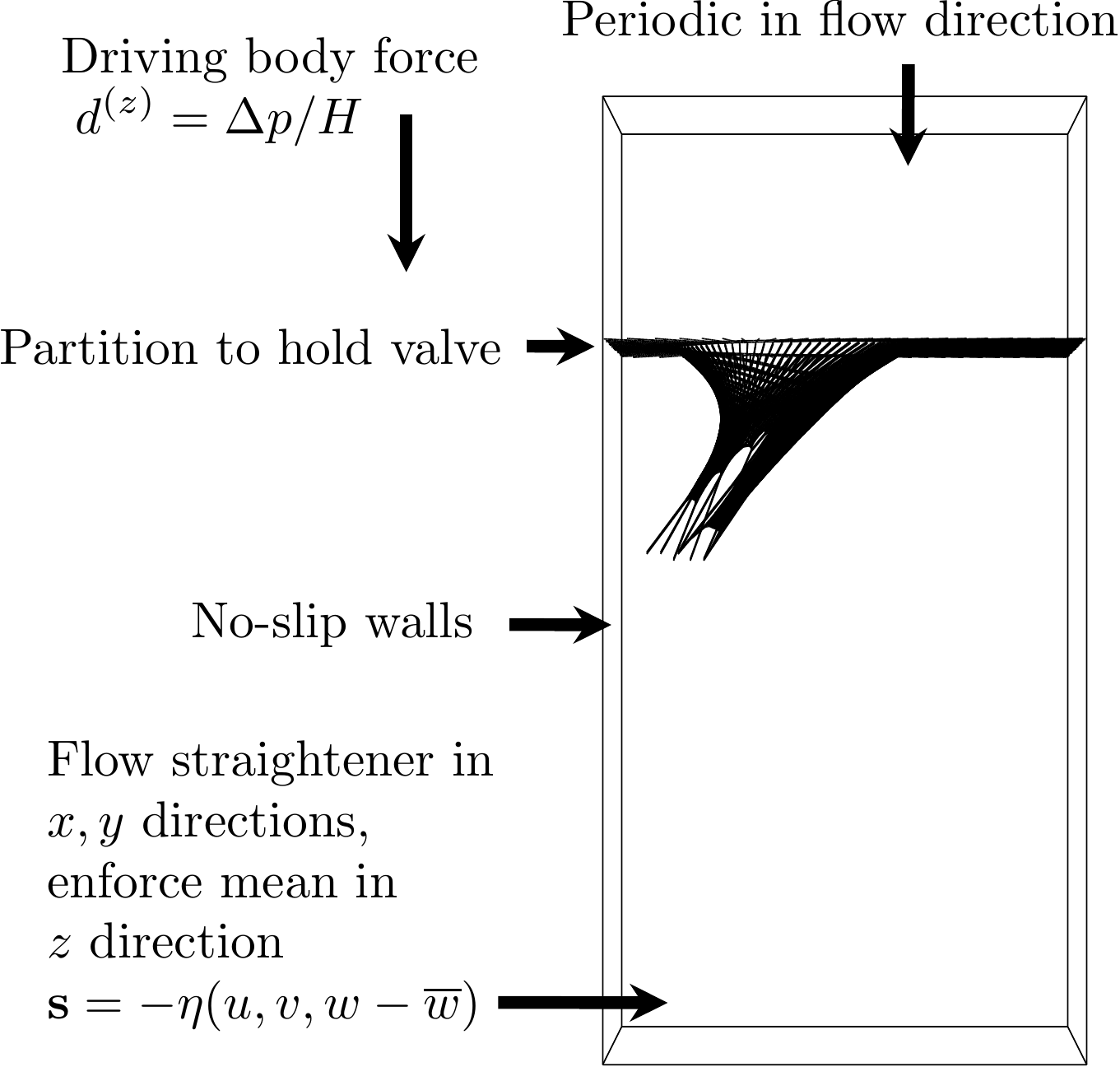}
\caption{Schematic of valve in computer test chamber. }
\label{simulation_schematic.pdf}
\end{figure}

To run simulations, the model valve is placed in a computer test chamber and mounted to a model partition. 
The test chamber is a rectangularly shaped box. 
This is meant to serve as a location to test the model's function and observe associated flows, and is is not meant to model a heart chamber.\footnote{Test chambers that are not meant to model the heart are regularly used in vitro studies of the isolated mitral valve or mitral valve prosthetics. For an example see \cite{Rabbah2013, IMBRIEMOORE2019}.}
The domain is taken to be periodic. 
A flow straightener, applied approximately with a penalty method force, serves to hide periodic effects from re-entering at the top of the chamber. 
To drive the flow, we prescribe a pressure difference across the chamber. 
We consider the pressure to be positive when the upstream (atrial) pressure is greater than the downstream (ventricular) pressure. 
A schematic of this is shown in Figure \ref{simulation_schematic.pdf}; details are provided in Sections \ref{setup} - \ref{layers}. 
All simulations are run with the software library IBAMR \cite{IBAMR, griffith2010parallel} using a staggered-grid discretization.

\subsection{Driving pressures}
\label{driving_pressures}

\begin{figure}[t]
\centering
\includegraphics[width=\columnwidth]{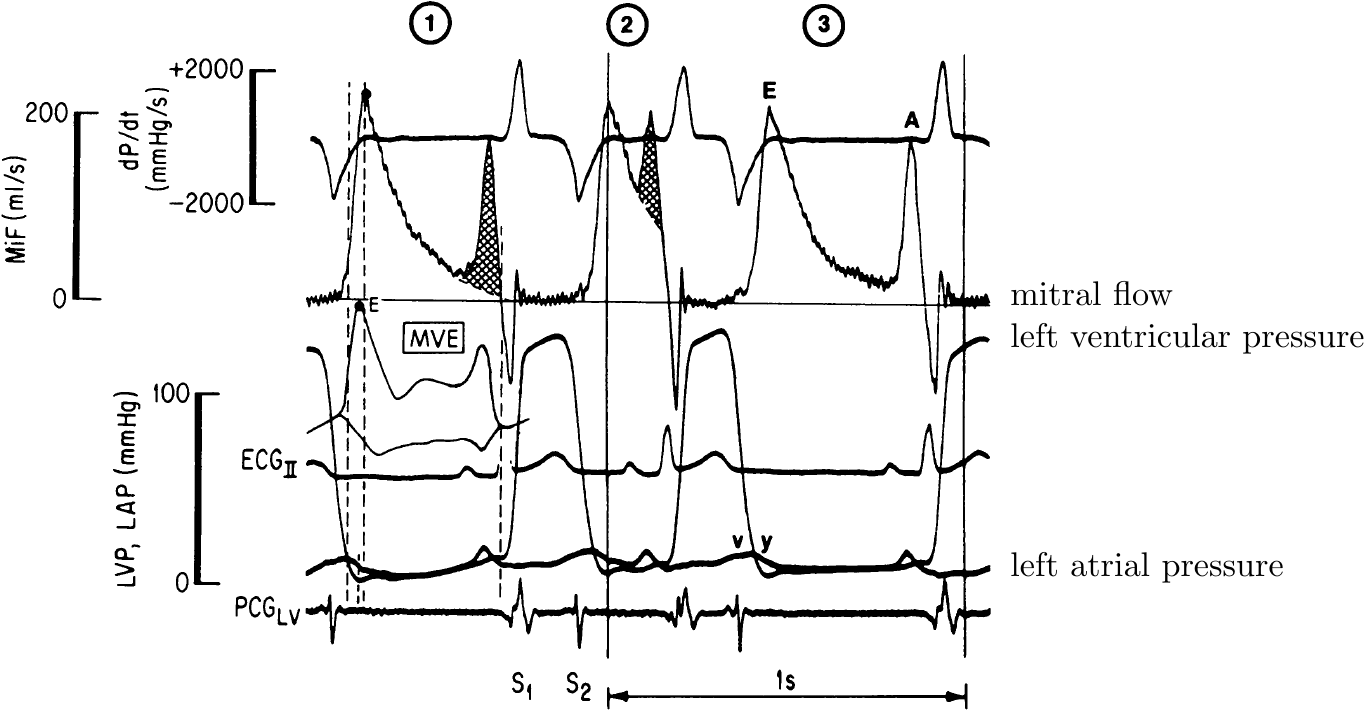}
\caption{Experimental records of mitral valve flow and surrounding pressures from \cite{yellin_book}. 
Additional labels for curves representing mitral flow, left ventricular pressure and left atrial pressure were added to the right of the original diagram.
Reprinted by permission from Springer-Verlag, 
\emph{Dynamics of Left Ventricular Filling} by Edward L. Yellin in 
\href{https://www.springer.com/us/book/9784431680208}{Cardiac Mechanics and Function in the Normal and Diseased Heart}, editors Hori et al., 1989. 
}
\label{yellin_dog_flow_stats}
\end{figure}

To drive simulations, we use experimental records of atrial and ventricular pressure, taken from \cite{yellin_book} and shown in Figure \ref{yellin_dog_flow_stats}. 
In these experiments, dogs were implanted with probes to measure left-heart function, including flow through the mitral ring, and left atrial and left ventricular pressures. 
These measurements are similar to those found in standard medical references on cardiovascular physiology \cite{mohrman2010cardiovascular}.

Active left ventricular filling, the second phase of diastole after isovolumic relaxation, starts at the leftmost dashed line. 
The left ventricular and left atrial pressures equalize and the mitral valve begins to open.
Following this time, the ventricular pressure continues to fall rapidly. 
This creates a transient forward pressure difference, and causes mitral flow to increase rapidly. 
Then for an extended period there is only a small forward pressure difference, and mitral flow correspondingly decreases. 
Just before the third dashed line, there is a bump in pressure caused by atrial systole, creating a second increase in mitral flow. 
Ventricular systole starts at the third dashed line. 
Ventricular pressure rapidly rises, and a large negative pressure difference occurs across the mitral valve. 
There is a brief, large spike of reverse flow through the mitral ring, representing fluid displaced through the mitral ring by the closure of the valve leaflets.
The volume loss represented by this transient is largely recovered later, during the movement of the closed valve towards the left ventricle as the valve unloads prior to opening.
Immediately following the closure transient, there is an oscillation in flow that is associated with the S$_{1}$ heart sound.  
This oscillation decays rapidly, and mitral flow stabilizes near zero. 
When the negative pressure difference across the valve begins to decrease in magnitude, the valve unloads and there is a small shoulder of forward flow through the mitral ring.

The negative pressure difference during ventricular systole, over 100 mmHg, is an order of magnitude larger than the peak forward pressure difference that occurs early in diastole, about 10 mmHg, and two orders of magnitude larger than the forward pressure difference that exists during most of diastole, about 1 mmHg.
Because negative pressure difference is so great, even the slightest failure to fully seal may cause significant regurgitation. 
These are demanding conditions.

From this record, we select the first beat to use in our simulations, as it has a duration of approximately $0.8$ seconds, corresponding to a typical heart rate of 75 beats per minute.
Note that the flow curves are \emph{not} used except for qualitative comparison later.  
To represent each of the two pressure curves we use a finite Fourier series as discussed in Section \ref{driving_pressures_appendix}.
Results are shown in Figure \ref{pressure_and_flow}, lower panel. 
The maximum systolic pressure difference is approximately 116 mmHg.

\subsection{Results -- the model valve in fluid}
\label{control_results}

\begin{figure}[t]
\centering 
\includegraphics[width=.5\columnwidth]{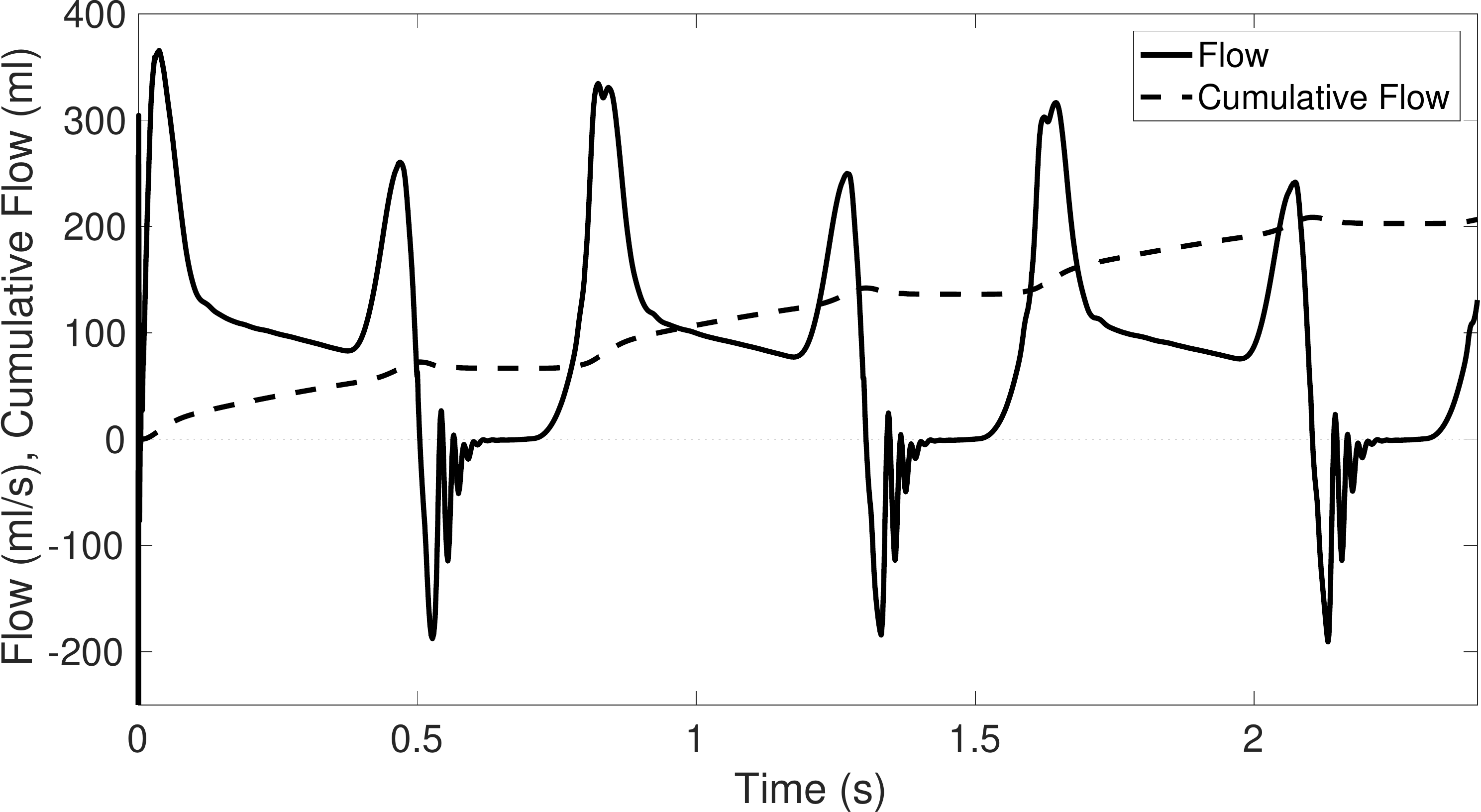}
\includegraphics[width=.5\columnwidth]{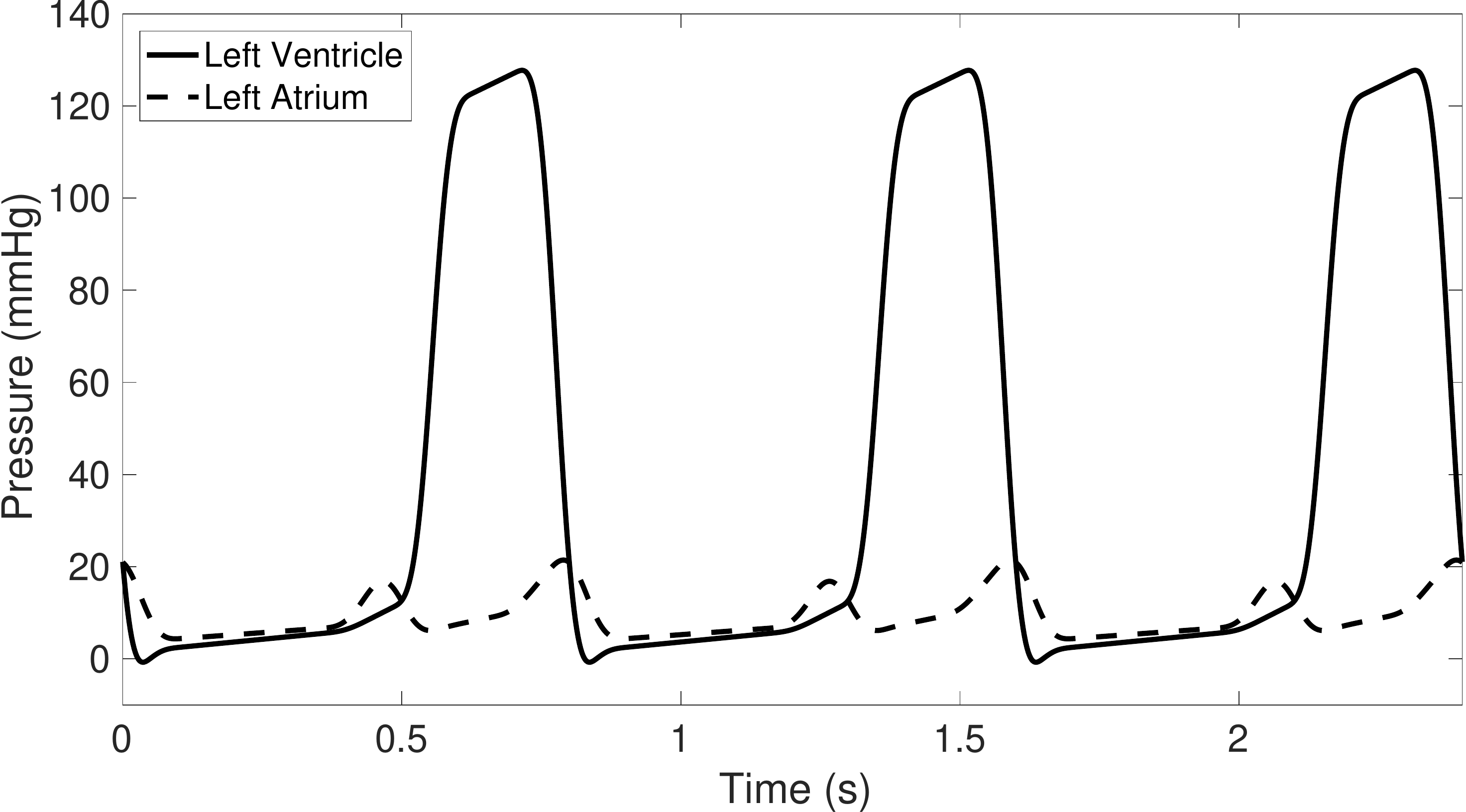}

\caption{Flow through mitral ring as emerges from the simulation (top) when driven by prescribed driving pressures (bottom).}
\label{pressure_and_flow}
\end{figure}

The simulation is driven by the pressures shown in Figure \ref{pressure_and_flow}, bottom panel. 
The top panel shows the emergent mitral flow (ml/s) and cumulative mitral flow (ml), which are output from the simulation. 
See also movies {\bf M1} (real time) and {\bf M2} (slow motion), which show the valve and pathlines of flow in this simulation, and movies {\bf M3} (real time) and {\bf M4} (slow motion), which show the valve and planar slices of the vertical component of velocity in this simulation. 
Movies {\bf M3} and {\bf M4}, as well as all further slice views, were created using the visualization software Visit \cite{VisIt}.

The emergent flow qualitatively matches the experimental flow; many features of the measured flow appear in the simulated flow. 
A rapid inflow occurs as ventricular pressure drops. 
Flow quickly decreases, then more slowly decreases in value until atrial systole. 
Pressure caused by atrial contraction causes a brief increase in the flow, after which the valve begins to close. 
We observe that there is a single large spurt of apparent backflow. 
This is followed by a rapid oscillation, which is quickly damped out. 
This causes the S$_{1}$ heart sound. 
Indeed, we have made a soundtrack directly from the computed flow waveform, and it sounds somewhat realistic. 
(Listen to the audio of the movies {\bf M1} and {\bf M3}.
To make these sounds, we take the flow only during systole, 
interpolate this flow to the standard audio sample rate of 44.1kHz by calling ``spline'' in Matlab \cite{MATLAB}, then call ``audiowrite'' to get a playable audio file.)
Following this oscillation, the valve is fully pressurized and closed. 
Then, there is a slow rise in forward flow as the valve unloads. 
Most of backflow that occurred during closure is recovered by the time the pressure difference becomes positive, see Table \ref{flow_table} and surrounding discussion. 
Finally, the increase in flow becomes rapid and the cycle repeats. 
The mean flow in this simulation is approximately 5.2 L/min, close to the nominal cardiac output of 5.6 L/min. 

\begin{figure*}[t!]
\begin{tabular}{c c c c}
\hspace{-3mm} 
\includegraphics[width=.32\textwidth]{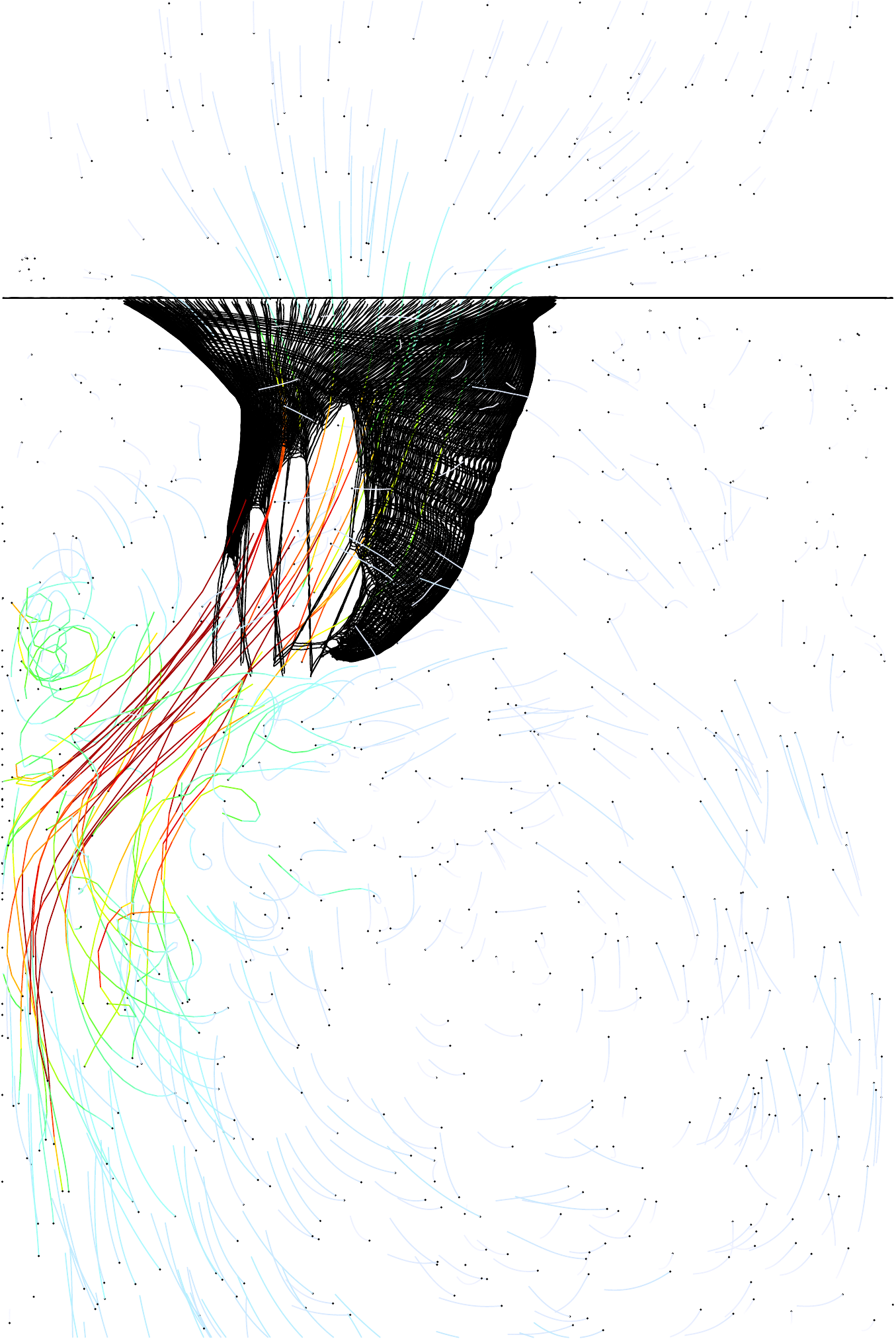}  & 
\hspace{-2mm} 
\includegraphics[width=.32\textwidth]{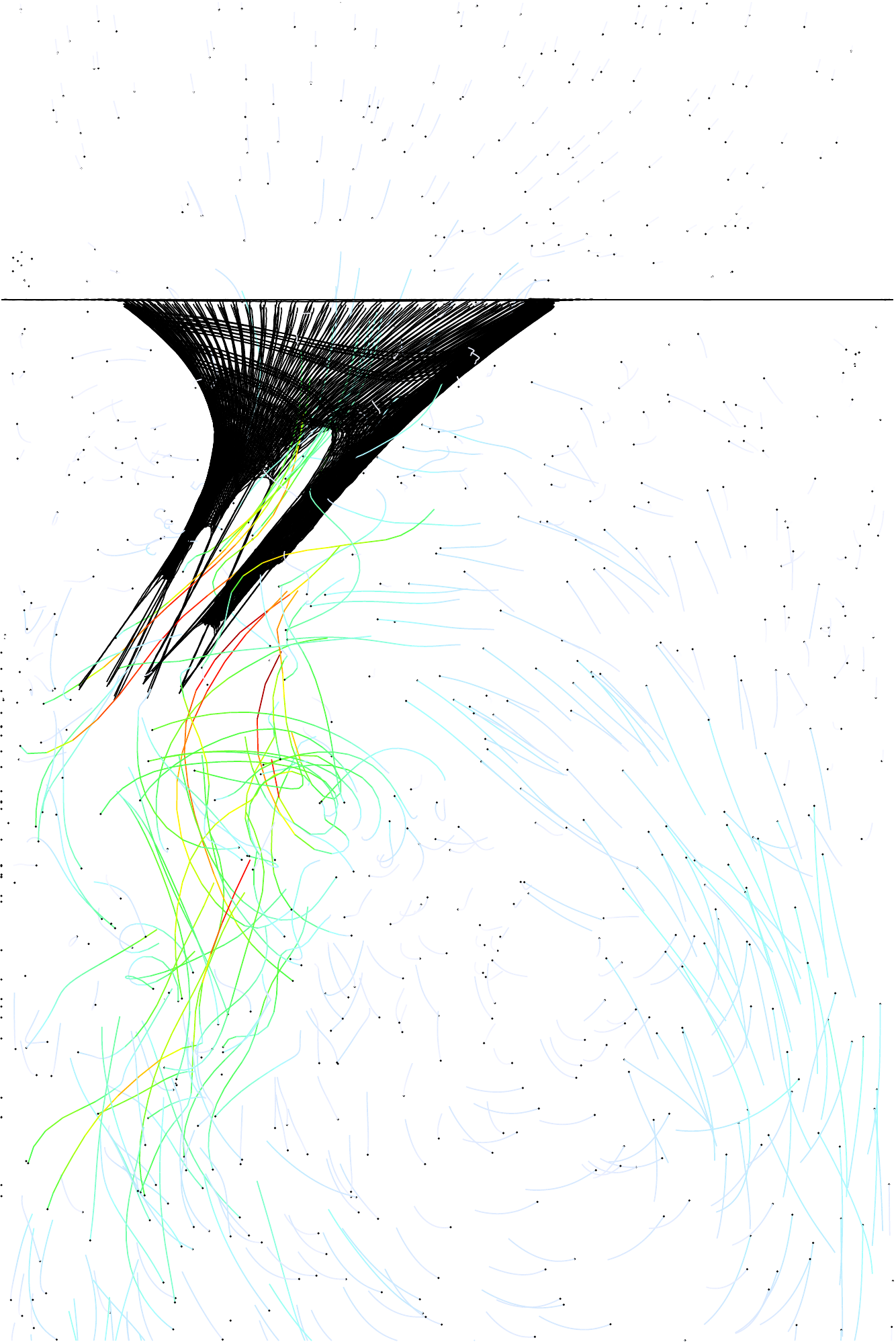}  & 
\hspace{-2mm} 
\includegraphics[width=.32\textwidth]{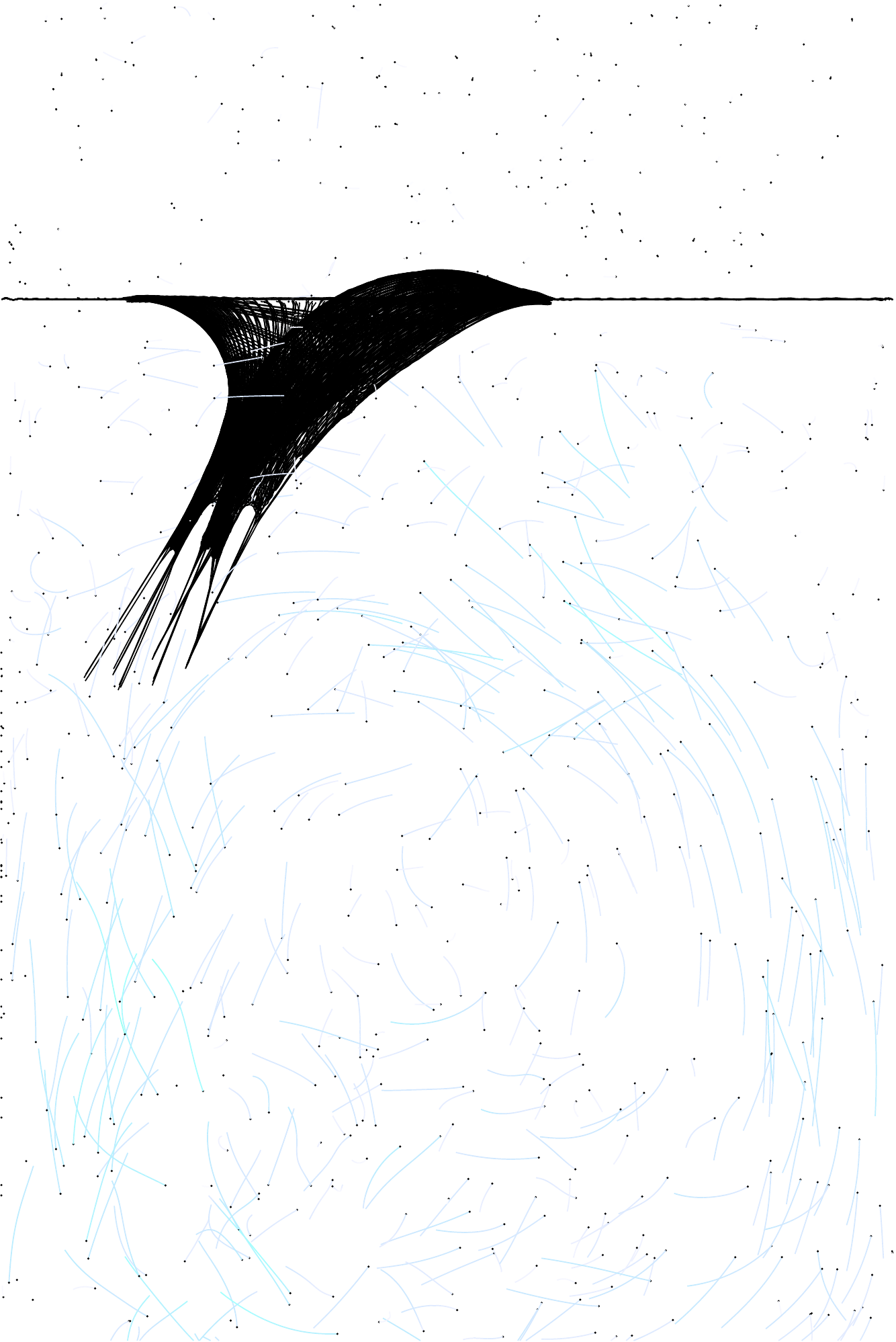}  & 
\raisebox{3.0cm}{ \hspace{-15mm} \includegraphics[width=.045\textwidth]{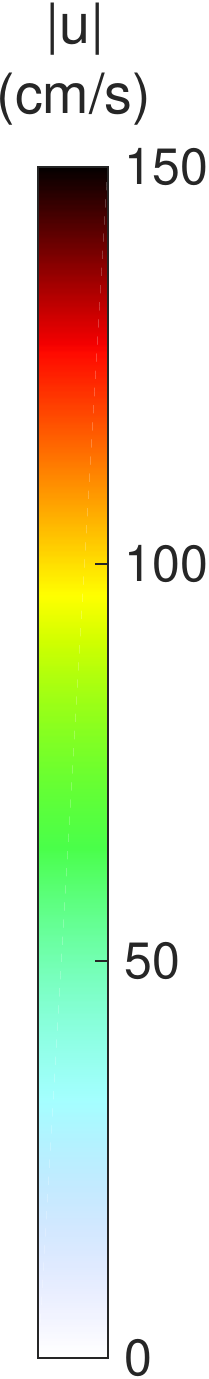} } \\ 
\hspace{-3mm} 
\includegraphics[width=.32\textwidth]{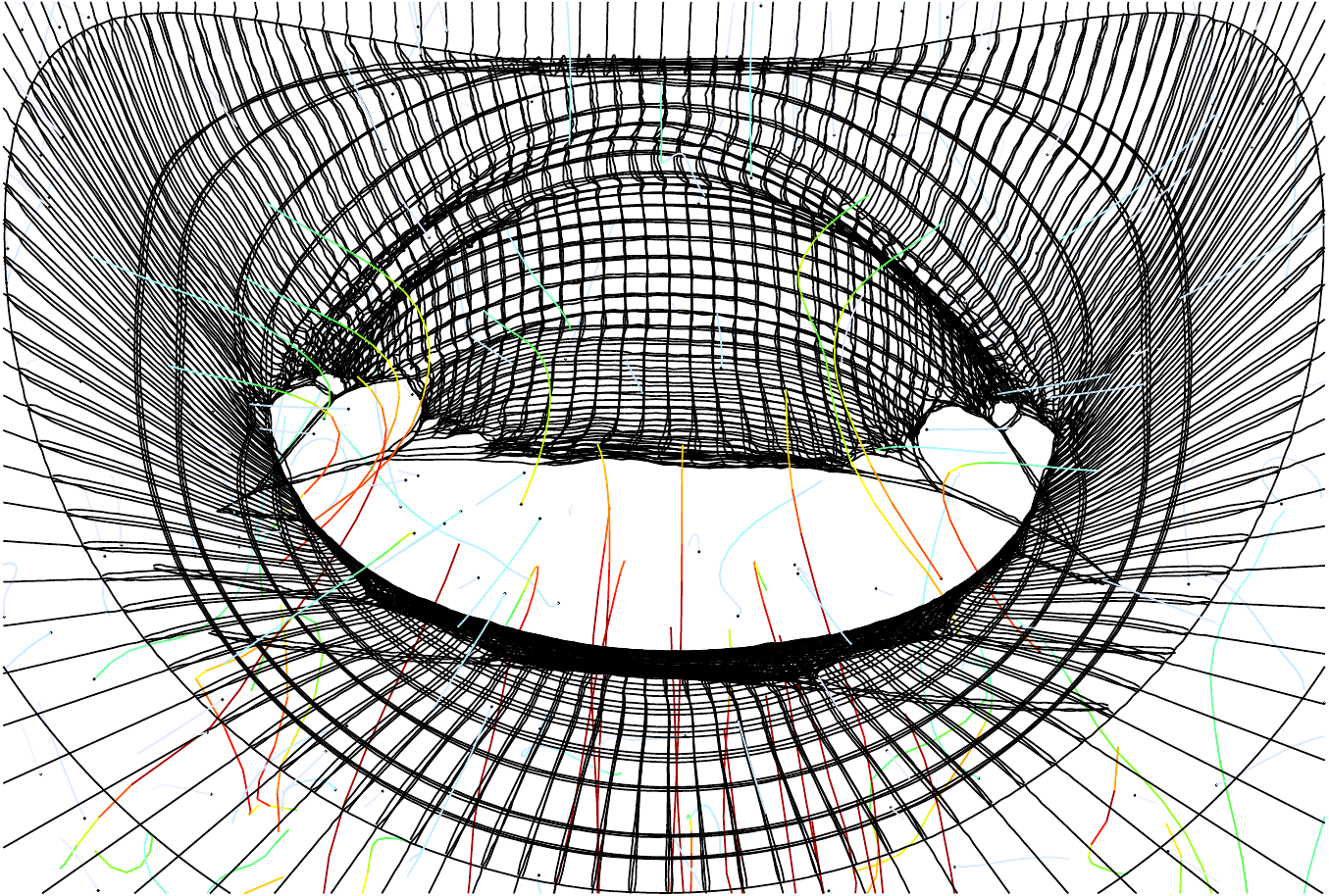}  & 
\hspace{-2mm} 
\includegraphics[width=.32\textwidth]{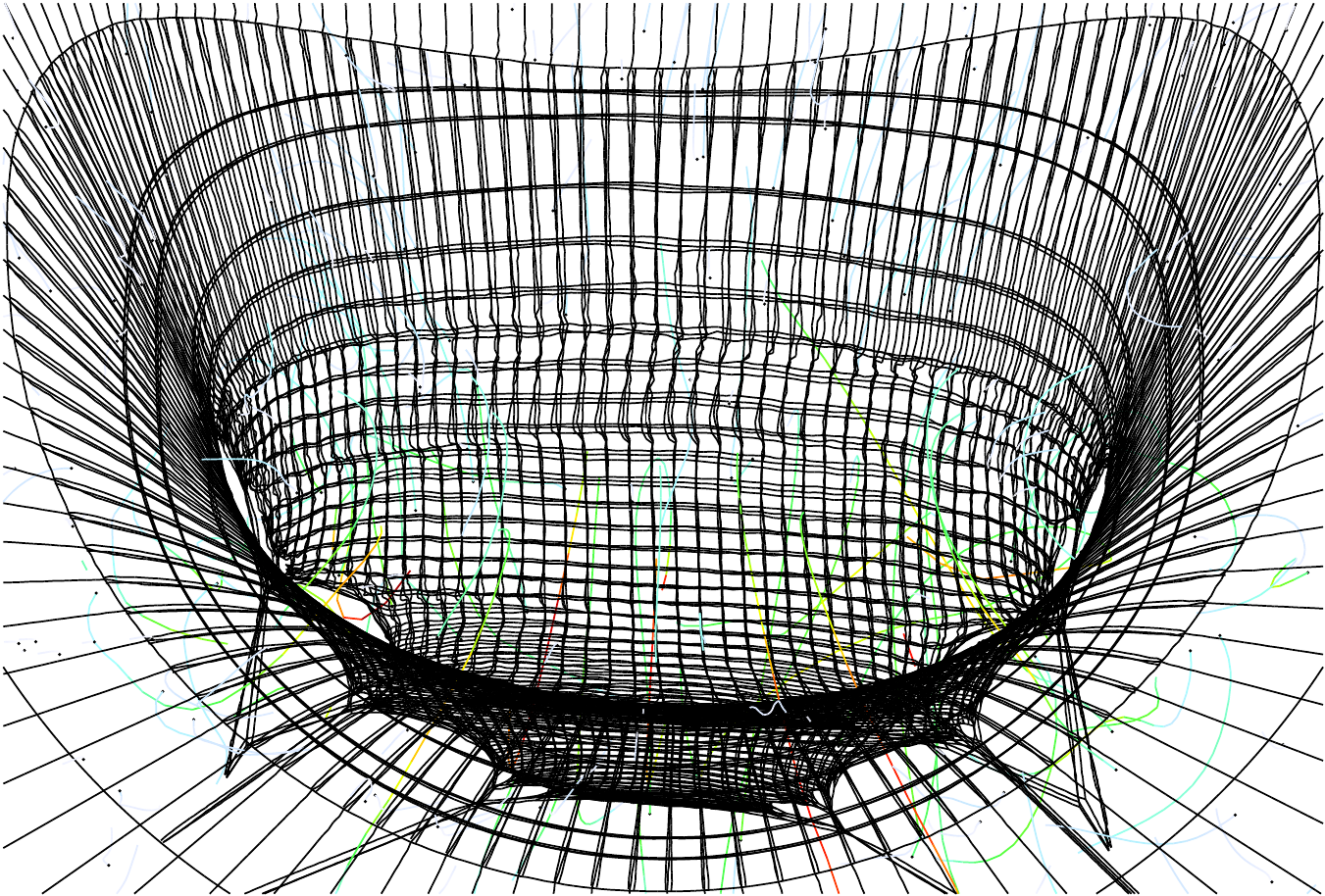}  & 
\hspace{-2mm} 
\includegraphics[width=.32\textwidth]{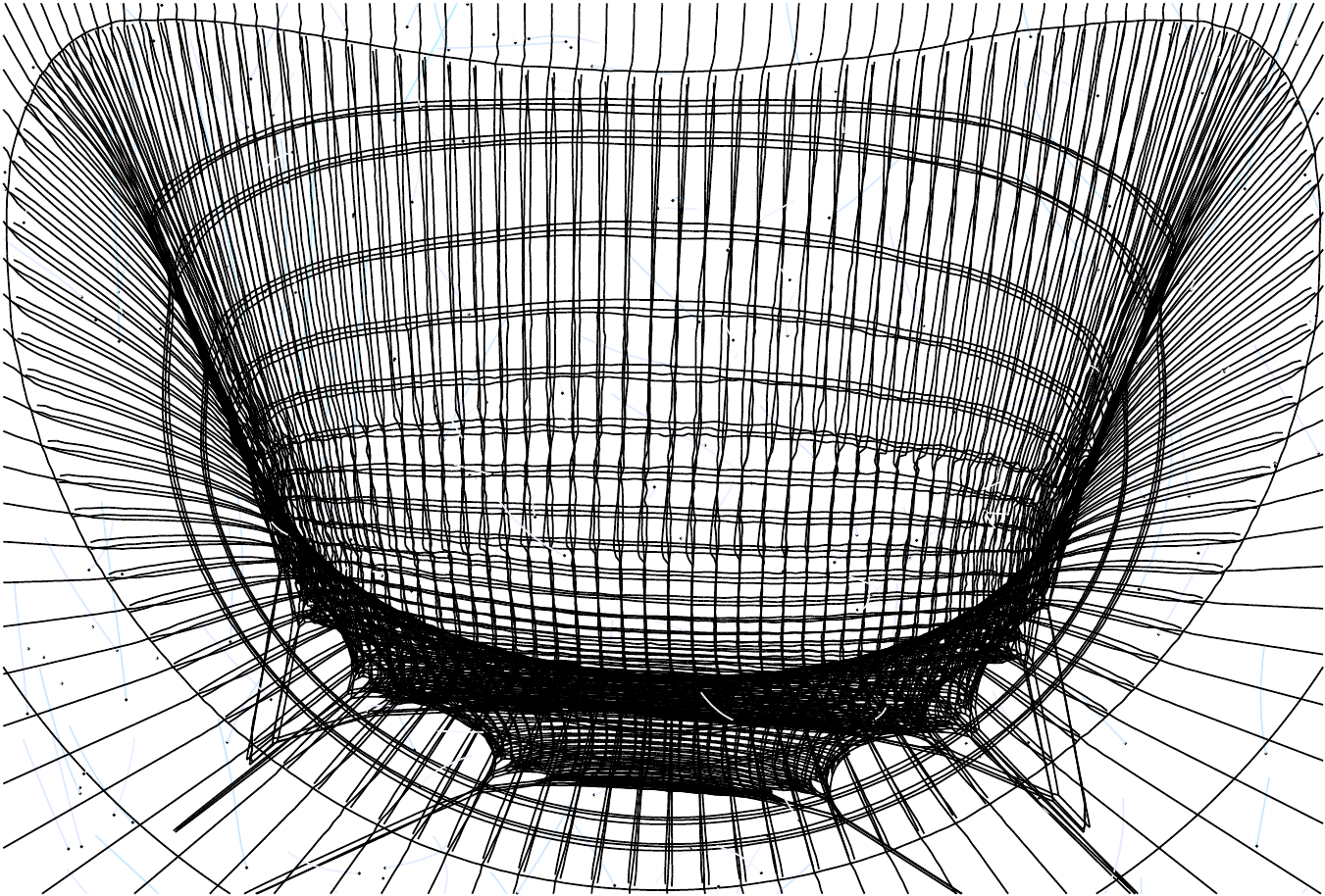}  & 
\end{tabular}
\caption{Views of the mitral valve in simulation showing filling (left), closing (center), and the fully closed position (right) viewed from the side (top row) and above (bottom row). Pathlines are drawn and colored with respect to the velocity magnitude. Every fourth fiber in the leaflets is shown for visual clarity.    $N = 512$.  See movies {\bf M1} (real time) and {\bf M2} for animations of these pathlines. }
\label{three_panels}
\end{figure*}

Three time steps of the simulation are shown in Figure \ref{three_panels}, which includes pathlines of fluid particles colored by velocity magnitude, along with the structure. 
The first frame shows diastole, near peak forward flow in the third beat of the simulation. 
The second shows the valve just after the onset of closure, at which time the leaflets have begun to come together. 
The third shows the valve in its fully closed position. 
Here the leaflets are pressurized and coapted, the chordae tendineae are tight, and there is no high-velocity flow. 

\begin{figure}[p!]  
\setlength{\tabcolsep}{0.5pt}
\centering
\makebox[0pt]{
\begin{tabular}{cc @{\hspace{10pt}} cc}
& 
\begin{tabular}{L{.15\columnwidth}L{.15\columnwidth}L{.15\columnwidth}}
velocity & modified pressure & vorticity \\ 
cm/s & mmHg & $s^{-1}$ \\ 
\includegraphics[width=.09\columnwidth]{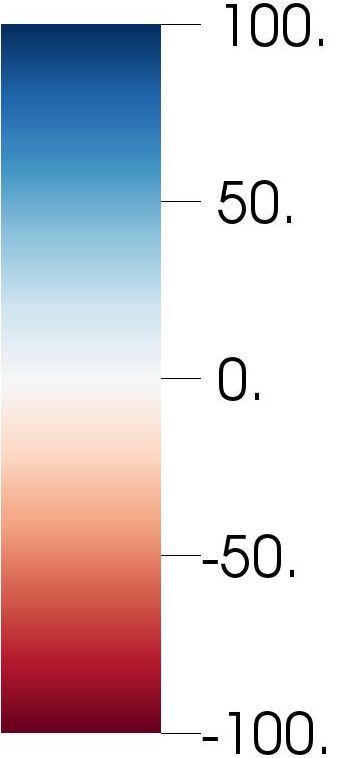} & 
\includegraphics[width=.09\columnwidth]{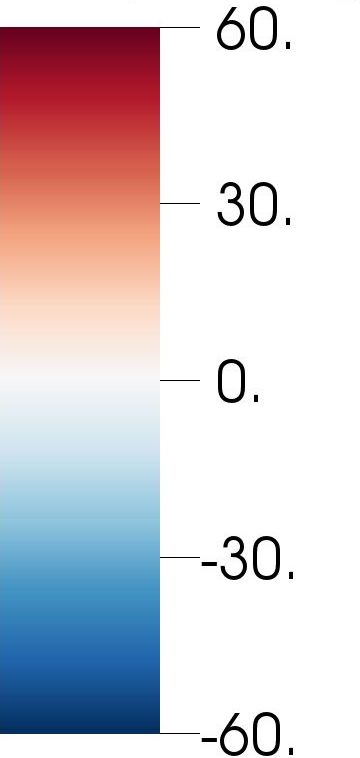} & 
\includegraphics[width=.09\columnwidth]{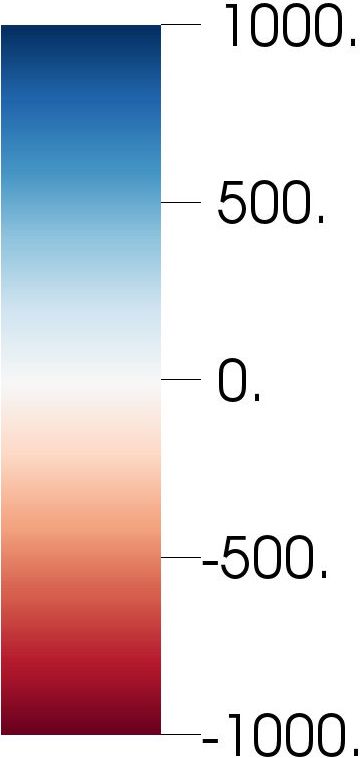} 
\end{tabular} 
 & & 
\begin{tabular}{L{.15\columnwidth}L{.15\columnwidth}L{.15\columnwidth}}
velocity & modified pressure & vorticity \\ 
cm/s & mmHg & s$^{-1}$ \\ 
\includegraphics[width=.09\columnwidth]{colorbar_velocity_kick_compare.jpeg} & 
\includegraphics[width=.09\columnwidth]{colorbar_pressure_basic.jpeg} & 
\includegraphics[width=.09\columnwidth]{colorbar_vorticity_kick_compare.jpeg} 
\end{tabular} 
\\ 
& & \\
\rotatebox[origin=l]{90}{ \parbox{4cm}{ $t = 1.675 $ s, early filling  } } & 
\includegraphics[width=.45\columnwidth]{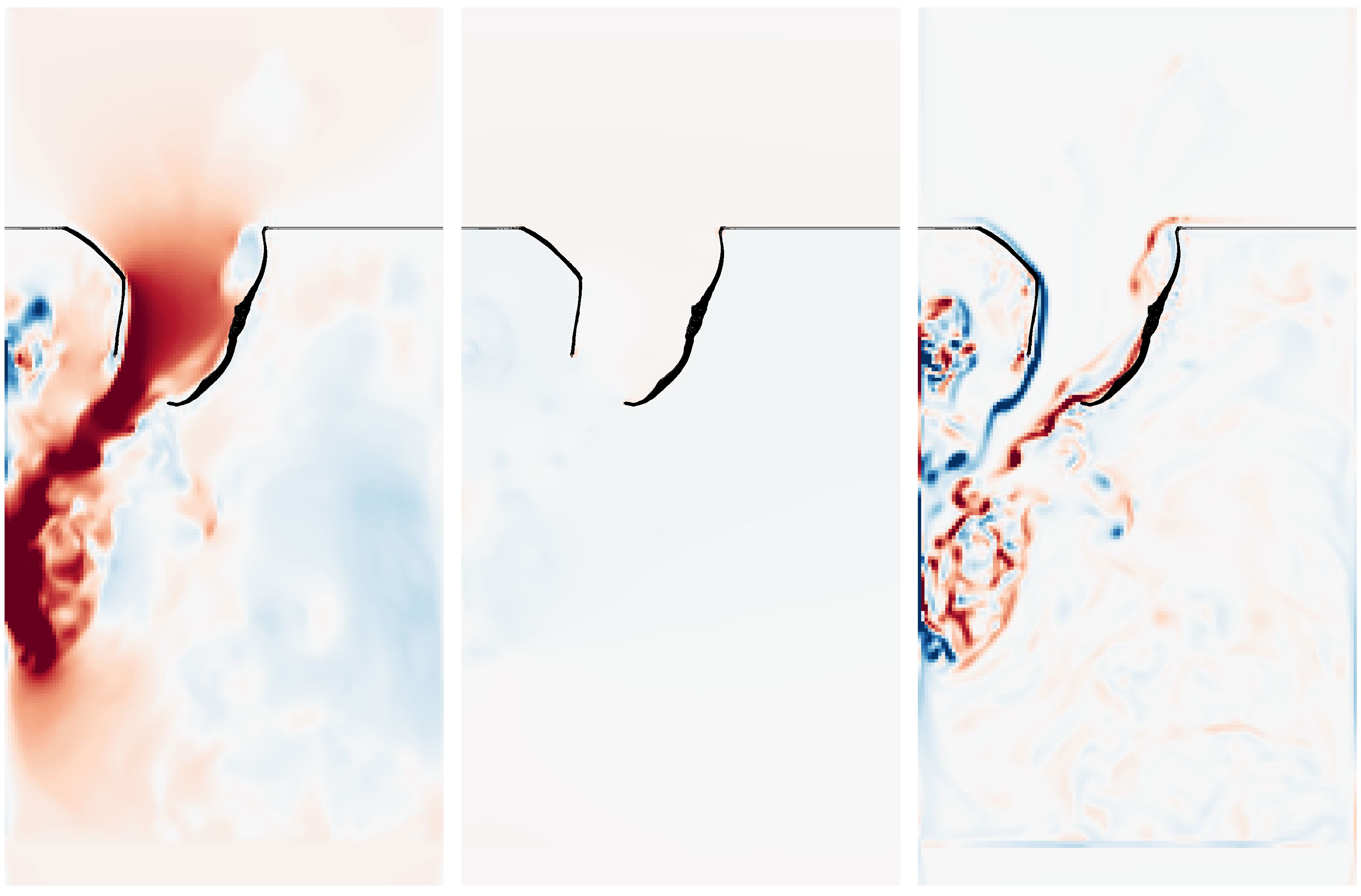}  & 
\rotatebox[origin=l]{90}{ \parbox{5cm}{ $t = 2.111 $ s, in process of closing  } } & 
\includegraphics[width=.45\columnwidth]{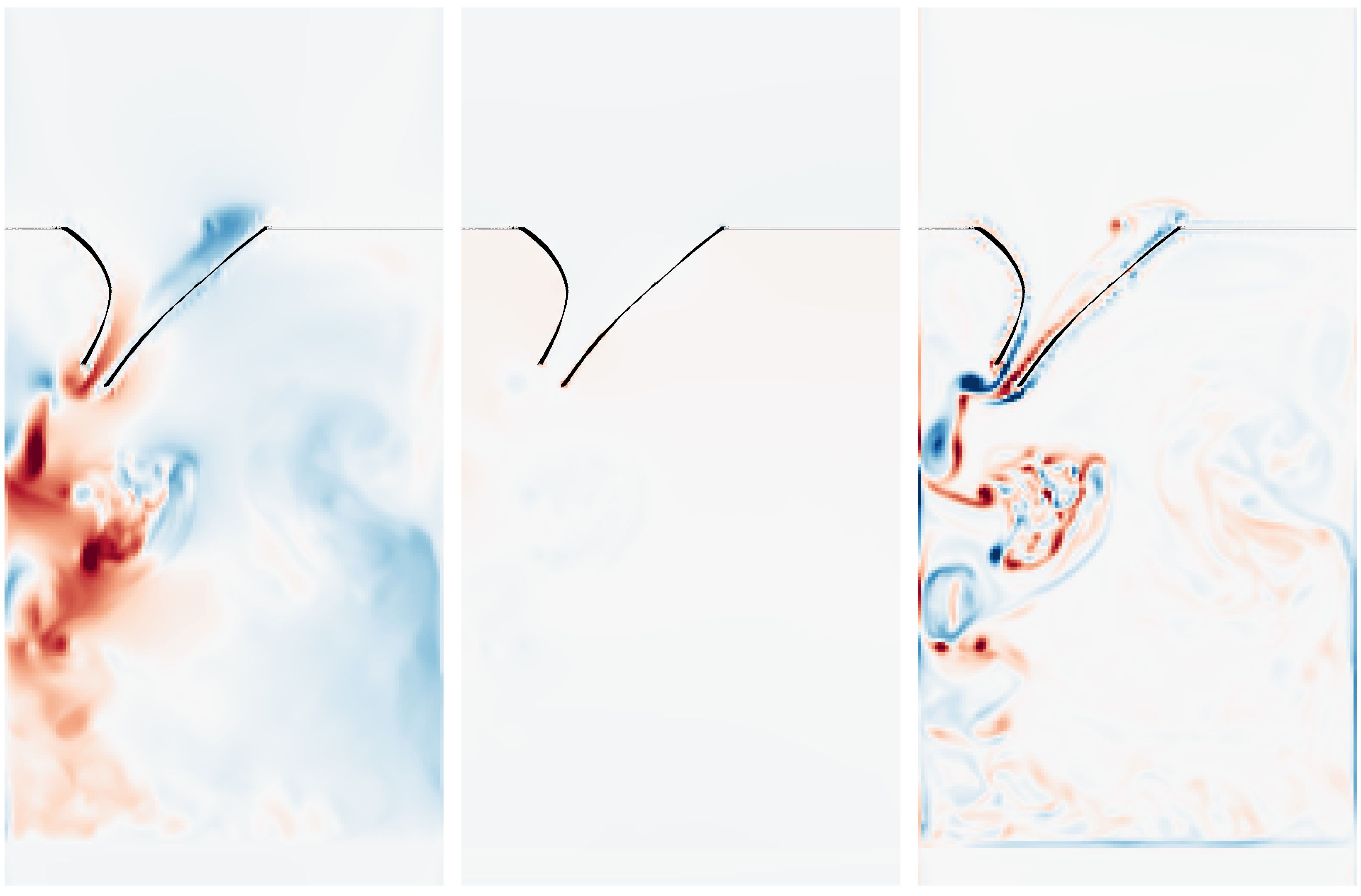} \\
\rotatebox[origin=l]{90}{ \parbox{4cm}{$t = 1.821 $ s, mid-diasole  }} & 
\includegraphics[width=.45\columnwidth]{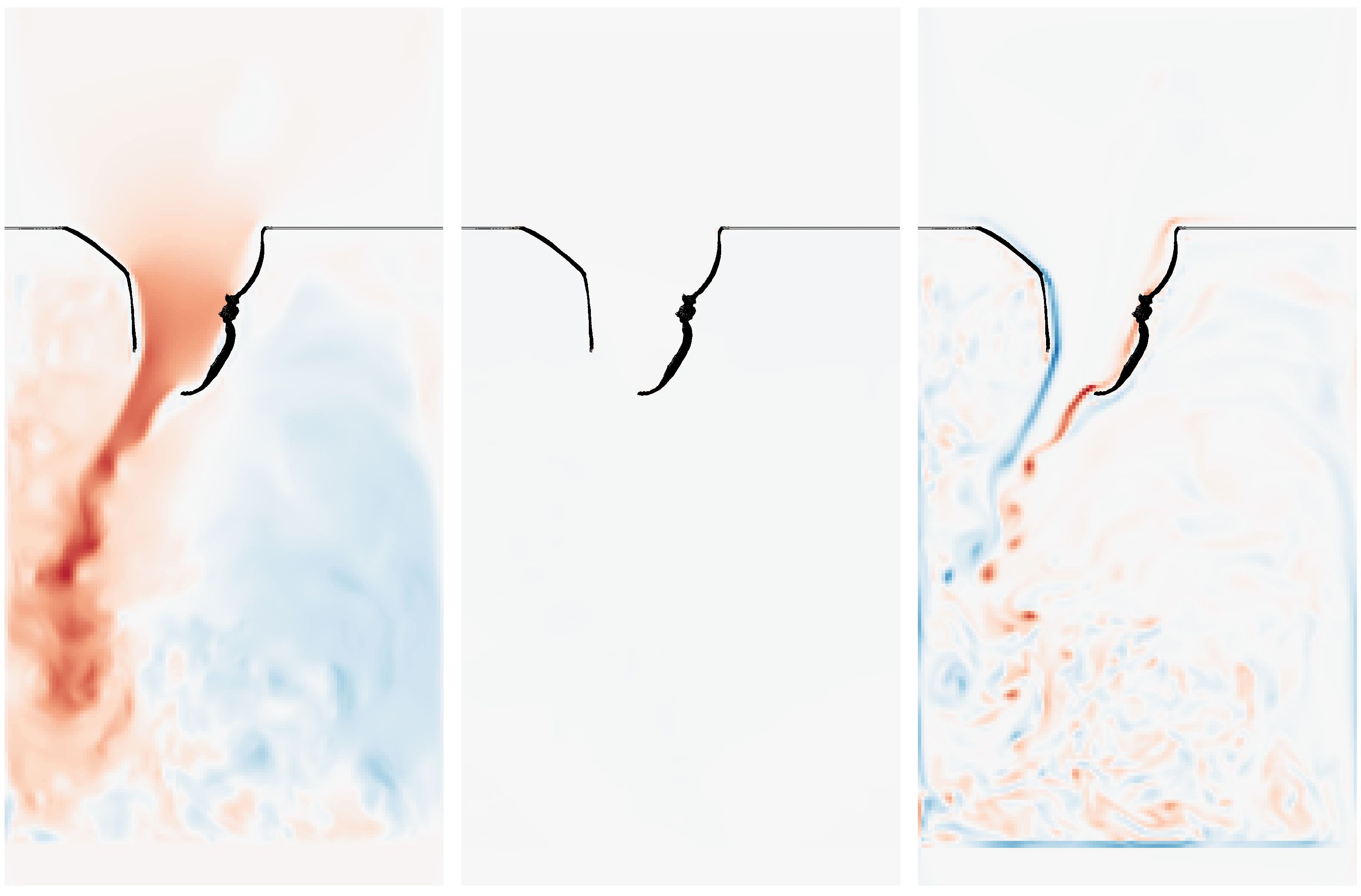} &
\rotatebox[origin=l]{90}{ \parbox{4cm}{ $t =  2.303 $ s, closed, mid-systole  } } & 
\includegraphics[width=.45\columnwidth]{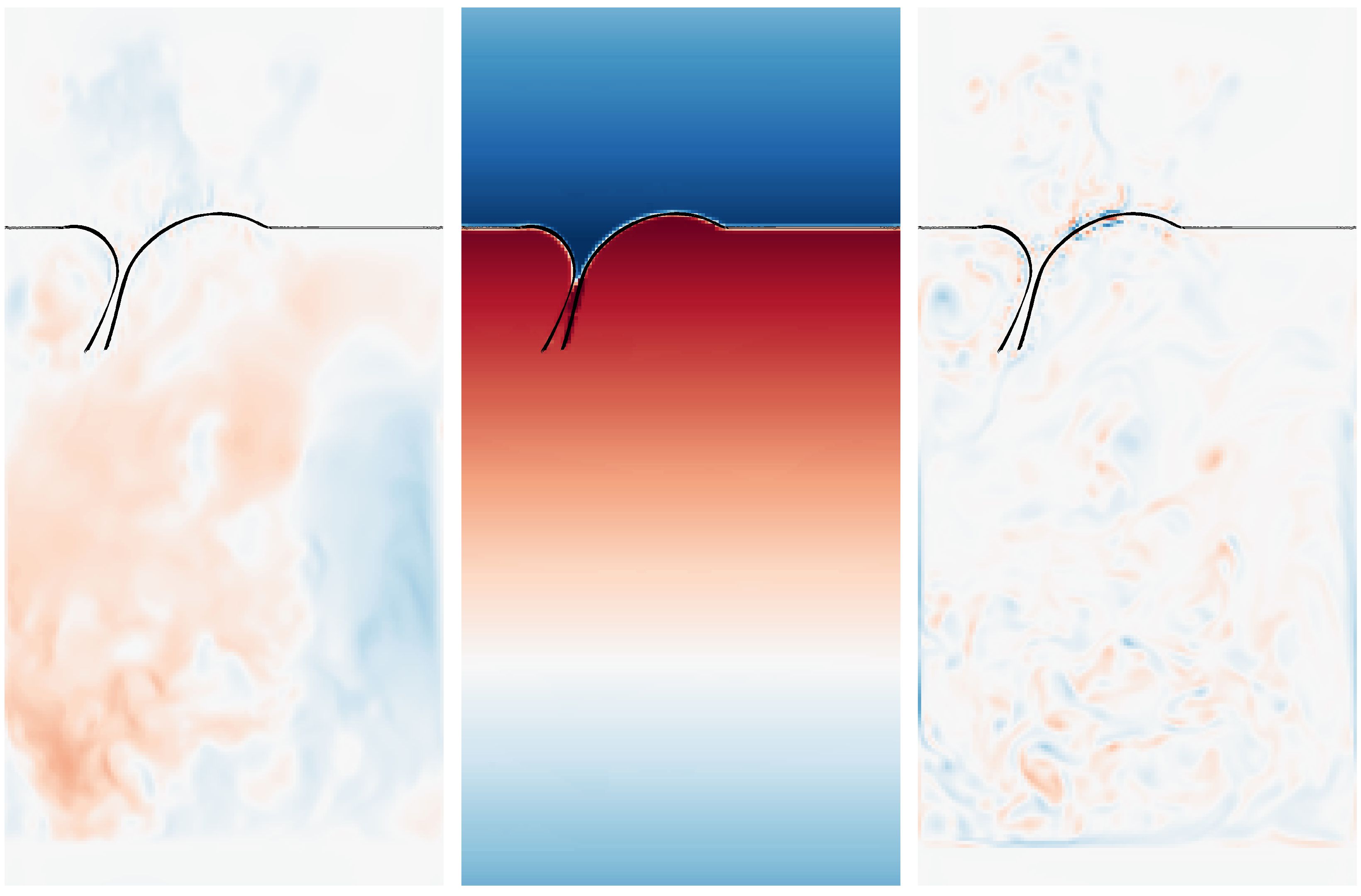}  \\
\rotatebox[origin=l]{90}{ \parbox{4cm}{ $t = 2.066 $ s, atrial systole   }} & 
\includegraphics[width=.45\columnwidth]{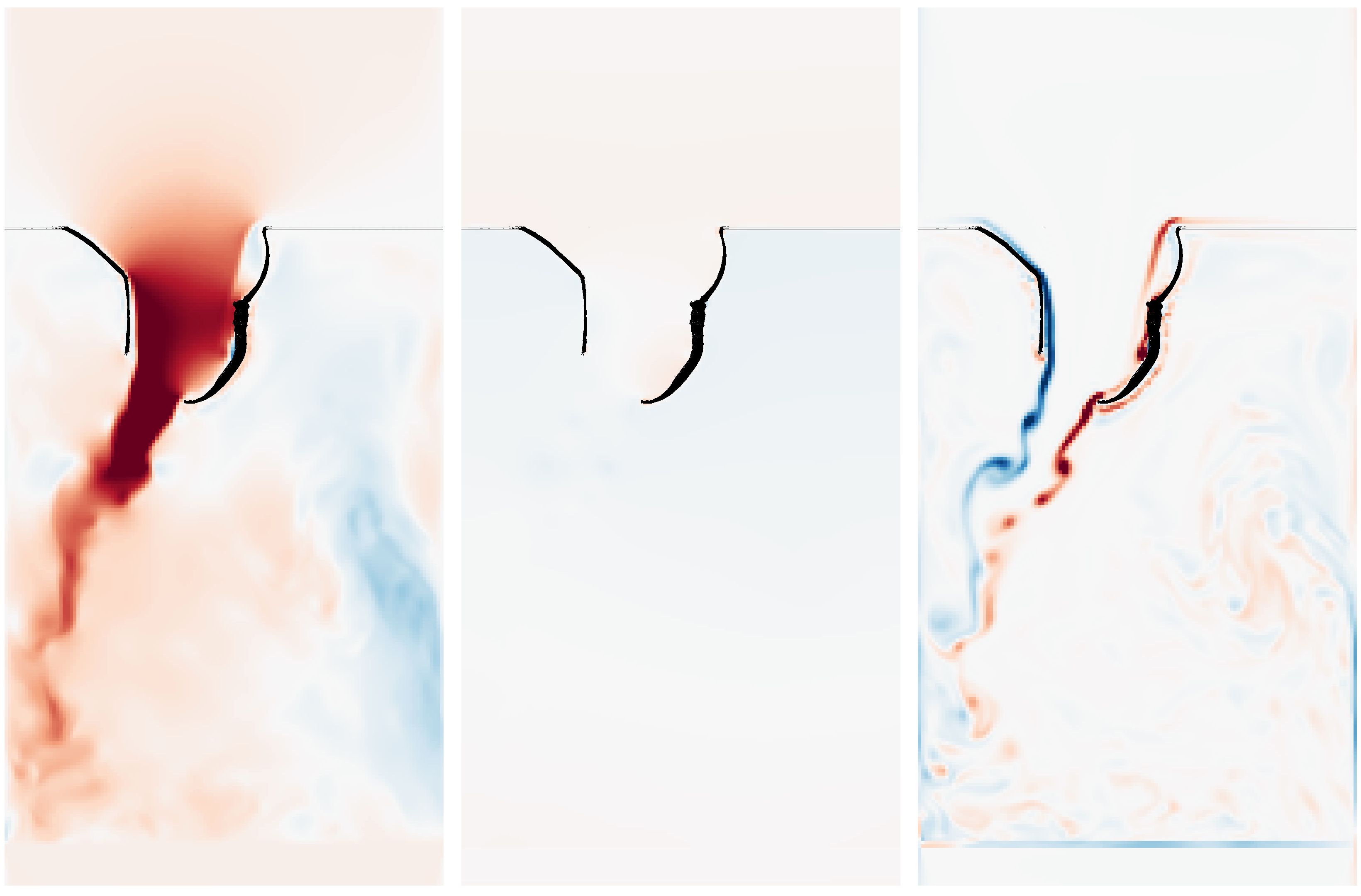} & 
\rotatebox[origin=l]{90}{ \parbox{4.5cm}{ $t =  2.386 $ s, closed and unloading \\ during isovolumic relaxation} } & 
\includegraphics[width=.45\columnwidth]{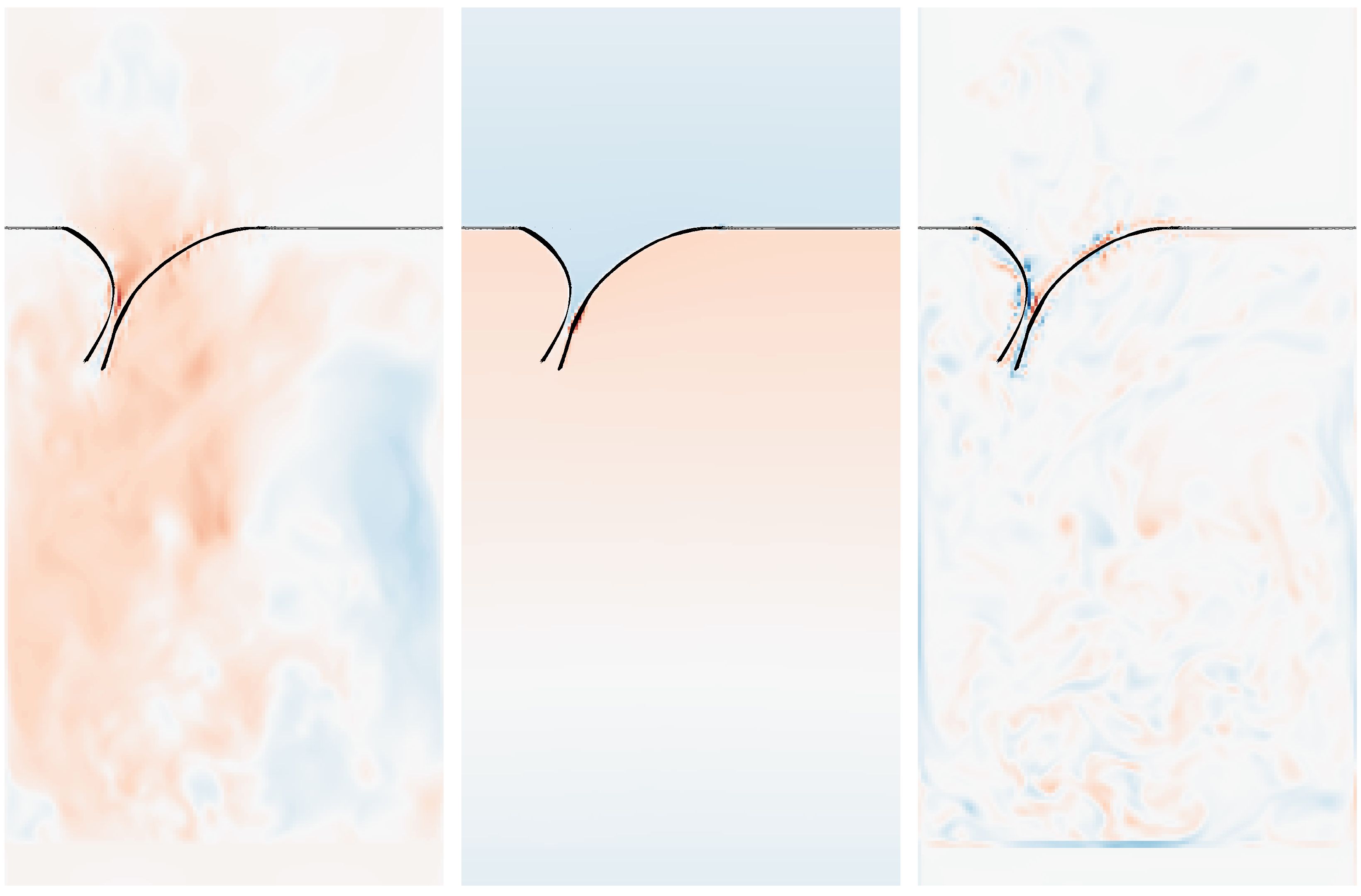} 
\end{tabular} 
}
\caption{
Mitral valve through the cardiac cycle in the third beat of the simulation. 
This figure shows slice views of the $z$ component of velocity, modified pressure and component of the vorticity vector that describes rotation within the plane of the figure.  
Since this component is normal to the plane, from now on we call it the out-of-plane vorticity.
This figure also shows a clipped cross section of the valve. 
The modified pressure is shown on a 120 mmHg scale centered on zero. 
From top to bottom, left to right, the frames depict early filling just after peak flow, filling during mid-diastole, peak flow during atrial systole, the valve in the process of closing immediately after the left atrial and left ventricular pressures equalize, the closed valve in mid-systole, and the closed valve in the process of unloading during isovolumic relaxation, the first phase of diastole. 
The first, fourth and fifth times are also shown as pathlines in Figure \ref{three_panels}. 
See movies {\bf M3} (real time) and {\bf M4} (slow motion) for animations of slice views of velocity. 
} 
\label{basic_through_cycle}
\end{figure}

Slice views of six time steps of the simulation are shown in Figure \ref{basic_through_cycle}.
This figure depicts the vertical component of velocity, the modified pressure and the out-of-plane (y-direction) component of vorticity. 
A clipped cross section of the valve is shown in all cases. 
The panels show these views during early filling, mid-diastole, atrial systole, the valve in the process of closing immediately after the start of ventricular systole, the closed valve mid-ventricular systole, and the closed valve in the process of unloading during isovolumic relaxation.

We estimate peak Reynolds number of this flow to be 2700, and the mean Reynolds number during forward flow to be 1000. 
The velocity used in this estimate is the spatially averaged velocity through the mitral ring as a function of time, i.e., the mitral flow (volume per unit time) divided by the area enclosed by the mitral ring; using the maximum would increase this estimate. 
Since the Reynolds number is much greater than one, the flow is inertially dominated. 
(See Section \ref{Reynolds_number} for computation.)

The simulations appear to be well resolved. 
During closure, flow rates are nearly identical at the resolution presented and resolution that is twice as coarse. 
Coarsening the resolution of the fluid and structure thickens the delta function used in interaction, and so reduces the effective orifice area of the valve. 
At lower fluid resolution, whether using fine or coarse structural resolution, we see a similar decrease in forward flow rate. 
This can be compensated for by adjusting the coefficients in the edge connector regions to allow a slightly wider opening area. 
See Section \ref{fluid_convergence} for discussion and figures.

\subsection{Results -- variations in driving pressures}
\label{further_results}
To further test the valve, we apply different driving pressures waveforms with lower ventricular systolic pressure, higher ventricular systolic pressure and the absence of atrial systole. 
Closure is robust in all three cases. 
The results suggest that a negative pressure difference alone, without any secondary mechanisms, is sufficient to close the mitral valve. 
Moreover, closure appears insensitive to changes in value of negative pressure differences.

\begin{figure}[ht]
\centering 
\includegraphics[width=.5\columnwidth]{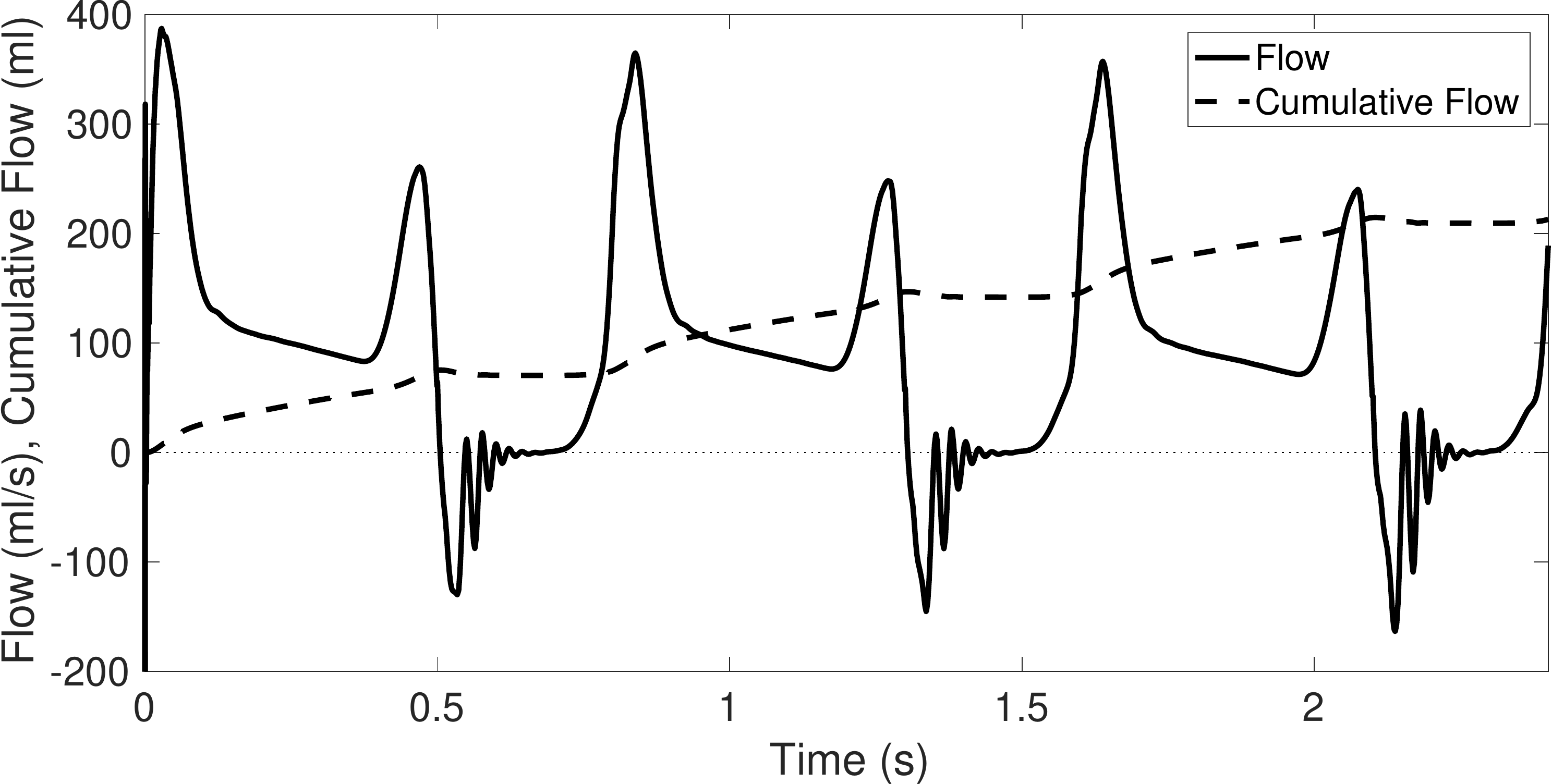}
\includegraphics[width=.5\columnwidth]{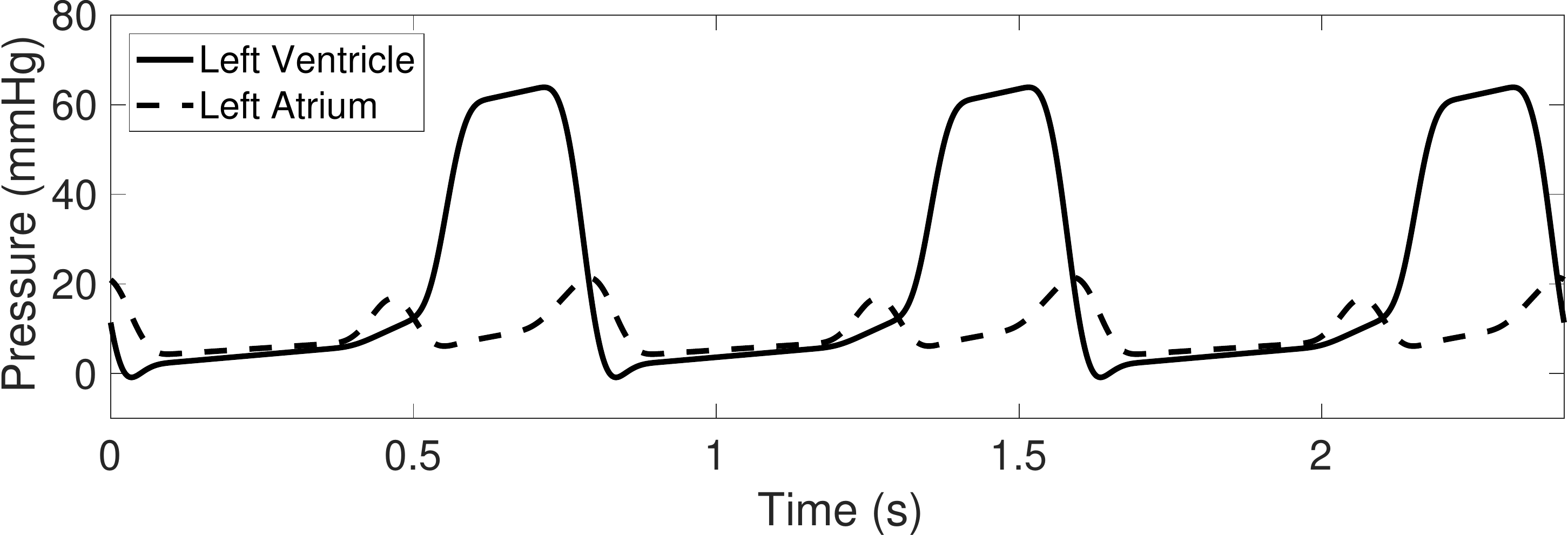}
\caption{Flow (top, emergent) and driving pressure (bottom, prescribed) with low systolic ventricular pressure.}
\label{pressure_flux_low}
\end{figure}

First, we prescribe a much lower ventricular systolic pressure, approximately half that of the standard.
This is shown in Figure \ref{pressure_flux_low}, along with the resulting flow. 
The lower pressure creates a smaller load on the model valve, and thus lower strains throughout. 
The initial spike of negative flow that occurs on closure is smaller in magnitude. 
It is possible at this pressure difference, which is much lower than the pressure difference for which it was tuned, the leaflets will not be strained enough to coapt properly. 
However, the flow shows this is not the case. 
The valve closes and seals effectively at these lower pressures as well. 
Otherwise, the resultant flow is quantitatively and qualitatively similar to the flow with standard pressures. 
See movie {\bf M5}. 

\begin{figure}[ht]
\centering 
\includegraphics[width=.5\columnwidth]{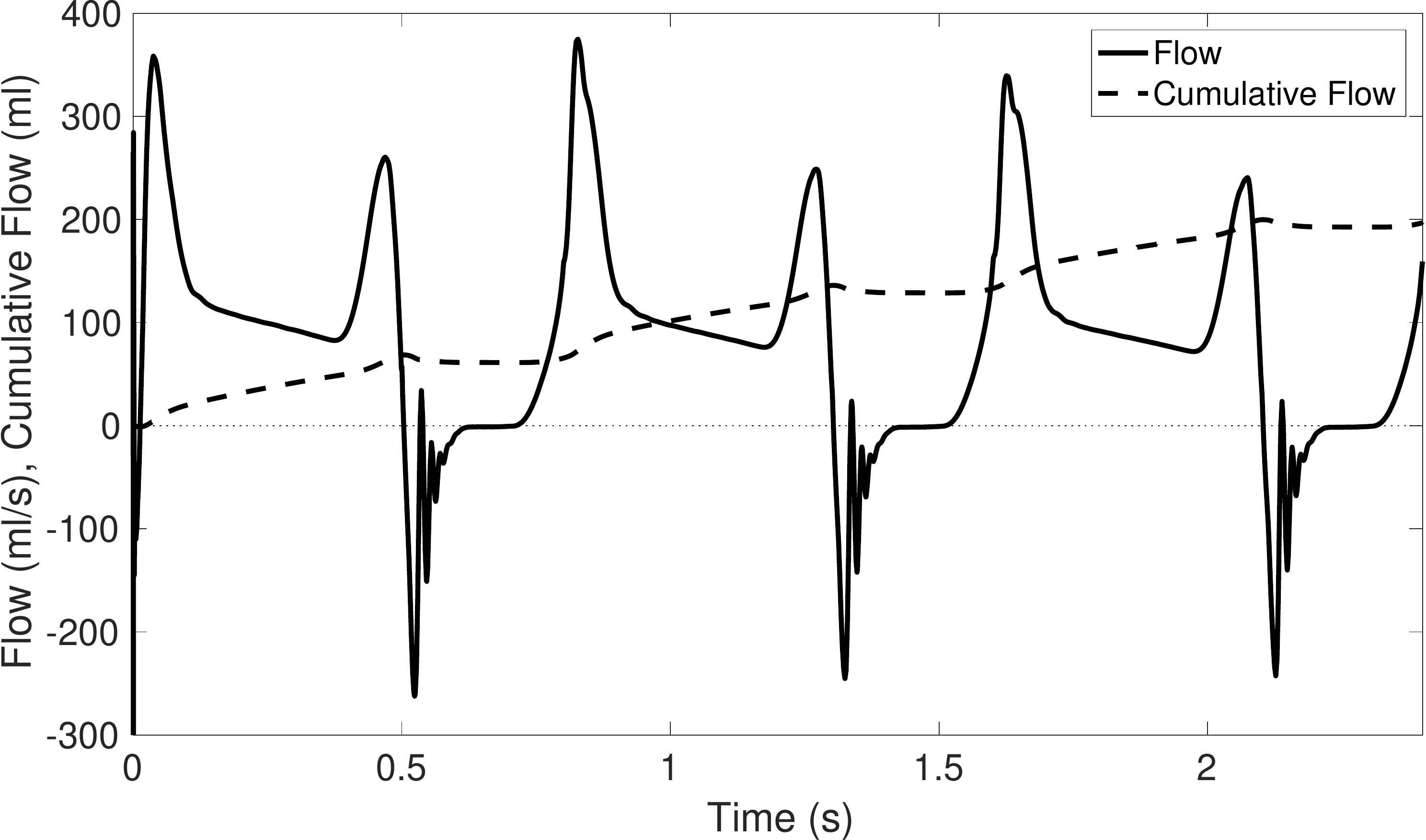}
\includegraphics[width=.5\columnwidth]{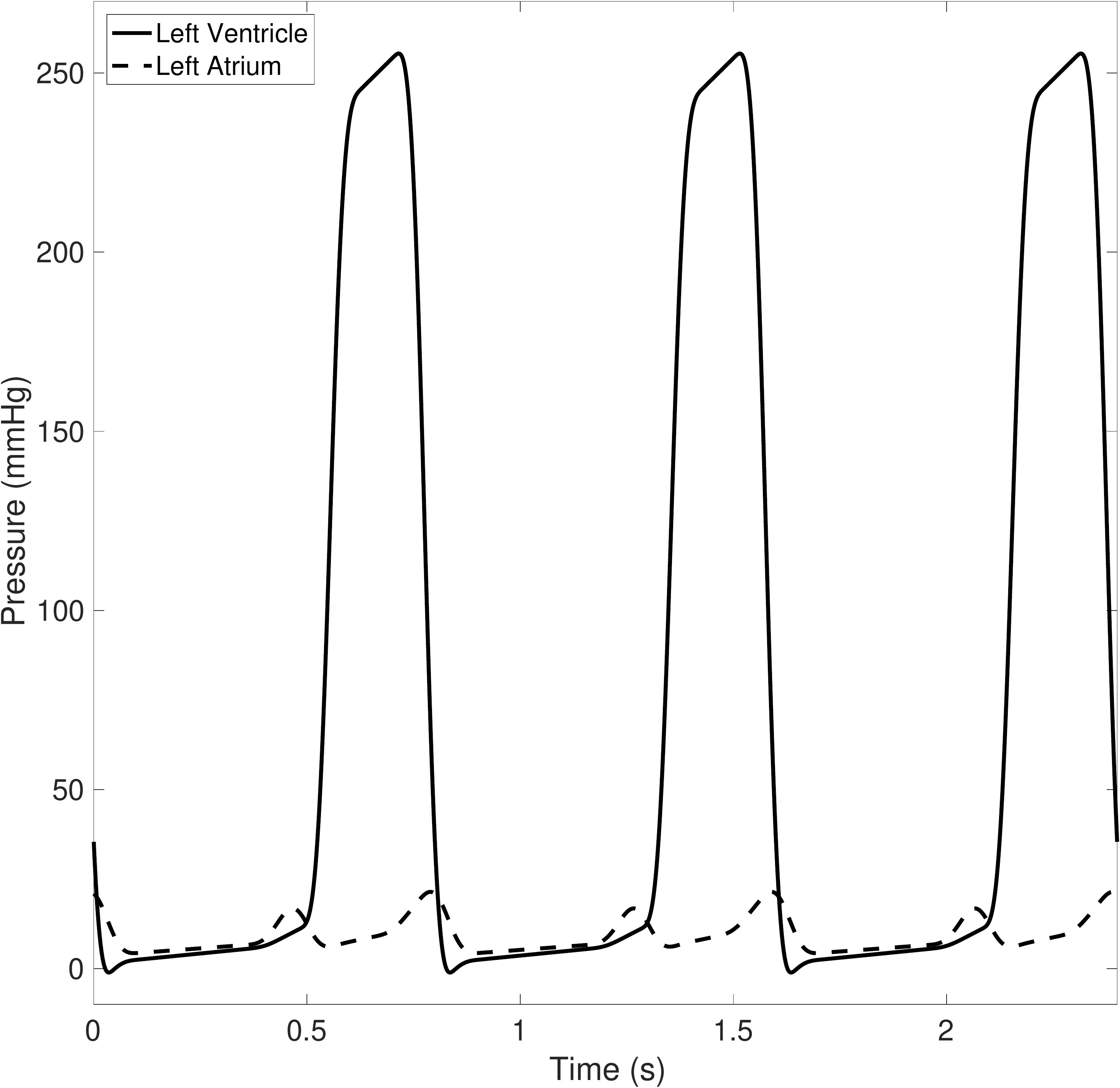}
\caption{Flow (top, emergent) and driving pressure (bottom, prescribed) with with high systolic ventricular pressure.}
\label{pressure_high}
\end{figure}

Next, we prescribe a much higher systolic pressure, and see if the valve still supports this pressure, or develops holes or leaks. 
High ventricular pressure during systole may occur in the body, for example in a patient with aortic stenosis, which restricts outflow from the ventricle, or systemic hypertension. 
Here, we set the ventricular systolic pressure to approximately twice that of the standard simulation. 
The driving pressures are shown in Figure \ref{pressure_high}. 
In the affine region of the constitutive law, the chord stiffness of every fiber in the valve model increases with increased strain. 
Thus, we expect the model to work effectively under higher pressures. 
The emergent flow is shown in Figure \ref{pressure_high}, which shows that this is the case. 
The most striking thing about this flow is that it is quantitatively similar to the flow with standard pressures. 
The initial spike of negative flow that occurs on closure is larger in magnitude, and the rest of the flow is similar to the flow with standard pressures and the valve seals well. 
Some of the oscillations appear to stay below zero, perhaps because the valve is settling into a more loaded state. 
See movie {\bf M6}.

\begin{figure}[ht]
\centering 
\includegraphics[width=.5\columnwidth]{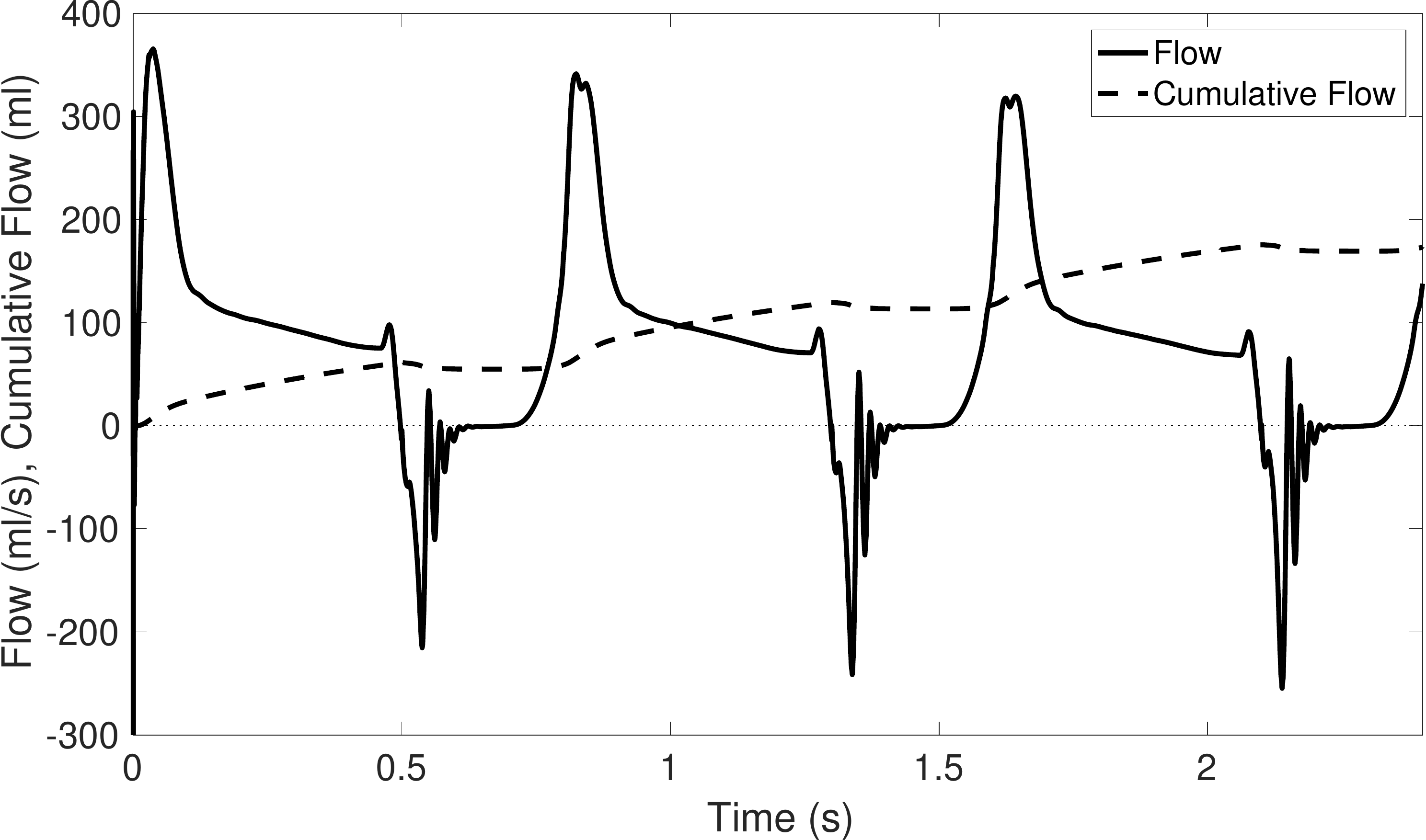}
\includegraphics[width=.5\columnwidth]{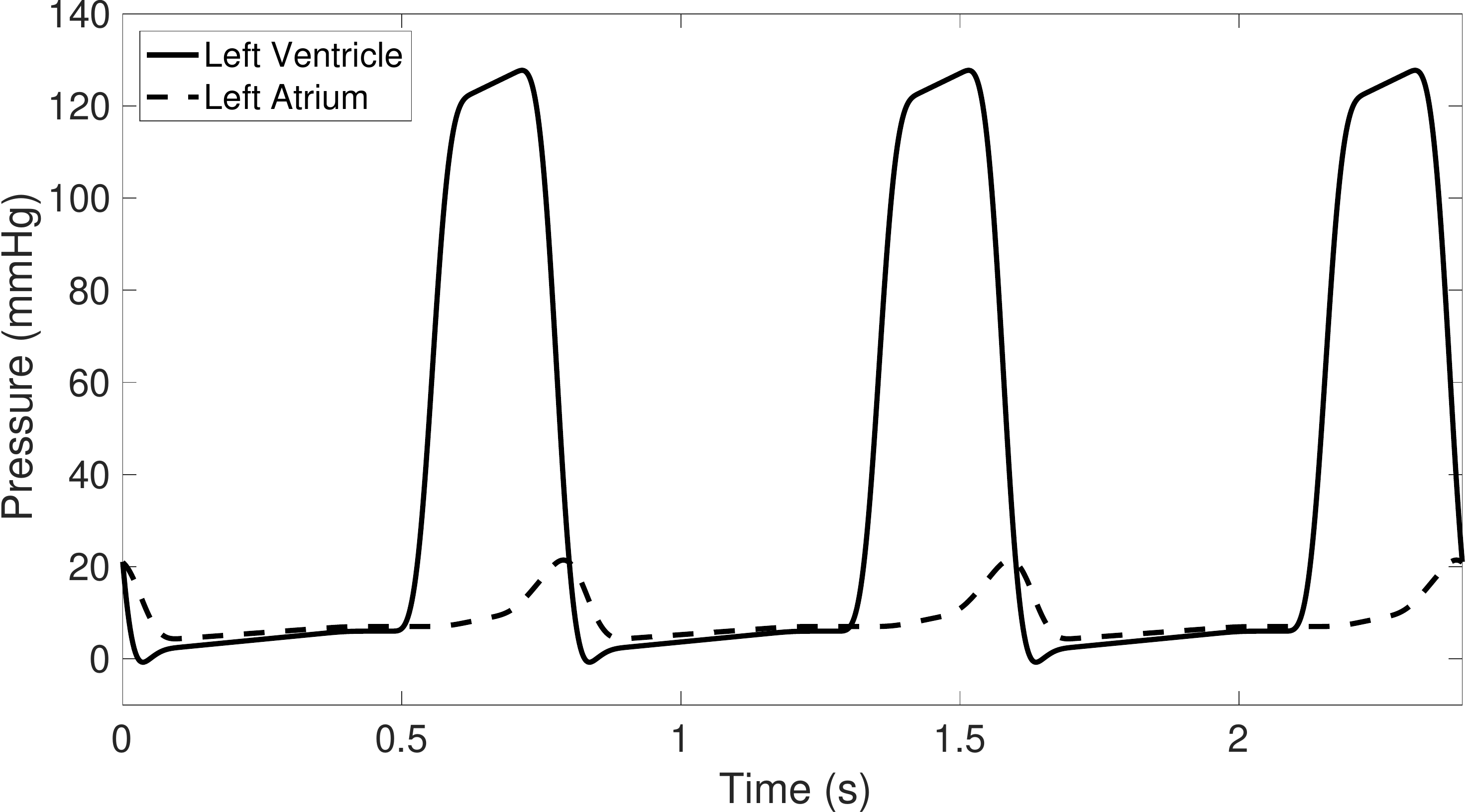}
\caption{Flow (top, emergent) and driving pressure (bottom, prescribed) with no atrial systole.}
\label{pressure_flux_nokick}
\end{figure}

Finally, to see what happens in the absence of atrial systole, e.g., as occurs in atrial fibrillation, we drive the flow with no atrial systole present in the pressure. 
Since it has been suggested that atrial systole affects closure of the valve, we wish to see if the model closes properly without it. 
The pressure and emergent flow are shown in Figure \ref{pressure_flux_nokick}. 
The spike in forward flow during the time in which atrial systole occurs is absent, as expected. 
The closure transient looks qualitatively similar to the closure transient following normal atrial systole.
It has been suggested that atrial systole affects closure of the valve, but the model here closes effectively without it. 
See movie {\bf M7}.

\begin{figure}[ht]
\setlength{\tabcolsep}{1.4pt}
\hspace{-8pt}
\begin{tabular}{cccc}
\raisebox{100pt}{
	\begin{tabular}{l}
	velocity \\ 
	cm/s \\ 
	\includegraphics[width=.07\columnwidth]{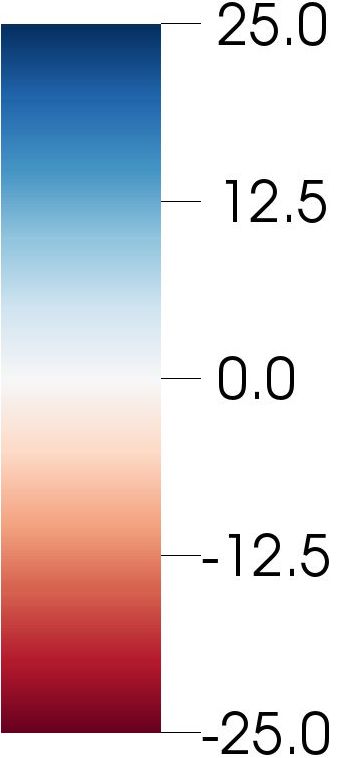} 
	\end{tabular}  
}
& 
\includegraphics[width=.3\columnwidth]{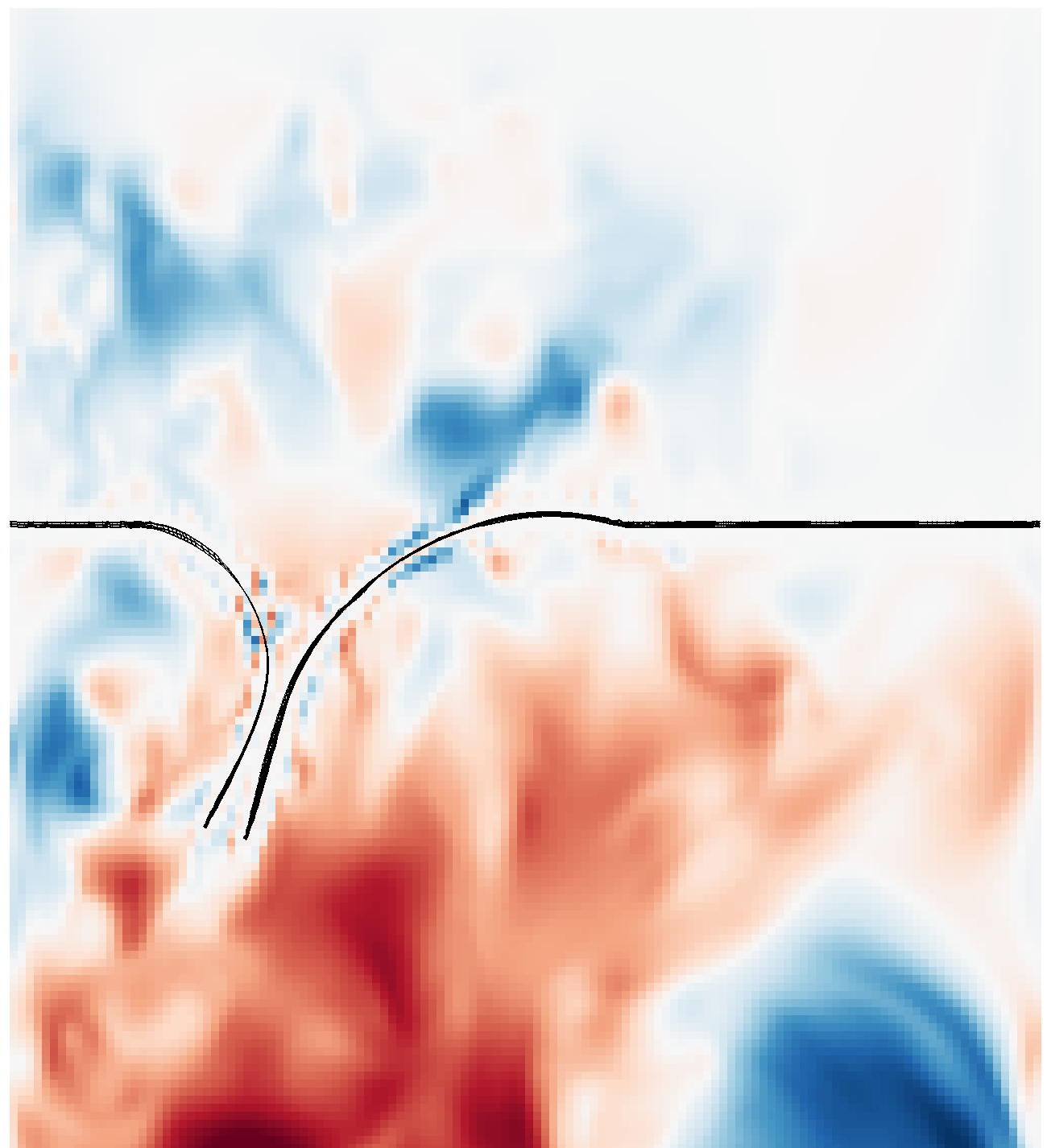}  &
\includegraphics[width=.3\columnwidth]{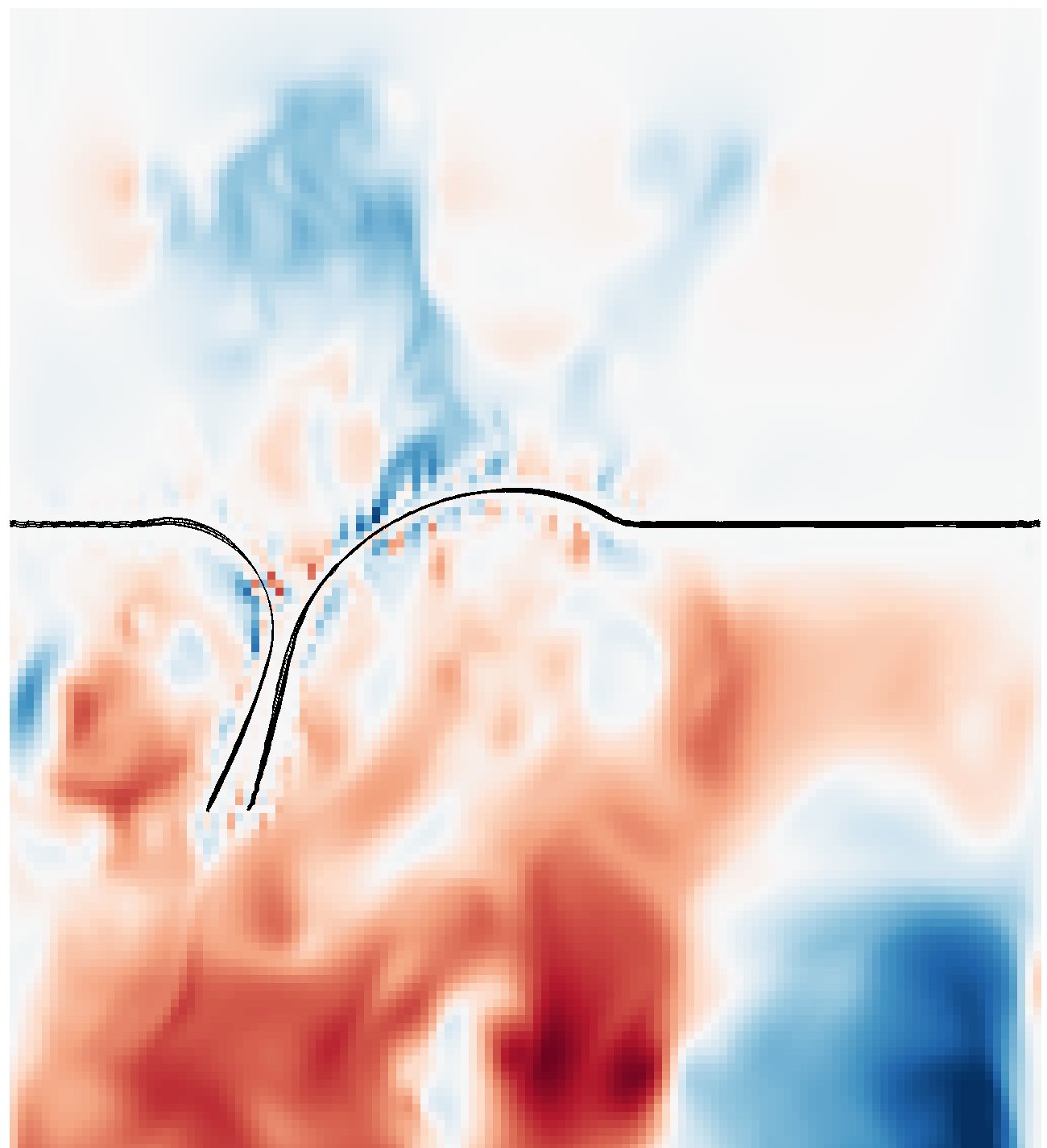}   &
\includegraphics[width=.3\columnwidth]{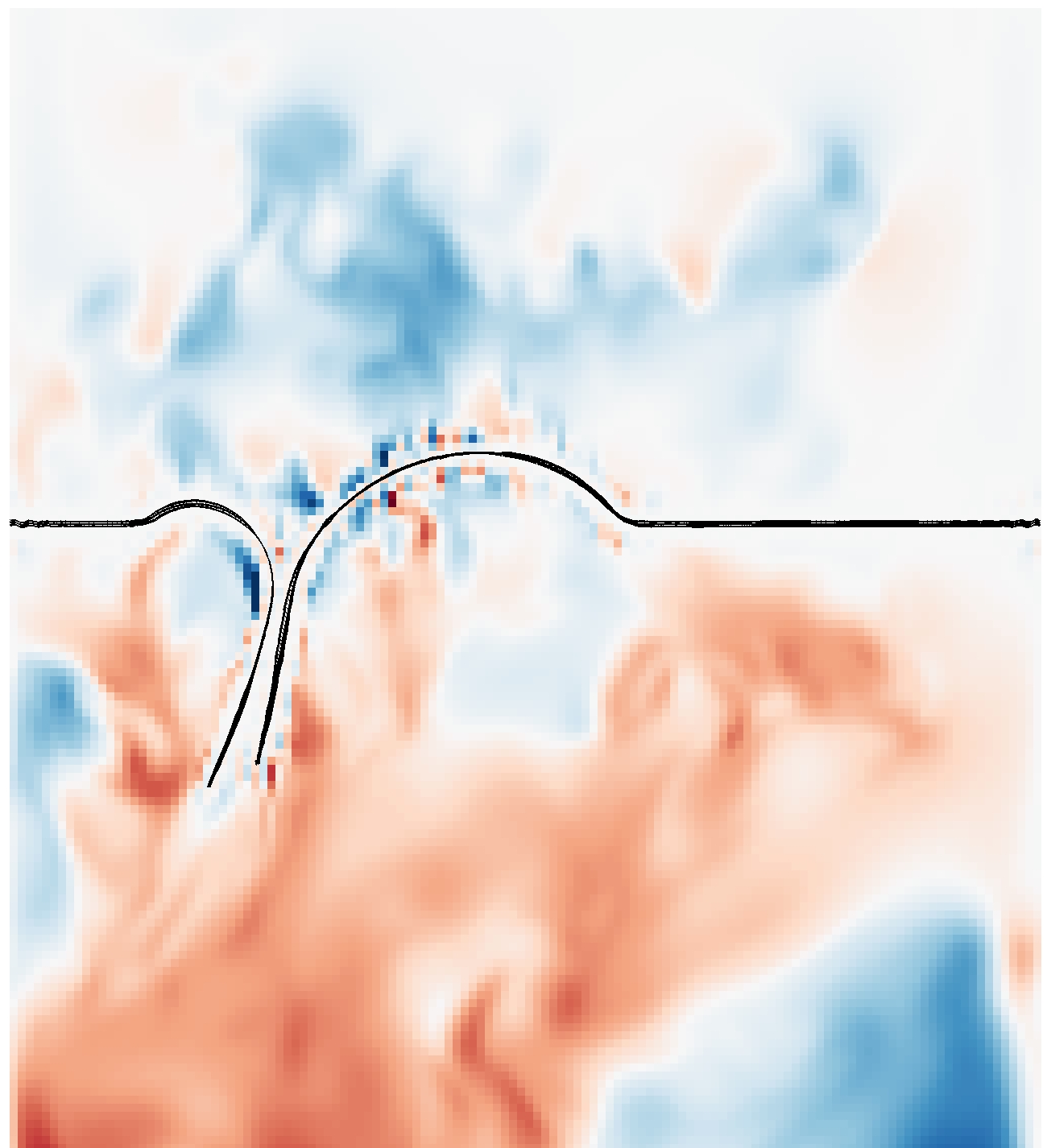}  
\end{tabular}
\caption{Slice view of the $z$ component of velocity with low, standard and high ventricular systolic pressure. 
The color scale is four times as fine as other velocity slice plots to show details during systole, when overall magnitudes of flows are lower.} 
\label{three_pressure_slices}
\end{figure}

To compare qualitatively, we view three slice views of the velocity field with the valve at $t = 2.32$, shown in Figure \ref{three_pressure_slices}. 
This is approximately peak pressure difference during the third beat of the simulation. 
The prescribed pressure difference in the left frame is 55.0 mmHg, the center frame is 114.9 mmHg, and the right frame is 242.5 mmHg. 
The valve appears to close well in all three frames. 
The higher the pressure, the more loaded the valve appears and the smaller radius of curvature we see in the leaflets. 
Despite very different loading pressures, each frame looks qualitatively similar. 

When the valve closes, two mechanisms have been proposed that assist it to do so effectively. 
For references on and discussion of the mechanisms of valve closure see \cite{doi:10.1146/annurev.fl.14.010182.001315}. 
 The first is called the ``breaking jet.''
During atrial systole, there is a jet of forward flow between the leaflets. 
This rapid flow produces a low pressure due to a Bernoulli effect, resulting in a pressure differential across the leaflets around the free edge. 
This then sucks the leaflets together. 
The second is a large vortex. 
Vortices are shed from the leaflet, and a large vortex forms in the left ventricle. 
This vortex then comes back around and pushes the valve, especially the anterior leaflet closed. 
Since the geometry of our model valve tester is not shaped like the ventricle, shed vortices may not form a large vortex to push the valve back into place. 
This means that the vortex may not be as effective in helping the valve close in the rectangular box geometry. 
However, the presence of the vortex would still suggest that this mechanism is plausible.

Here, we look at three views of the simulation, showing the velocity, pressure and vorticity in a plane. 
The pressure is shown on a 10 mmHg scale so that fine details are visible. 
The component of vorticity normal to the viewing planes is shown. 
Note that this is also normal to the approximate symmetry plane of the valve. 
These features suggest that both the breaking jet and vortex are plausible mechanisms to assist valve closure. 

\begin{figure}[p!]  
\setlength{\tabcolsep}{0.5pt}
\centering
\makebox[0pt]{
 \begin{tabular}{cc @{\hspace{10pt}} c}
& {\large standard driving pressure} &  {\large driving pressure without atrial systole} \\ 
& & \\
& 
\begin{tabular}{L{.15\columnwidth}L{.15\columnwidth}L{.15\columnwidth}}
velocity & modified pressure & vorticity \\ 
cm/s & mmHg & $s^{-1}$ \\ 
\includegraphics[width=.09\columnwidth]{colorbar_velocity_kick_compare.jpeg} & 
\includegraphics[width=.09\columnwidth]{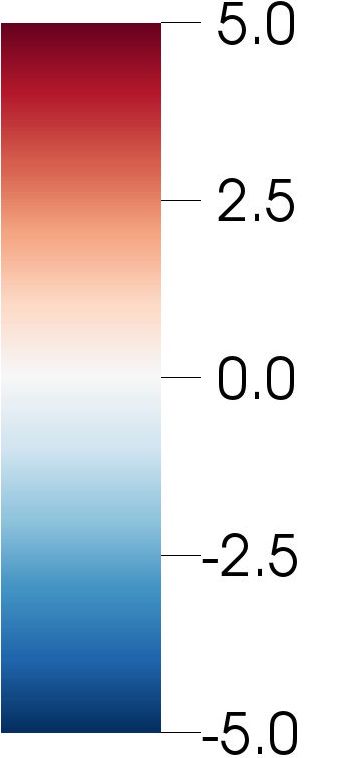} & 
\includegraphics[width=.09\columnwidth]{colorbar_vorticity_kick_compare.jpeg} 
\end{tabular} 
& 
\begin{tabular}{L{.15\columnwidth}L{.15\columnwidth}L{.15\columnwidth}}
velocity & modified pressure & vorticity \\ 
cm/s & mmHg & s$^{-1}$ \\ 
\includegraphics[width=.09\columnwidth]{colorbar_velocity_kick_compare.jpeg} & 
\includegraphics[width=.09\columnwidth]{colorbar_pressure_kick_compare.jpeg} & 
\includegraphics[width=.09\columnwidth]{colorbar_vorticity_kick_compare.jpeg} 
\end{tabular} 
\\ 
& & \\
\rotatebox[origin=l]{90}{ \parbox{4cm}{ $t = 2.09146$ s \\ near the end of atrial systole, pressures about to cross } } & 
\includegraphics[width=.45\columnwidth]{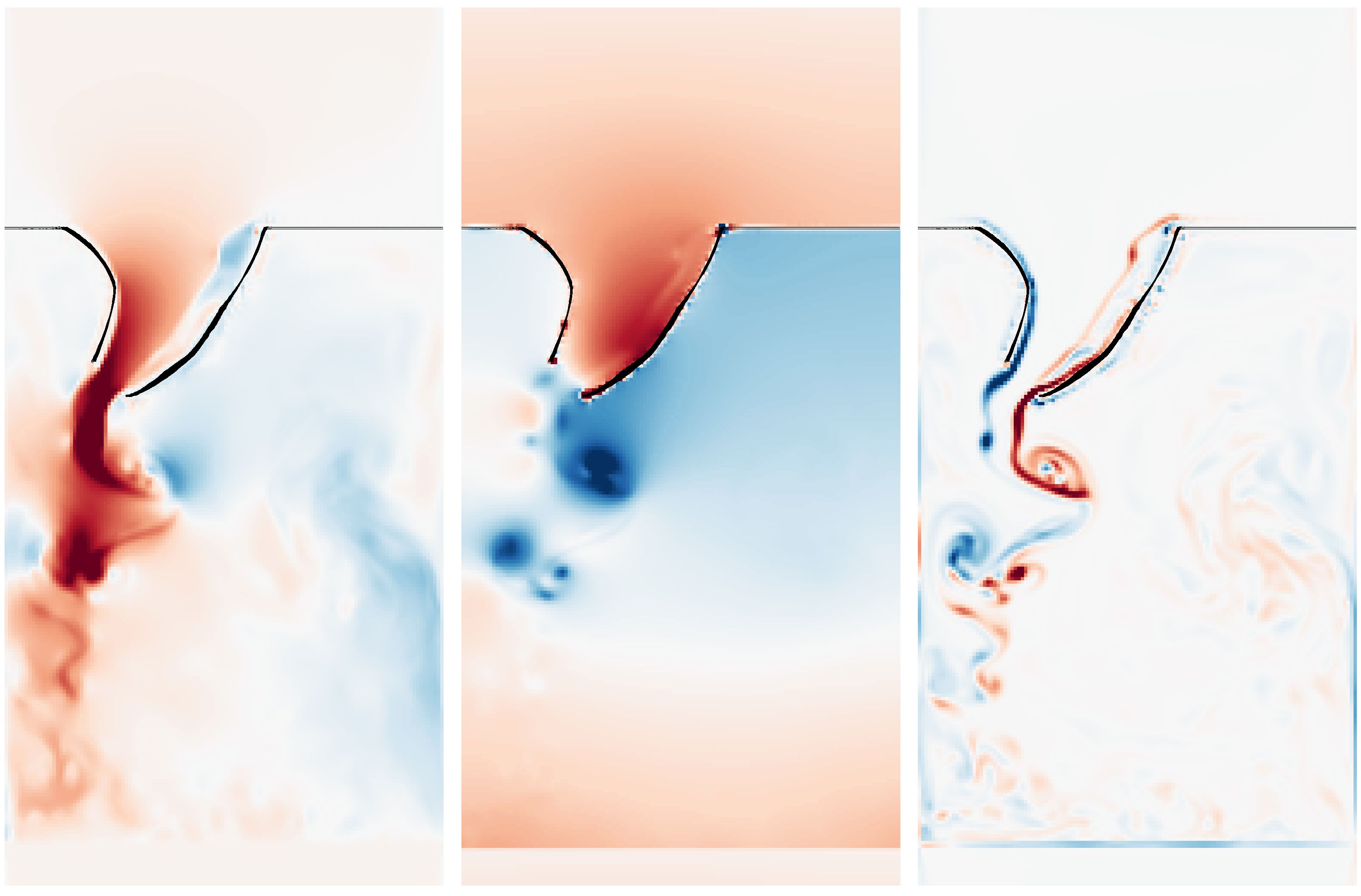}  & 
\includegraphics[width=.45\columnwidth]{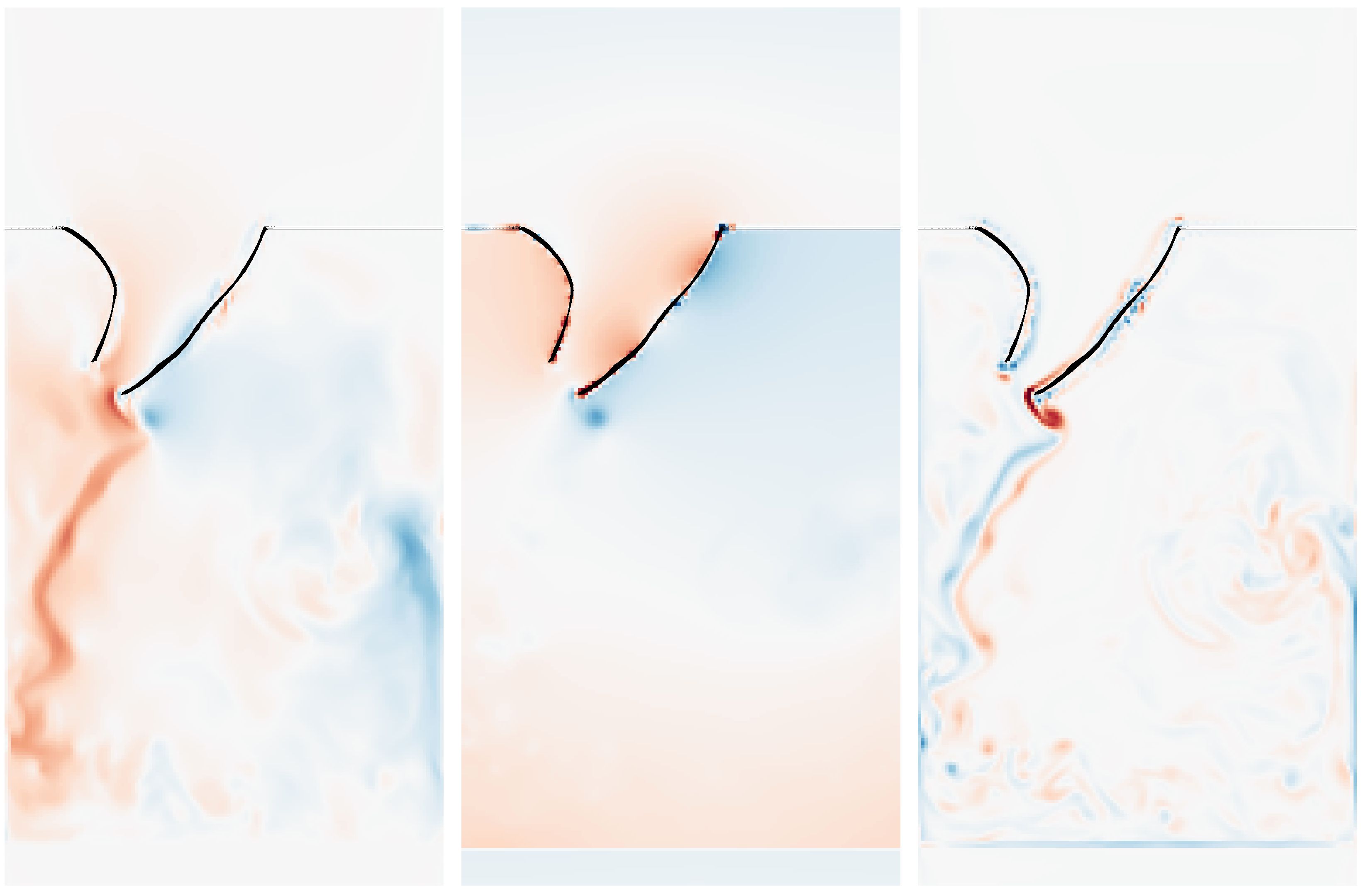} \\
\rotatebox[origin=l]{90}{ \parbox{4cm}{$t = 2.10646$ s \\ valve closing, immediately \\ after pressures cross}} & 
\includegraphics[width=.45\columnwidth]{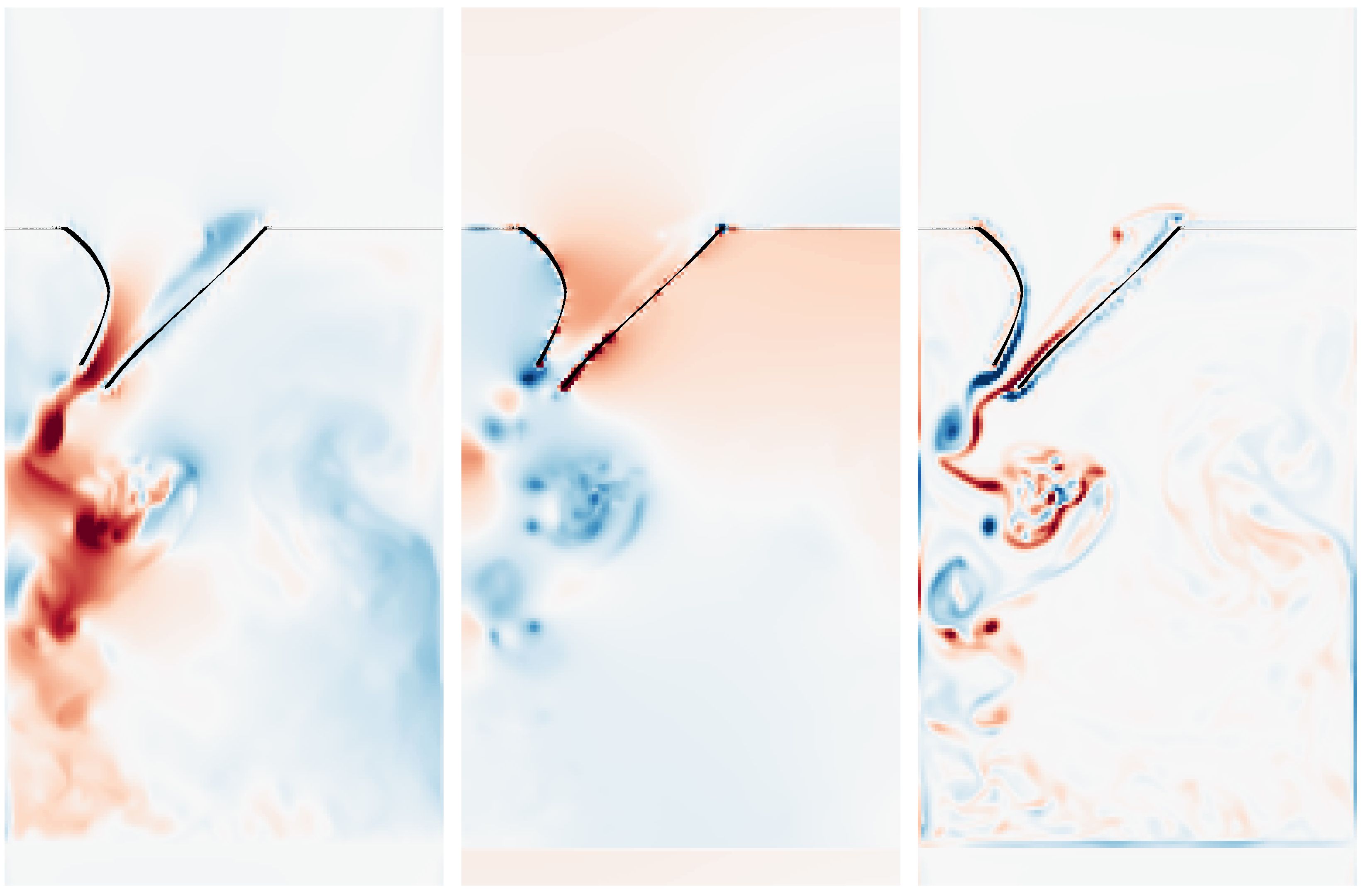} &
\includegraphics[width=.45\columnwidth]{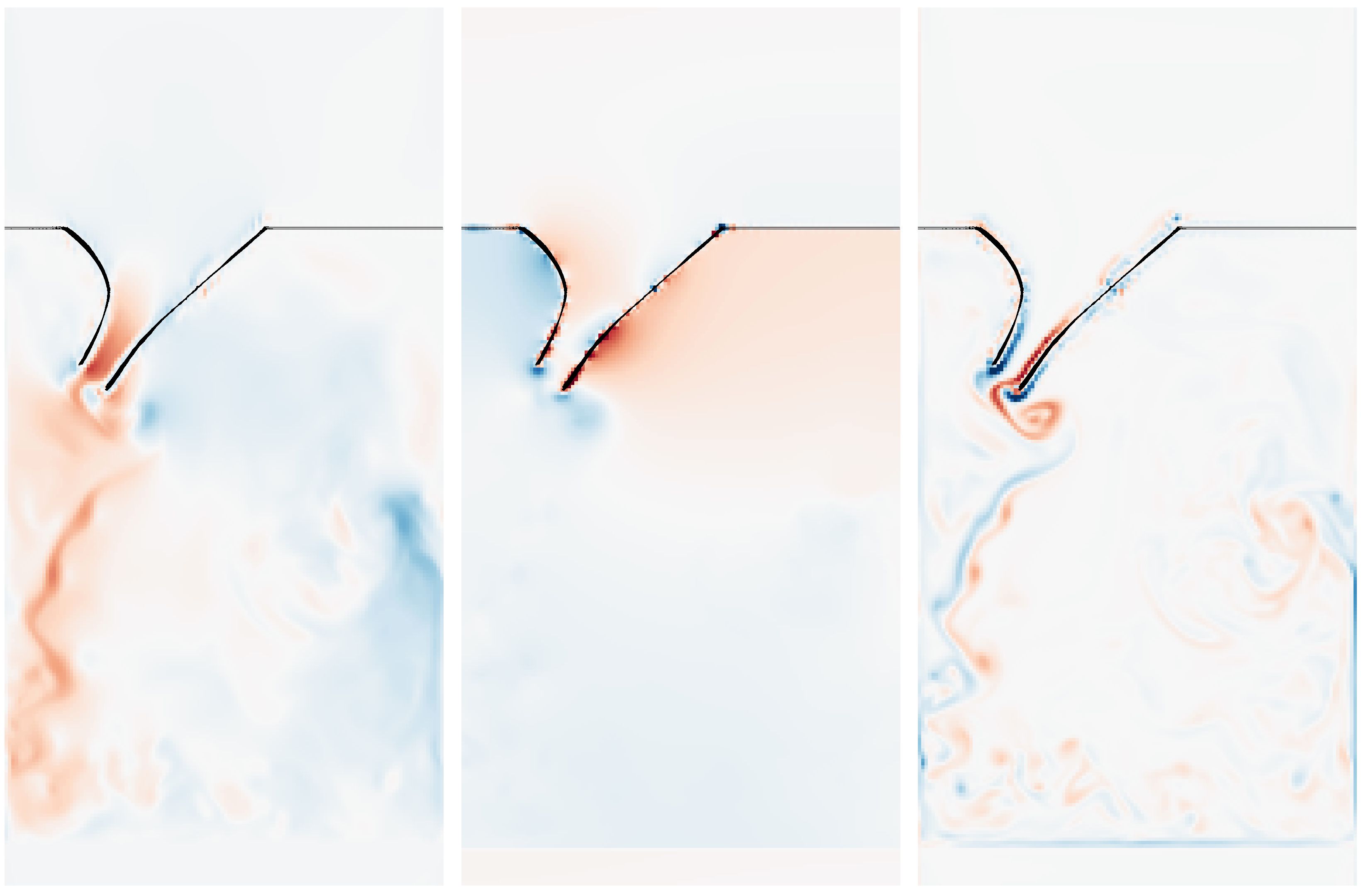}  
\\
\rotatebox[origin=l]{90}{ \parbox{4cm}{ $t = 2.12145$ s \\ valve continuing to close, \\ coaptation imminent }} & 
\includegraphics[width=.45\columnwidth]{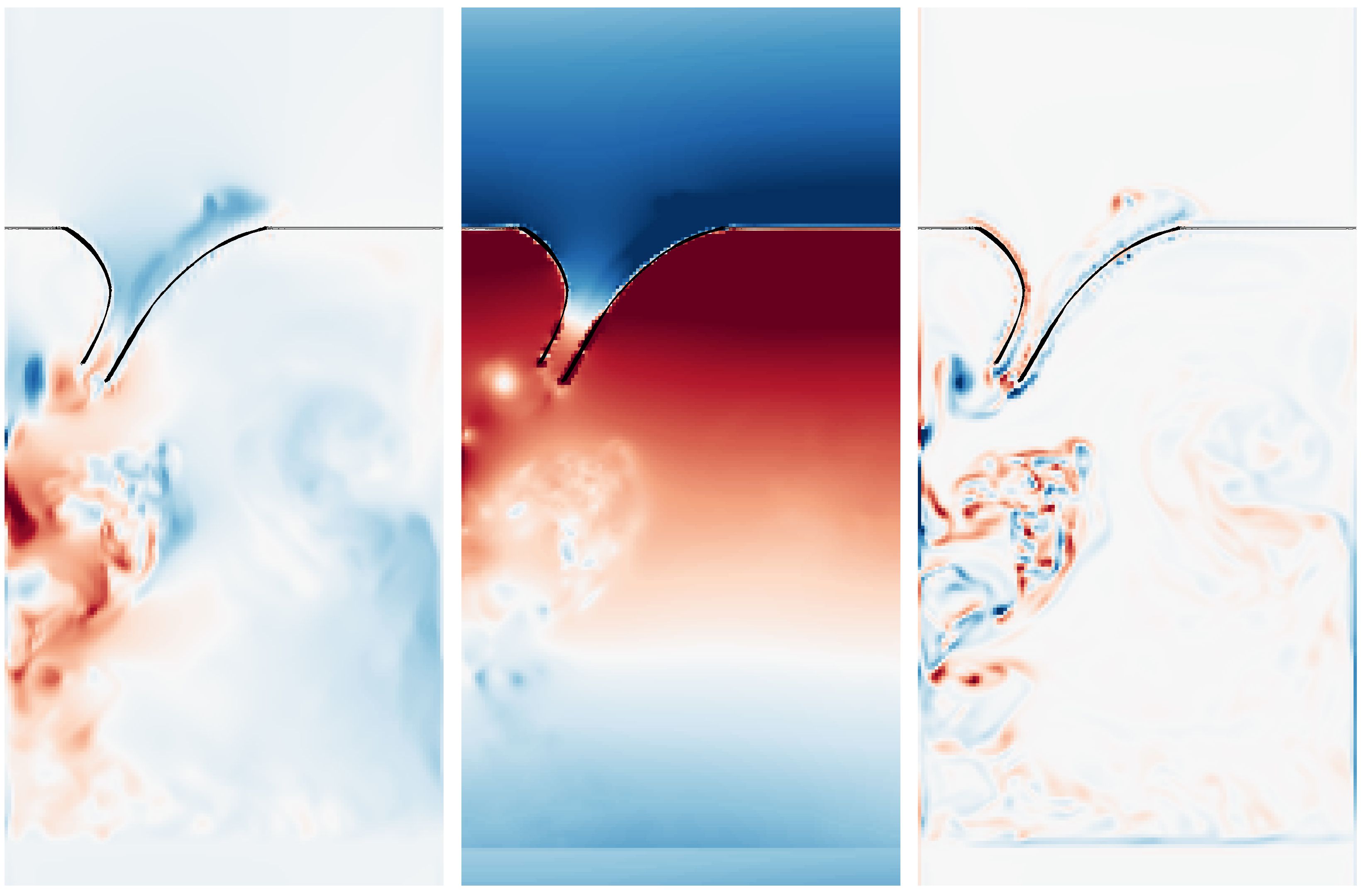} & 
\includegraphics[width=.45\columnwidth]{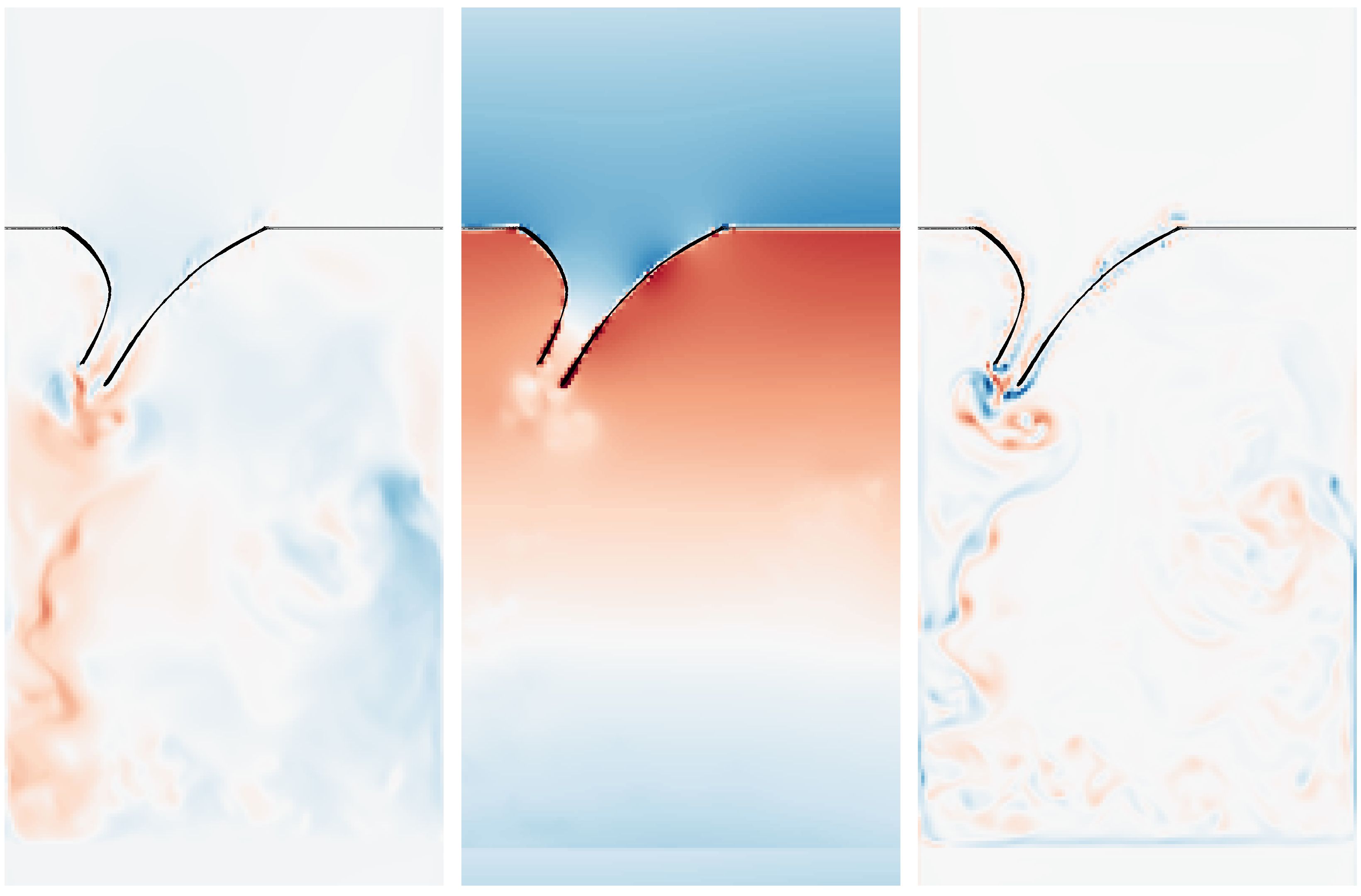} 
\end{tabular} 
}
\caption{
Mitral valve in the process of closing, showing results with standard driving pressure (left), and driving pressure without atrial systole (right). 
This figure shows slice views of the $z$ component of velocity, modified pressure and the component of the vorticity vector that describes rotation within the plane of the figure, which we call out-of-plane vorticity early (top), in the middle (center) and late (bottom) in the period when we expect the breaking jet. } 
\label{three_velocities_breaking_one}
\end{figure}

The first collection displays the flow at $t = 2.09146$ s, near the end of atrial systole during the third beat of the simulation, and is shown in Figure \ref{three_velocities_breaking_one}, top panel, left column. 
The velocity field contains an unbroken jet of forward flow between the leaflets. 
The pressure has a forward pressure drop across the valve. 
A number of tight vortices are visible below the leaflet in the vorticity field. 
These vortices are also visible in the pressure field, in which they appear as localized low pressures. 
Fifteen milliseconds later, at $t = 2.10646$ s, the jet appears to have begun to break, as shown in the central-left panels. 
Near the valve ring, there is an absence of forward flow. 
But in a small region between the free edges of the leaflets, there is a local high velocity and low pressure. 
We believe that this shows the jet in the process of breaking; that is, inertial effects carry the jet forward, even as the valve has otherwise begun to close. 
This creates a pressure differential across the anterior leaflet that may help the suck the leaflets together. 
There is a still lower pressure on the ventricular side of the posterior leaflet, which we suspect may be related to the geometry of the chamber. 
A larger but less organized set of vortices are present below the valve, connected to vorticity shed from the anterior leaflet. 
This structure cannot hit the ventricular walls and return to push the anterior leaflet closed, because there are no ventricular walls here.
However, were this field to occur in a heart, this structure that could plausibly create a vortex that would assist in valve closure. 
Additionally, a small but prominent vortex has just been shed from the posterior leaflet. 
Another fifteen milliseconds later, at $t = 2.12145$ s, a slight remnant of the jet remains between the free edges, as shown in the bottom-left panels. 
By this point, there is a back pressure differential across the entire valve and partition. 
There is a region between the free edges of the leaflets in which the pressure is lower than the pressure on the ventricular side of both leaflets. 
The large region of vorticity below the free edges remains, and appears less organized in this frame.

For comparisons, we examine the same views in the simulation without atrial systole. 
To see if we can identify the breaking jet and vortex, which are now possibly missing, we view this simulation at three time steps immediately before closure.
Figure \ref{three_velocities_breaking_one}, top panels, right column, shows the simulation without atrial systole at $t = 2.09146$ s, during the time when the kick would normally occur. 
Forward flow is much reduced, there is little evidence of a large jet between the leaflets at this time. 
There is a small pressure drop over the anterior leaflet at this time. 
This may be because the papillary muscles are in motion, pulling the leaflet to the left in the frame. 
There is a small vortex being shed from the anterior leaflet, but no larger structures appearing below the valve. 
At time $t = 2.10646$ s, shown in the central-right panels, there is forward flow remaining between the leaflets near the free edges, and a pressure difference across the anterior leaflet. 
This appears to be some amount of smaller breaking-jet-type phenomena than was observed when atrial systole is present. 
The vortex shed from the anterior leaflet remains tight, and does not form a larger structure below the valve. 
At time $t = 2.12145$ s, shown in the bottom-right panels, there is still some forward flow between the leaflets when the flow at the ring is no longer forward.
The pressure may be lower between the leaflets, but this is not obvious. 
There is much less visible structure in the vorticity than was present with atrial systole. 

Both the breaking jet and vortex phenomena are less present in this model without atrial systole. 
This suggests that these phenomena are indeed helped by atrial systole, and lends plausibility to the argument that these effects help the valve close. 
This supports, though does not prove, the assertion that atrial systole assists in mitral valve closure. 
However, since the valve closes effectively in this case, this also suggests that the primary mechanism of closure, the negative pressure difference, is sufficient to close the mitral valve on its own.

To make these comparisons quantitative, we examine the flows when driven by each of the previously described pressures. 
We compute the cumulative flow per beat, i.e., the net volume of blood that passes through the mitral ring during one cardiac cycle.
We also compute the cumulative ``positive flow'' (ignoring flow with a negative sign) to determine the total forward flow, and the cumulative  ``negative flow'' (ignoring flow with a positive sign).  
We compute the cumulative ``systolic flow'' as the cumulative flow during the period in which the pressure difference is negative. 
These values are shown in Table \ref{flow_table}. 
Under standard pressures, the negative flow is approximately three times larger than the systolic flow. 
When the pressure difference reduces, the valve unloads, and blood that is captured in the valve moves forward relative to the ring, recovering some of the apparent backflow in early closing. 
With low systolic ventricular pressure, the cumulative negative flow is slightly less negative, reflecting less loading.
With high systolic ventricular pressure, the cumulative negative flow is slightly more negative, as is the systolic flow. 

\begin{table}[htb]
\centering 
\begin{tabular}{c | c | c | c | c | c | }
Simulation      & \multicolumn{4}{c|}{ Cumulative flow per beat (ml) } \\ 
    & Total & Positive  & Negative & Systolic \\ 
 \hline 
Normal & 68.87 & 74.96 & -6.09 & -1.81 \\ 
 \hline 
Low & 71.00 & 76.49 & -5.49 & -2.58 \\ 
 \hline 
High & 65.63 & 73.38 & -7.75 & -2.17 \\ 
 \hline 
No atrial systole & 57.71 & 64.35 & -6.64 & -2.29 \\ 
 \hline 
\end{tabular}
\caption{Flows under various driving pressures. Systolic flow (cumulative flow under back pressure) is larger in magnitude than ``negative flow,'' the total flow in the negative direction because the valve unloads as negative pressure goes to zero and apparent regurgitation is recovered.}
\label{flow_table}
\end{table}

The simulation with no atrial systole has reduced forward flow, as expected.  
The total negative flow and systolic flow are somewhat more negative. 
Thus, the absence of atrial systole appears to have small but detrimental effects on valve closure.  
In our simulations, this seems to be because the breaking jet mechanism is less prominent in the absence of atrial systole. 
We emphasize that these effects are quantitatively small, and that the valve seals well even in the absence of atrial systole.

\section{Conclusions}

Using a design-based elasticity approach we have built a model mitral valve. 
The model incorporates many realistic anatomical details. 
The form of the constitutive law for this model is taken from experiments. 
The geometry of the valve and the material constants emerge from the requirement that the model supports, through tensile forces, a physiological pressure when the valve is closed.

When simulated under physiologically realistic driving pressures, this model produces flows that match experimental records. 
Features of the flow that are observed, such as the vibration leading to the S$_{1}$ heart sound, emerge. 
Since the simulations are driven by pressures, neither the flows nor the valve motions are prescribed in advance, so the resulting physiological responses under the drastically different conditions of systole and diastole are emergent properties of the model. 
These responses are robust to changing conditions such as hypertension, hypotension and the absence of atrial systole. 
We hope that this work has contributed to the understanding of the basic principles and mechanisms underlying the form and function of the mitral valve.

\section{Acknowledgements}
ADK was supported by the National Science Foundation Graduate Research Fellowship Program, grant DGE 1342536, and a Henry M. MacCracken Fellowship through New York University. 
This work was supported in part through the NYU IT High Performance Computing resources, services, and staff expertise. 
Research on anatomy was performed in collaboration with Mark Alu and Cynthia Loomis of the Experimental Pathology Research Laboratory at the New York University Langone Medical Center. 
The Experimental Pathology Research Laboratory is partially supported by the Cancer Center Support Grant P30CA016087 at NYU Langone's Laura and Isaac Perlmutter Cancer Center.

\bibliographystyle{acm}
\bibliography{mv_refs.bib}


\section*{Appendix}

\setcounter{figure}{0} 
\renewcommand{\thefigure}{A.\arabic{figure}}

\setcounter{section}{0} 
\renewcommand{\thesection}{A.\arabic{section}}

\setcounter{table}{0} 
\renewcommand{\thetable}{A.\arabic{table}}

This appendix includes technical details on the mitral valve anatomy, boundary conditions, and numerical methods used in the article. 

\section{Mitral Valve Anatomy }

\label{anatomy_appendix}

\begin{figure}[ht]
\centering 
\includegraphics[width=.6\columnwidth]{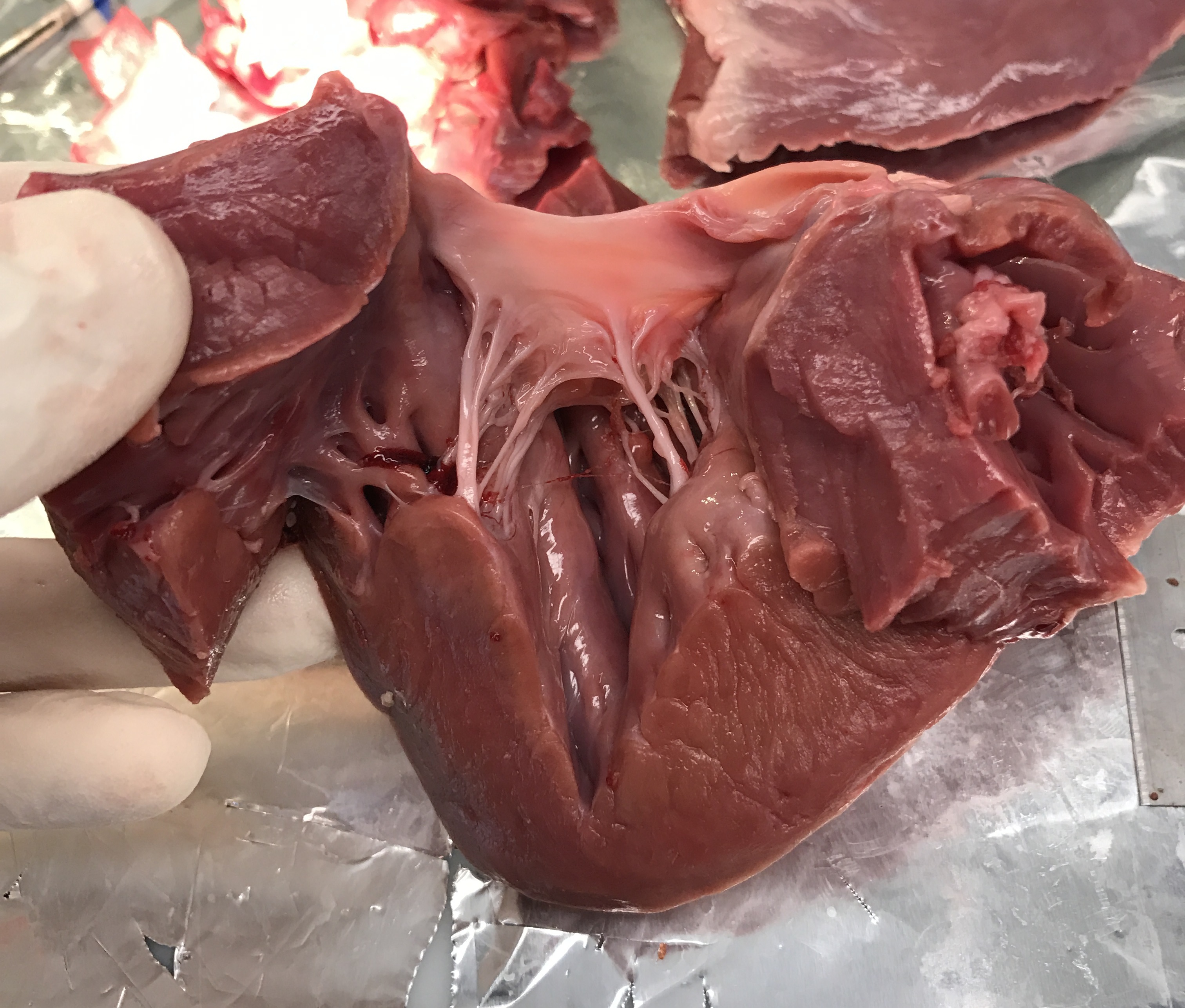}
\caption{Porcine mitral valve in left ventricle. }
\label{valve_in_heart}
\end{figure}

To orient the reader, first we show the mitral valve in the left ventricle of a partially dissected porcine heart in Figure \ref{valve_in_heart}. 

\begin{figure}[htb]
\centering 
\includegraphics[width=.6\columnwidth]{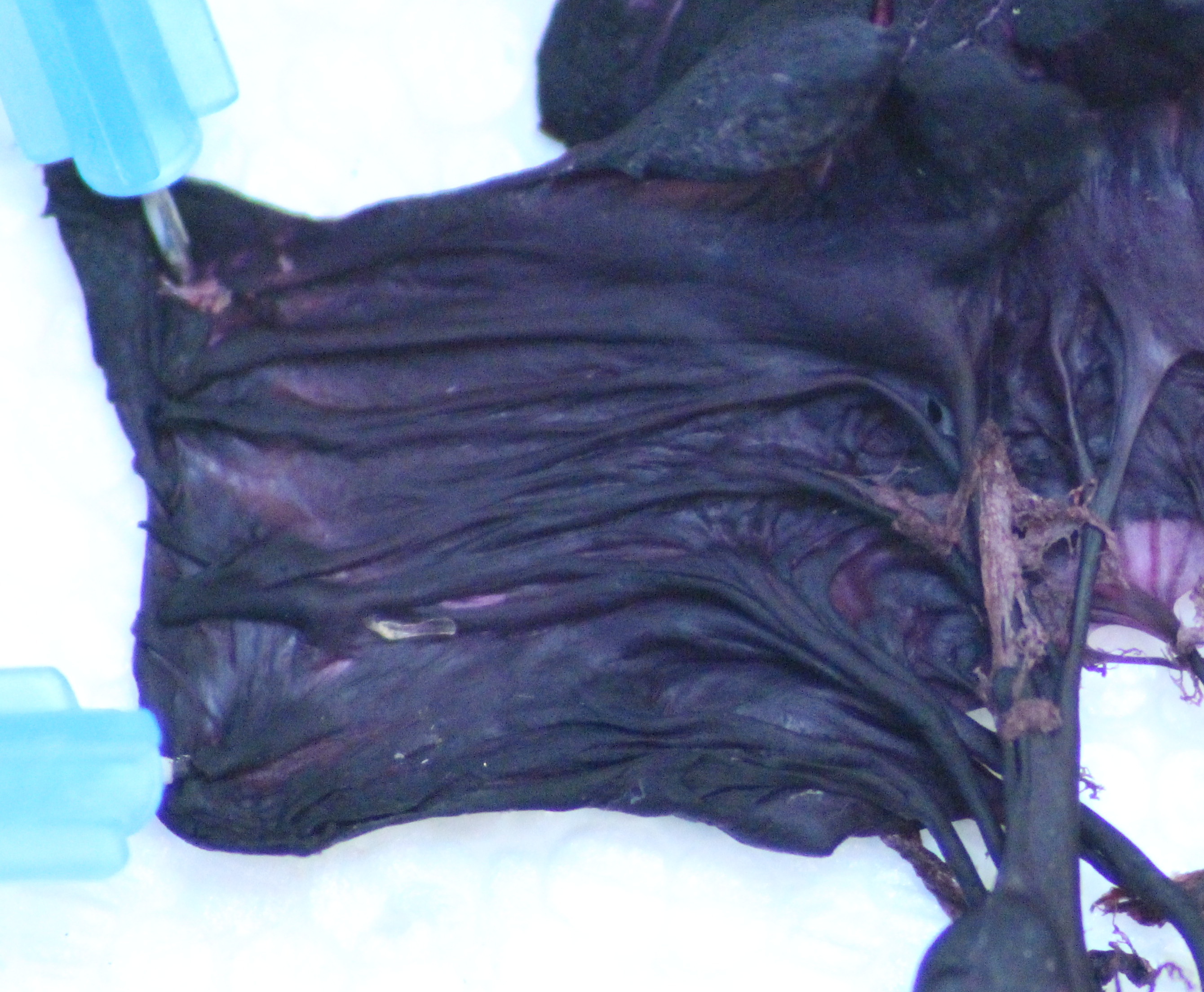} 
\caption{Detail of a portion of the posterior leaflet showing circumferential fibers and branching structure. The valve ring is at the top of the frame; circumferential orientation is horizontal.}
\label{pinned_posterior_detail}
\end{figure}

To examine valve fiber structure, an entire porcine mitral valve was stained with van Gieson's picrofuchsin, following \cite{SAUREN198097}. 
The tissue is highly reactive to the stain, suggesting that there is collagen throughout the leaflets and chordae, as expected. 
Figure \ref{valve_labeled} in Section \ref{anatomy} shows the full valve from the ventricular side. 
At locations where the chordae insert into the leaflet, there is frequently the appearance of a split or fanning out of the chordae. 
At times, the fibers appear to go two ways. 
One large fiber turns towards the opposing papillary muscle, taking on a circumferential orientation. 
The other turns up, taking on a radial orientation in the direction of the valve ring. 
This branch is frequently less conspicuous when viewed from the ventricular side. 
It sometimes appears to sink into the leaflet, perhaps going under other fibers. 
An example is shown in Figure \ref{pinned_posterior_detail}. 
Literature claims that there is a large region of leaflet in which there appear to be two distinct fiber families, and supports the observation that the chordae branch into radial and circumferential fibers \cite{fenoglio1972canine}.

We hypothesize that there are radially oriented fibers, possibly concentrated on the atrial side or below the surface of the ventricular side of the valve leaflets. 
Since, at their insertions, some of the branches of the trees of chordae tendineae appear to point in the radial direction, these fibers may continue towards the valve ring. 
This is less visible, perhaps because the fibers may duck ``under'' the circumferential fibers that are visible on the ventricular side of the leaflet. 
We use this observation in building models of the mitral valve.
The foregoing has been challenging to confirm, since the atrial side has a smoother, more uniform appearance than the ventricular side of the valve. 
Literature supports the concept that the atrial side of the valve has more radially oriented fiber structure than the ventricular side, see Figure 10.7c, right panel in \cite{Lee2015}, which supports, but does not not prove, this hypothesis. 

\begin{figure}[htb]
\centering 
\includegraphics[width=.8\columnwidth]{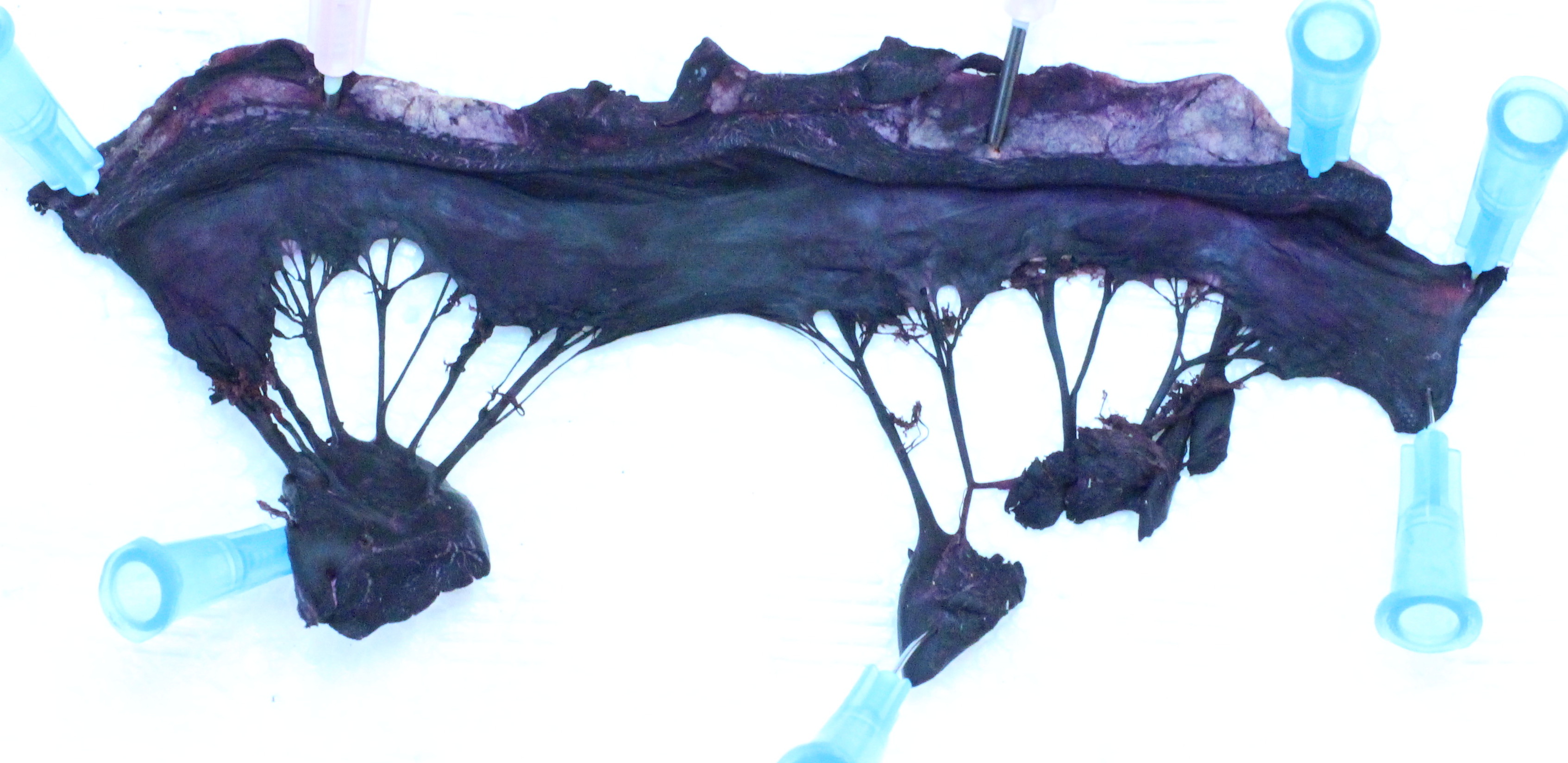}
\caption{Stained porcine valve pinned, atrial side.}
\label{pinned_atrial_zoom_out}
\end{figure}

Figure \ref{pinned_atrial_zoom_out} shows the same specimen from the atrial side. 
There is a band of tissue below and approximately parallel to the valve ring in which there is no apparent separation into distinct leaflets. 
Below this band, separate anterior, posterior and commissural leaflets appear
The texture of the atrial side is smoother than that of the ventricular side, making observations about the fiber structure there challenging. 
At some locations, especially around the commissures, there are small, thin pieces of membranous tissue that lie below some of the insertions of the chordae.

\begin{figure}[ht]
\centering 
\includegraphics[width=.6\columnwidth]{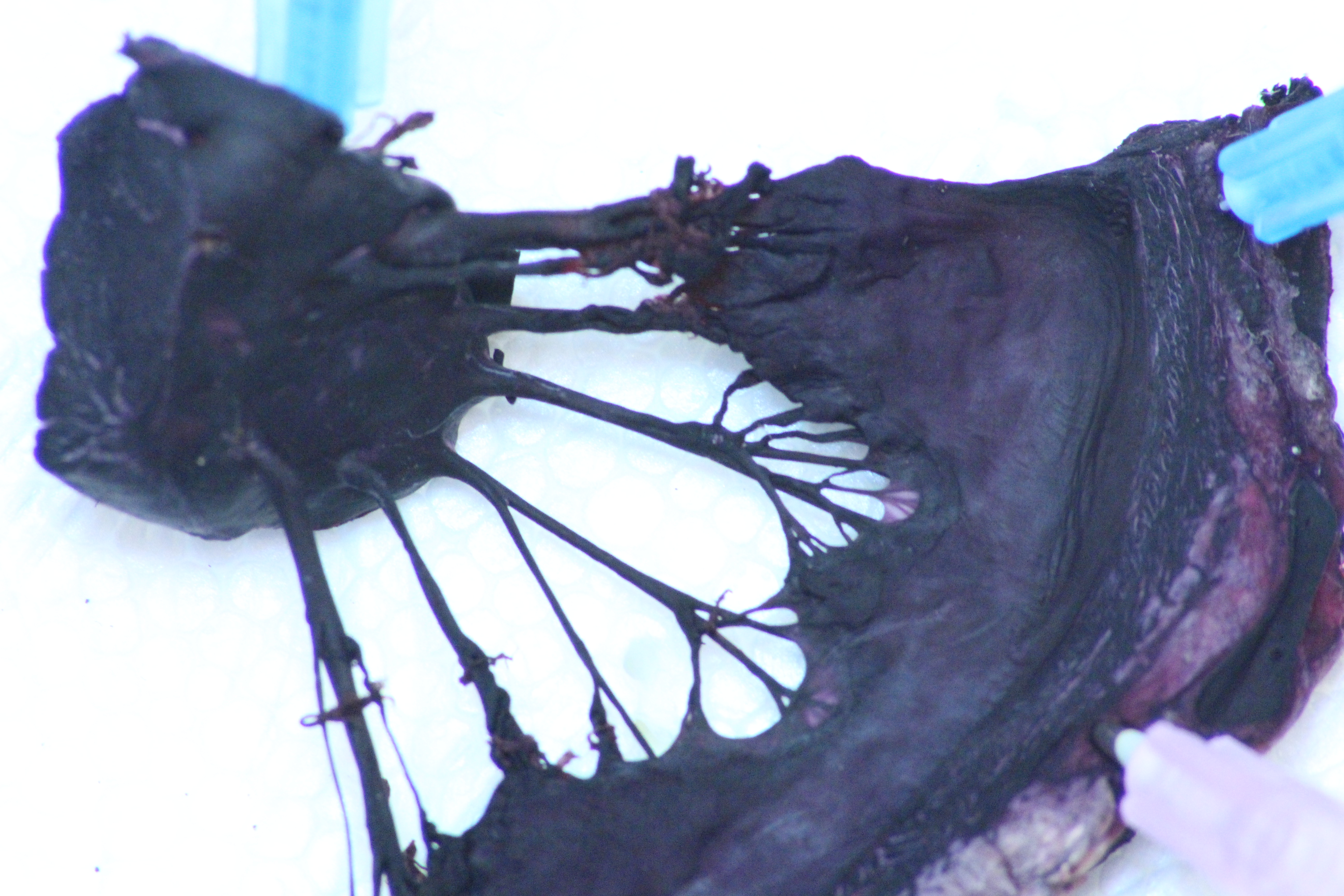}
\caption{Detail of papillary muscle and chordae attachments.}
\label{papillary_detail}
\end{figure}

Figure \ref{papillary_detail} shows details of the attachment of the chordae tendineae to one papillary muscle. 
The attachments are arranged along an approximately circular arc; they occur over about three fourths of the circle. 
The remaining fourth, with no attachment, is roughly pointed at the opposing papillary muscle. 
See \cite{thesis} for additional dissection images and commentary. 

\subsection{Material properties, models in the literature and valve fiber structure}
\label{models_literature}

There appears to be no consensus on the material properties of mitral valve tissue, associated values of coefficients for models and methods to determine local fiber orientation. 
Heterogeneous variations in coefficient properties and differences across individuals are even less explored. 
After much discussion, a recent review notes that ``Still, there is no definite answer to the question [of] which strain energy functions should be used'' when creating a constitutive law for the mitral valve, and that ``we still have a long way to go to identify the coherent stress and strain patterns from different experimental and numerical studies'' \cite{doi:10.1002/cnm.2858}.

Another major complication is the recruitment of collagen fibers, which are wavy and therefore slack at low strains but become straight and therefore stiff as the strain increases.
This may occur at a variety of strains, and the recruitment can take place over a narrower or more broad range of strains. 
Collagen recruitment makes the stress-strain relation nonlinear over
 the range of strains in which collagen recruitment is occurring,
 and this creates major differences between the tangent (or slope)
 stiffness and the chord stiffness of the stress-strain relation.
Differences in the strains required to recruit collagen fibers may lead to the appearance of anisotropy when the fully-recruited tangent moduli may be more similar. 

Our design-based methodology provides an alternative approach to the selection of material properties and parameters for modeling the mitral valve, motivated by the ideal function of the valve and mathematics derived from that idealization, rather than measurements conducted on specific specimens. 
It remains to be seen to what extent the predictions of our design-based approach are confirmed by structural and mechanical measurements on excised mitral valves.
 
We review here some relevant literature on experimental material testing, and on the implications of these studies for valve fiber structure.
We refer to strains at which collagen fibers are fully recruited as \emph{post-transition}. 
Stress/stain curves are reported to be approximately linear in this region in many studies. 
Extrapolating the linear part of the stress-strain curve to zero stress gives an alternate reference point from which to compute strains; we refer to the intersection as the \emph{extrapolated reference strain}. 

May-Newman and Yin reported the ratio of circumferential stress to radial stress under an equibiaxial strain test to be $5.7 \pm 4.38$  in the porcine anterior leaflet \cite{doi:10.1152/ajpheart.1995.269.4.H1319}. 
They compute strains with respect to an excised reference configuration. 
They do not take into account the extrapolated reference strain, which appears different in the radial and circumferential directions. 
They compute the tangent modulus computed at a given membrane stress of $2 \cdot 10^{4}$ dynes/cm, and report for the anterior leaflet a circumferential tangent modulus of $(9.0 \pm 8.8) \cdot 10^{7}$ dynes/cm$^{2}$, radial tangent modulus of $(2.4 \pm 1.4) \cdot 10^{7}$ dynes/cm$^{2}$. 
The post-transition tangent modulus is much less anisotropic; it is closer to 1:1 by visual inspection of their Figure 4.

Pham et al. tested mechanical properties of human valve leaflets \cite{PHAM2017345}.
They report post-transition tangent modulus in the anterior leaflet of $ (12.8 \pm 3.9) \cdot 10^{7}$ dynes/cm$^{2}$ (circumferential) and $ (6.9 \pm 2.2) \cdot 10^{7}$ dynes/cm$^{2}$ (radial), giving an approximately 2:1 ratio of post-transition tangent modulus in the circumferential to radial directions. 
For the posterior leaflet they report post-transition tangent moduli of $(4.1 \pm 0.1) \cdot 10^{7}$ dynes/cm$^{2}$ (circumferential) and $(0.6 \pm 0.1) \cdot 10^{7}$ dynes/cm$^{2}$ (radial).
They refer to the extrapolated reference strain as the ``extensibility'' of the valve in the given direction, and report mean extensibility in the anterior leaflet of $0.04$ (circumferential) $0.09$ (radial) and posterior leaflet of $0.03$ (circumferential) and $0.12$ (radial).
These are dramatic differences in the strains at which fibers are recruited. 

Kunzelman and Cochran performed uniaxial tests on porcine mitral valve tissue, and reported an approximately 48\% difference in post-transition tangent modulus in the anterior leaflet depending on whether the test strip is taken from center of the leaflet or near the free edge \cite{kunzelman_experimental}. 
In the anterior leaflet, they reported a ratio of circumferential to radial post-transition tangent modulus of 2.5-3.7:1, depending on which part of the anterior leaflet is used for comparison. 
For the posterior leaflet they reported approximately equal post-transition tangent moduli of $(2.0 \pm 0.2) \cdot 10^{7}$ dynes/cm$^{2}$ and $(1.9 \pm 0.2) \cdot 10^{7}$ dynes/cm$^{2}$ in the circumferential and radial directions, respectively.
Their measurements also vary with the rate at which tissue is strained; see their Table 1 for more information.

We believe that these data rule out a mitral valve model in which collagen fibers run exclusively in the circumferential direction, as suggested by the highly cited study \cite{PMID:1751231} based on light-scattering data, since such a model would predict much more extreme anisotropy  ratios than 2:1 and 1:1.  
Indeed, the aortic valve is supported primarily by circumferential collagen fibers, and its stiffness anisotropy ratio is 20:1 \cite{SAUREN1983327}.

Simulations have produced a broad ranges of stresses using a broad range of models. 
One simulation study shows highly heterogeneous principal stresses in the leaflets during valve closure ranging from 10$^{4}$ to 10$^{7}$ dynes/cm$^{2}$, see Figure 7 in \cite{Toma2016}. 
A review summarizes stresses that emerge in many modeling studies; they report a difference of well over one order of magnitude among the modeled stresses \cite{doi:10.1002/cnm.2858}. 
One study reports stress of $7.2 \cdot 10^{5}$ dynes/cm$^{2}$ \cite{wenk2010first}. 
Another has a mean stress of $1.5 \cdot 10^{7}$ dynes/cm$^{2}$ (maximum over time, average over animal-specific models); the maximum on a single animal-specific model is $3.0\cdot 10^{7}$ dynes/cm$^{2}$ \cite{KRISHNAMURTHY20091909}. 
Another modeling study claims that ``leaflet as a whole thus acts essentially as an isotropic, nonlinear-elastic membrane'' \cite{doi:10.1002/cnm.3142}.

One systematic study compared a number of discrete fiber models with fiber-based models that include distributions of fibers. 
When they compare a ``Bimodal fiber distribution versus two classes of reinforcing fiber'' they conclude that ``a two-fiber class model is an appropriate simplification from a bimodal continuous distribution for physiologically relevant modes of planar biaxial deformation'' \cite{BISCHOFF2006167}.

Estimates of strain experienced in vivo or in vitro vary widely between studies. 
One in vivo study reported strains of ``circumferential strains are $5.0 \pm 2.7\%$, and radial strains are $7.8 \pm 4.3\%$'' \cite{RAUSCH20111149}. 
Another study by a different group computed stretches in vivo with a similar technique, then compared to excised geometry of the same specimens \cite{Amini2012}. 
They they use Euler-Almansi strains in their work, so we report their values of stretch, the ratio of current to initial length, here.
They first compute stretches relative to the in vivo configuration at minimum left ventricular pressure, and report circumferential stretches of 1.10 $\pm$ 0.01 and radial stretches of 1.31 $\pm$ 0.03 during peak systole. 
They next compute stretches relative to an excised configuration of the same specimens, and report circumferential stretches of 1.22 $\pm$ 0.07 and radial stretches of 1.65 $\pm$ 0.08. 
The in vivo configuration at minimum left ventricular pressure has circumferential stretch of 21\% and radial stretch of 41\% compared to the excised state. 
This suggests that the valve is residually stressed through the cardiac cycle in vivo.

In one study it was concluded that the anterior mitral valve leaflet, when measured in vivo, behaves much more linearly than reported in other in vitro studies \cite{KRISHNAMURTHY20091909}. 
One possible explanation for this behavior is that the mitral valve is strained and stressed throughout the cardiac cycle so that collagen fibers remain fully recruited (i.e., straight) at all times.
Excising pieces of the valve leaflets would relive this residual stress and strain, thus suggesting a no- or low-stress reference state exists on the laboratory table that is not achieved in a living animal. 
In a followup to this study, the authors show that including prestrain when deriving constitutive laws changes material stiffness parameters by multiple orders of magnitude, and dramatically alters material response on simulated uniaxial tests \cite{Rausch2013}.

\section{Construction of the Model Mitral Valve }

\subsection{Inertio-elastic timescale}
\label{Inertio-elastic_timescale}

To justify our construction, we estimate the inertio-elastic timescale of the system. 
This determines, very roughly, the time that the valve will take to reach equilibrium under the specified forcing. 

Because the valve has a complex geometry, we estimate the timescale using the model of a pressurized cylinder. 
Let $\sigma$ denote the hoop stress, $h$ the thickness, $\ell$ the length, $r$ the radius and $p$ the pressure. 
Consider mechanical equilibrium of half of the cylinder, sliced in a plane containing the axis of the cylinder. 
The pressure acts on an area of $2r \ell$. 
The stress acts in the opposite direction over an area of $2h \ell$. 
After canceling the factor of $2 \ell$, this gives the following relationship between hoop stress and pressure 
\begin{align}
\sigma h = p r . 
\end{align}
This is referred to as \emph{Laplace's law}. 
Diagrams and further discussion can be found in \cite{PRESSURE_VESSELS}. 
The force on a patch of the cylinder with area $A$ is $pA$. 
Applying Laplace's law gives the force as 
\begin{align}
F = \frac{\sigma h A}{r}  
\end{align}
The stress is locally related to the strain $E$ and local elastic modulus $\eta$ by the linear relation $\sigma = \eta E$. 
Suppose that the radius increases under this load by some value $\delta$. 
Let $C$ denote the circumference of the hoop. 
The strain is then  
\begin{align}
E = \frac{\Delta C}{C} = \frac{2\pi \delta}{2\pi r} = \frac{\delta}{r} . 
\end{align}
Combining these, we have 
\begin{align}
F = \frac{\eta h \delta A}{r^{2}}  
\end{align}
The mass $m$ of the relevant portion of the cylinder is given as 
$m = \rho h A$,
where $\rho$ is the density. 
The acceleration $a$ is 
$a = \delta/t^{2}$, 
where $t$ is the timescale of interest.
Substitute these quantities into Newton's second law, $F = ma$ to obtain 
\begin{align}
\frac{\eta h \delta A}{r^{2}}  = \frac{\rho h A \delta}{t^{2}} . 
\end{align}
Cancelling $h \delta A$ and rearranging shows that 
\begin{align}
t = r \sqrt{  \frac{\rho}{\eta} } . 
\end{align}
We estimate the elastic modulus from the slope of the affine portion of a uniaxial test, see Figure 3a in \cite{sacks_collagen_constitutive}. 
Its value is approximately $\eta = 2.6 \cdot 10^{8}$ dynes / cm$^{2}$. 
The radius is $r = 2.19$ cm and the density $\rho = 1$ g/cm$^{3}$. 
Using these values, we obtain
\begin{align}
t = 2.19 \sqrt{ \frac{1}{2.6 \cdot 10^{8}}} s \approx   1.36 \cdot 10^{-4} s 
\end{align}
This time scale is approximately three orders of magnitude shorter than the duration of closure, which is approximately 0.2 s. 
Thus, we can expect the analysis made using the assumption of equilibrium to be a good predictor of the dynamics in general.

\subsection{Boundary conditions for the mitral valve ring}
\label{Boundary-conditions-mitral-valve-ring}

To specify boundary conditions at $v = 1$, we first comment on the anatomy of the mitral ring. 
It is observed that the valve ring takes on a ``rounded-D'' or ``lima-bean'' shape, especially in systole.  
The anterior leaflet is centered on a slight depression in the ring, as shown in an anatomy study, Figure 1 in \cite{DEGANDT20071250}. 
One review of valve anatomy describes this as a ``mild concave form, because it is directly related to the circular aspect of the aortic orifice'' \cite{Misfeld1421}. 
That is, the center of the anterior leaflet corresponds to the location at which the mitral valve ring is adjacent to the aorta.
The slight concavity may be caused by the valve ring being squeezed against the aorta during ventricular contraction. 
It is also observed that the anterior leaflet takes up less than half of the valve ring in terms of angle \cite{mccarthy2010anatomy}.

\begin{figure}[ht]
\centering
\includegraphics[scale=.6]{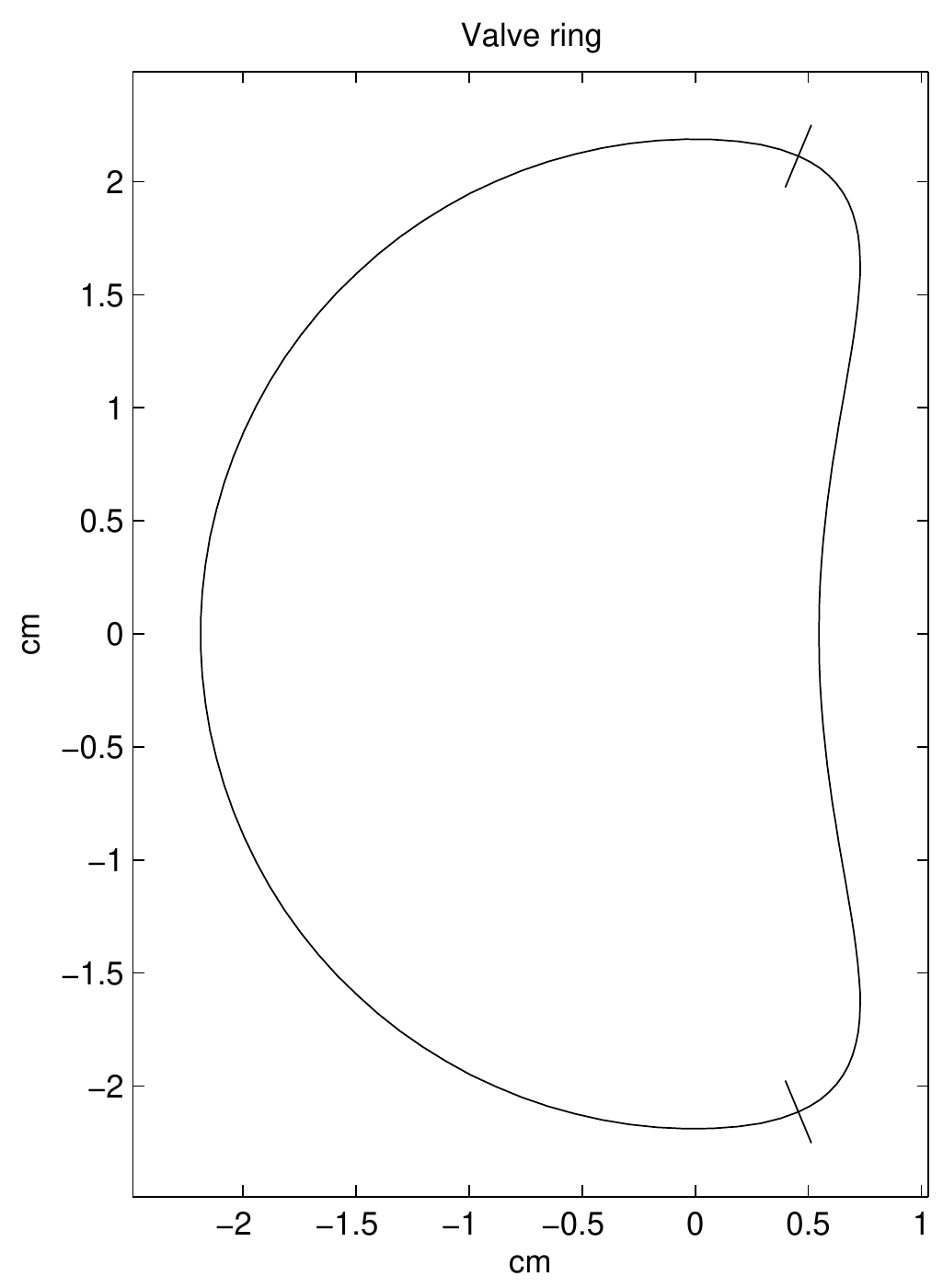}
\caption{Valve ring. The anterior leaflet attaches on the right of the frame, between the two ticks, and the posterior on the remainder of the ring.}
\label{valve_ring_2d}
\end{figure}

Thus, we specify that the shape of the ring is a semicircle on the posterior side, and is modified from a semicircle to be slightly concave on the anterior side. 
We specify that each leaflet takes up a prescribed angle of this lima-bean shaped ring, the anterior taking up under half of the total angle. 
The ring is shown in Figure \ref{valve_ring_2d}. 
To implement this, we define a number of parameters and functions.
Let $\theta$ denote the angle around the valve ring, and arrange the coordinates such that $\theta = 0$ corresponds to the center of the anterior leaflet.
This implies that $\theta = \pm \pi$ corresponds to the center of the posterior leaflet.   
Let  
\begin{align}
\varphi(\theta) = 
\begin{cases}
\cos^{2}(\theta)   & : \; | \theta | < \pi/2 \\ 
0                         & : \; | \theta | \geq \pi/2 , \\ 
\end{cases}
\end{align}
where $\theta$ is restricted to $\theta \in [-\pi,\pi]$.
Let $a_{ant} = 5\pi/6$ and $a_{post} = 2\pi - a_{ant} = 7\pi/6$ denote the angles that the anterior and posterior leaflet, respectively, occupy on the ring. 
The anterior leaflet is fixed to the ring for $\theta \in (-a_{ant}/2, a_{ant}/2)$, and the posterior leaflet is fixed to the ring for $\theta \in (\pi -a_{post}/2, \pi + a_{post}/2)$. 
Let $\rho_{dip}$ be a tunable parameter that controls the slight concavity of the valve ring around the anterior leaflet, which we select as $\rho_{dip} = 0.75$. 
The anterior leaflet attaches to the ring on $u \in [0,1/2], v = 1$, and the posterior attaches to the leaflet on $u \in (1/2,1), v = 1$. 
The value of $\theta$ is expressed as a linear function of $u$, which is given as 
\begin{align}
\theta(u) = 
\begin{cases}
a_{ant}2u - a_{ant}/2   & : \;  u \in [0,1/2]  \\ 
a_{post}2(u - 1/2) - a_{post}/2 + \pi                        & : \;   u \in (1/2,1)  \\ 
\end{cases}
\end{align}
We define the valve ring boundary condition as 
\begin{align}
\bb X(u,1) = r ( \cos(\theta(u)) -  \rho_{dip} \varphi(\theta(u)) , \sin(\theta(u)), 0)
\label{valve_ring_function}
\end{align}
We estimate the radius $r$ of the semicircular arc as 2.19 cm. 
Note that if $\rho_{dip} = 0$ the valve ring would be a circle, and for $\rho_{dip} > 0.5$ the ring has the desired concavity.

Commentary from medical device manufacturers suggests a 3:4 ratio of the anterior-posterior diameter to the transverse (commissure to commissure) diameter is desirable \cite{Annuloplasty_ring}, which in our framework implies $\rho_{dip} = 0.5$ and no concavity in the center of the anterior leaflet. 
However, this device is used for patients with annular dilation and it may be that a lower ratio creates other problems such as reducing forward flow during diastole. 
Using $\rho_{dip} = 0.75$ gives a ratio of 5:8, which is slightly narrower in the anterior-posterior diameter than specifications from this manufacturer suggest, but we believe that the slight concavity given by this value is more anatomical for modeling a normal valve.

\subsection{Further details of discretization and their relationship to anatomy}
\label{further_discretization}
Regions of the mesh that correspond to different parts of the valve have different properties. 
We view the anterior leaflet as $u \in [0,1/2]$, and the posterior as $u \in (1/2,1)$. 
The anterior and posterior regions have location dependent values of the coefficients $\alpha, a, \beta, b$. 
This is discussed further in Section \ref{Values_coefficients} and values are shown in Table \ref{coefficients_table}. 
Circumferential fibers with $v > 1 - 1/64$ are seen as rings, and are topologically circles.
The value of $1/64$ is selected to match, at least approximately, the anatomy, and accordingly to create 8 rings when $N = 512$. 
They too have their own $\alpha, a$. The parameter $\beta$ does not have a distinct value in this region, since radial fibers in that region are continuous with radial fibers in the leaflets.
However, $b$ does have a distinct value, so that we may use it to influence fiber spacing in the rings. 

Finally, we add a small section of ``edge connectors'' at the commissures. 
We found that many variants of our model were prone to leaking in this region,, and adding a patch of extra tissue helped to prevent this. 
The edge connector region also has its own constants. 
Radial fibers in this region simply end at the free edge, rather than connect to chordae tendineae. 
Whatever tension they support must be supported by curvature in the circumferential fibers in that region. 
Adding these fibers makes more fibers into topological circles, but we still think of the regions as being separate because their parameters are different from those of the rings and leaflets. 
Fibers that are not included in other regions with $v \in [ 1 - 2/64, 1 - 1/64]$ are seen as edge connectors, which creates 8 edge connectors when $N = 512$.
This range was selected to visually match the anatomy and reduce leaks at the commissures. 

Actual chordae tendineae have complicated and highly variable branching structures.  
In our model, however, we use only binary trees, in which every branch has exactly two descendants.  
(For the terminal branches, these descendants are a radial and a circumferential fiber of a valve leaflet.)  
The root of each tree is a point on the tip of a papillary muscle (see below) from which the ancestral branch, perhaps better called the trunk, of the tree emanates.  
The location of the root of each tree is prescribed (see below), but the locations all of the other tree junction points are determined as part of the solution of the equilibrium equations for the structure of the model valve.
We observe that there are one to two chordae anchored in each papillary muscle that support the anterior leaflet, another one to two that support the central posterior leaflet. 
We observe that there are two to four chordae that support either side of the posterior leaflet near the commissures (also referred to as the commissural leaflets). 

The model contains eight trees. 
In the model, we place one tree per papillary muscle for the anterior leaflet, one for the posterior leaflet, and two trees, attached to the posterior leaflet, for the commissural region. 
(See Figure \ref{mesh_schematic_II.eps}.)
It is possible that these numbers should be increased slightly, as many specimens have more chordae emerging from the papillary muscles than this. 
Indeed, we tried models that did so, and incorporated explicit commissural leaflets. 
These models are even more complicated to tune, and were generally less robust in simulation, and thus we leave them for future work. 
See \cite{thesis} for discussion. 
The locations of the roots of the trees are prescribed as boundary conditions, and spread across a portion of a small circle surrounding the center of the papillary tip.
Note that we specify the topology of the trees, but do not specify the lengths of any links. 
As in the leaflets, the physical configuration emerges from solving the equilibrium equations.

\subsection{Scaling of coefficients in trees of chordae tendineae}
\label{scaling_laws}

Here we derive a rule to determine how $\gamma$ and $c$ scale in the trees of chordae, the number of leaves of which scale with $N$ but the number of roots of which do not scale at all. 
For each tree in the model, we would like the total force applied by the tree as a whole to the leaflets to be approximately constant as the mesh is refined. 
Let $L$ be the number of leaves in a tree, and let $l$+1 be the number of generations, so that $L = 2^l$. 
We number the generations by the index $g = 0 \dots l$, where $g$=0 denotes the generation containing the finest branches, which connect to the leaves, and $g=l$ denotes the generation containing the trunk of the tree, which connects to the root.  
We denote the tension coefficient in generation $g$ of a tree with $L$ leaves by $\gamma_{g,L}$.
Now consider $\gamma_{0,L}$.
Under a refinement of $N$ by a factor of two, there are twice as many leaves. 
Recall that $N$ is the number of points on the valve ring, and that $1/N = \du = \dv$. 
Note that $L$, the number of leaves in any tree, is proportional to $N$. 
This means that each leaf should transmit half the force of the leaves in the coarser resolution tree, so the coefficient $\gamma_{0,M}$ should scale as 
\begin{align}
\gamma_{0,2L} = \frac{\gamma_{0,L}}{2} . 
\end{align}
Applying this relationship repeatedly, we find the relationship of $\gamma_{0,L}$ to $\gamma_{0,1}$ the coefficient in the degenerate case of a tree with one leaf. 
\begin{align}
\gamma_{0,L} = \frac{\gamma_{0,1}}{2^{l}} = \frac{\gamma_{0,1}}{L} 
\end{align}
The constant $\gamma_{0,1}$ is a tunable free parameter, representing the total of the coefficients in the leaf generation. 

Each tree has a single link that is connected to its root, and the tension coefficient in this link, $\gamma_{root,L}$ should remain constant as the mesh is refined. 
Thus we set $\gamma_{root,L}$ equal to $\gamma_{root}$, independent of $L$; this is a tunable, free parameter. 

Finally, to determine the coefficients throughout, we assume that the tension coefficient at each generation in the tree is a constant multiple of the tension coefficient of the next generation. 
That is, 
\begin{align}
\gamma_{g+1,L} = m \gamma_{g,L}
\label{coeffscale1}
\end{align}
for any generation index $g$ in the tree. 
The factor $m$ is selected to preserve the prescribed values of $\gamma_{0,1}$ and $\gamma_{root}$. 
Since there are $L = 2^{l}$ leaves in the tree, there are $l + 1$ levels. 
Then 
\begin{align} 
\gamma_{root} = m^{l} \gamma_{0,L} = m^{l} \frac{ \gamma_{0,1} }{2^{l}}  . 
\end{align}
This implies that 
\begin{align} 
m = \left(  \frac{\gamma_{root}}{ \gamma_{0,1} / 2^{l} } \right)^{1/ l} = 2 \left(  \frac{\gamma_{root}}{\gamma_{0,1}} \right)^{1/ l }   . 
\label{coeffscale3}
\end{align}

The coefficient $c_k$, which has units of length, is scaled relative to the tension coefficient $k$ (hence the notation $c_k$). 
This has the effect that when the mesh is refined by doubling $N$, the branches of the new finest level of the chordal trees are about half as long as the branches of the old finest level, that the branch that connects to the root of the tree maintains its length under mesh refinement, and that the other generations of the tree interpolate between these behaviors.
We write $c_{0}$ for the coefficient corresponding to $\gamma_{0,1}$ in the leaves, and $c_{root}$ for the coefficient corresponding to $\gamma_{root}$. 
We scale $c_{0}$ as 
\begin{align}
c_{0} \propto \frac{1}{N}
\end{align}
regardless of $L$. 
This ensures that the finest level of the trees is approximately the same length scale as the leaflet mesh. 
The root coefficients $c_{root}$ are constants. 
Intermediate coefficients $c_{g}$ are scaled analogously to $\gamma_{g}$, as derived in formulas \eqref{coeffscale1}-\eqref{coeffscale3}.

\subsection{Values of coefficients}
\label{Values_coefficients}

The model requires many coefficients. 
They are selected manually, by trial and error and inspection of solutions by eye, to achieve a closed shape after solving the equations of equilibrium. 
We define a base tension $\tau = 2.4240 \cdot 10^{6}$ dynes. 
For our nominal value of $p = 100$ mmHg = $1.3332 \cdot 10^{5}$ dynes/cm$^{2}$, this number was computed by inspecting solutions visually for closure and subsequently modifying the ratio $p/\tau$. 
This ratio was tuned to the value 0.055/cm$^{2}$.  
We call the coefficients in the decreasing tension model that determine the shape of the curve, $a,b,c$ the decreasing tension coefficients. 
We define a base length for the decreasing tension coefficients $\ell = 4.6$ cm, which was selected by trial and error. 
Other coefficients are scaled from these values. 
Table \ref{coefficients_table} summarizes their values. 

To select these parameters, we built a tool to change these parameters interactively. 
This allows the user to change a single parameter in a converged solution, then use the previous solution as an initial guess.
This proved essential to hand tune the many parameters here. 
In the future, automated tuning of the parameters with optimization algorithms would be a significant step forward for this line of research. 

\begin{table*}[ht]
\centering 
\begin{tabular}{ | c  |  c |  c  |  c |  c | c |  c |  c |  c | }
Fiber Type	& Max tension   &  Max tension  & Max tension & Dec tension & Dec tension   & Dec tension & Leaves \\ 
	&   name  &   $\cdot 10^{6}$ dynes &  $ \cdot 10^{6}$ dynes &  name  &  cm & cm  & number   \\ 
	&    &   (unscaled)  &  (scaled)  &   &  (unscaled)  & (scaled)   &  (if any) \\ 
\hline 
\multicolumn{8}{|c|}{ Anterior  } \\ 
\hline 
Circumferential & $\alpha$ & 1.75 $\tau$ &  4.24 & $a$ &  2.5 $\ell$  & 11.5 & \\ 
\hline
Radial & $\beta$ & 1.10 $\tau$ &  2.67 & $b$ &  1.5 $\ell$  & 6.9 & \\ 
\hline
Rings  & $\alpha$ & 0.50 $\tau$ &  1.21 & $a$ &  2.0 $\ell$  & 9.2 & \\ 
(Circumferential) &  &  &  &  &    &  & \\ 
\hline
Rings & $\beta$ & 1.10 $\tau$ &  2.67 & $b$ &  0.5 $\ell$  & 2.3  & \\ 
(Radial) &  &  &  &  &    &  & \\ 
\hline 
Chordae - Leaves & $\gamma_{0,1}$ & 0.50 $\tau$ &  1.20 & $c_{0}$ &  (1.0/$N$) $\ell$  & 4.6/$N$ & $N$/4 \\ 
\hline
Chordae - Roots & $\gamma_{root}$ & 0.29 $\tau$ &  0.71 & $c_{root}$ & 0.0039 $\ell$ & 0.018 &  \\ 
\hline 
\multicolumn{8}{|c|}{ Posterior  } \\ 
\hline 
Circumferential & $\alpha$ & 1.00 $\tau$ &  2.42 & $a$ &  1.0 $\ell$  & 4.6 & \\ 
\hline
Radial & $\beta$ & 1.00 $\tau$ &  2.42  & $b$ &  1.5 $\ell$  & 6.9 & \\ 
\hline
Rings & $\alpha$ & 0.50 $\tau$ &  1.21 & $a$ &  2.0 $\ell$  & 9.2 & \\ 
(Circumferential) &  &  &  &  &    &  & \\ 
\hline
Rings & $\beta$ & 1.00 $\tau$ &  2.67 & $b$ &  0.5 $\ell$  & 2.3 & \\ 
(Radial) &  &  &  &  &    &  & \\ 
\hline
Chordae - Leaves & $\gamma_{0,1}$ & 0.18 $\tau$ &  0.44 & $c_{0}$ &  (1.0/$N$) $\ell$ & 4.6/$N$ & $N$/8 \\ 
(central) &  &  &   &  &   &  &  \\ 
\hline
Chordae - Roots & $\gamma_{root}$ & 0.11 $\tau$ &  0.26 & $c_{root}$ & 0.0039 $\ell$ & 0.018 &  \\ 
(central) &  &  &   &  &   &  &  \\ 
\hline
Chordae - Leaves & $\gamma_{0,1}$ & 0.11 $\tau$ &  0.27 & $c_{0}$ &  (1.0/$N$) $\ell$ & 4.6/$N$ & $N$/16 \\ 
(commissure) &  &  &   &  &   &  &  \\ 
\hline
Chordae - Roots & $\gamma_{root}$ & 0.07 $\tau$ &  0.16 & $c_{root}$ & 0.0039 $\ell$ & 0.018 &  \\ 
(commissure) &  &  &   &  &   &  &  \\ 
\hline 
\multicolumn{8}{|c|}{ Edge connector} \\ 
\hline
Circumferential & $\alpha$ & 1.00 $\tau$ &  2.42 & $a$ &  2.0 $\ell$  & 9.2 & \\ 
\hline
Radial & $\beta$ & 0.01 $\tau$ &  0.02 & $b$ &  2.0 $\ell$  & 9.2 & \\ 
\hline
\end{tabular}
\caption{Values of coefficients in model}
\label{coefficients_table}
\end{table*}

\subsection{Numerical method for construction of the model mitral valve}
\label{Numerical-method}

In this subsection, we describe the details of solving the nonlinear system that arises in building the model mitral valve. 
The equilibrium equations for the leaflets and chordae form a nonlinear system of difference equations. 
This nonlinear system is solved by Newton's method, with a number of modifications to increase robustness. 

Let $\Phi_{i}$ denote the vector of all unknowns on iteration $i$, let $F(\Phi_{i}) = 0$ denote the entire system of equations, and let $J(\Phi_{i})$ be the Jacobian of $F(\Phi_{i})$. 
The iteration for Newton's method with line search is given by 
\begin{align} 
\Phi_{i + 1} = \Phi_{i} - s_{i} J(\Phi_{i})^{-1} F(\Phi_{i}) , 
\label{newtons_line_search} 
\end{align} 
in which we solve a linear system instead of explicitly inverting $J$. 
The scalar parameter $s_{i} \in (0,1]$ is the line search parameter. 
At each step the value of $s_{i}$ is initialized to one. 
If the Euclidean norm of $F$ decreases with this value of $s_{i}$, then the value of $s_i$ is accepted. 
Otherwise the value of $s_{i}$ is decreased by half repeatedly until the Euclidean norm of $F$ decreases, which is guaranteed to occur for $s_i$ sufficiently small, since $|F(\Phi_{i+1})|$ in \eqref{newtons_line_search} is a decreasing function of $s_i$ for $s_i$ sufficiently small.  
This is shown by differentiating $|F(\Phi_{i+1})|^2$ with respect to $s_i$ and then setting $s_i$ = 0.

\begin{figure}[ht]
\centering 
\includegraphics[width=.5\columnwidth]{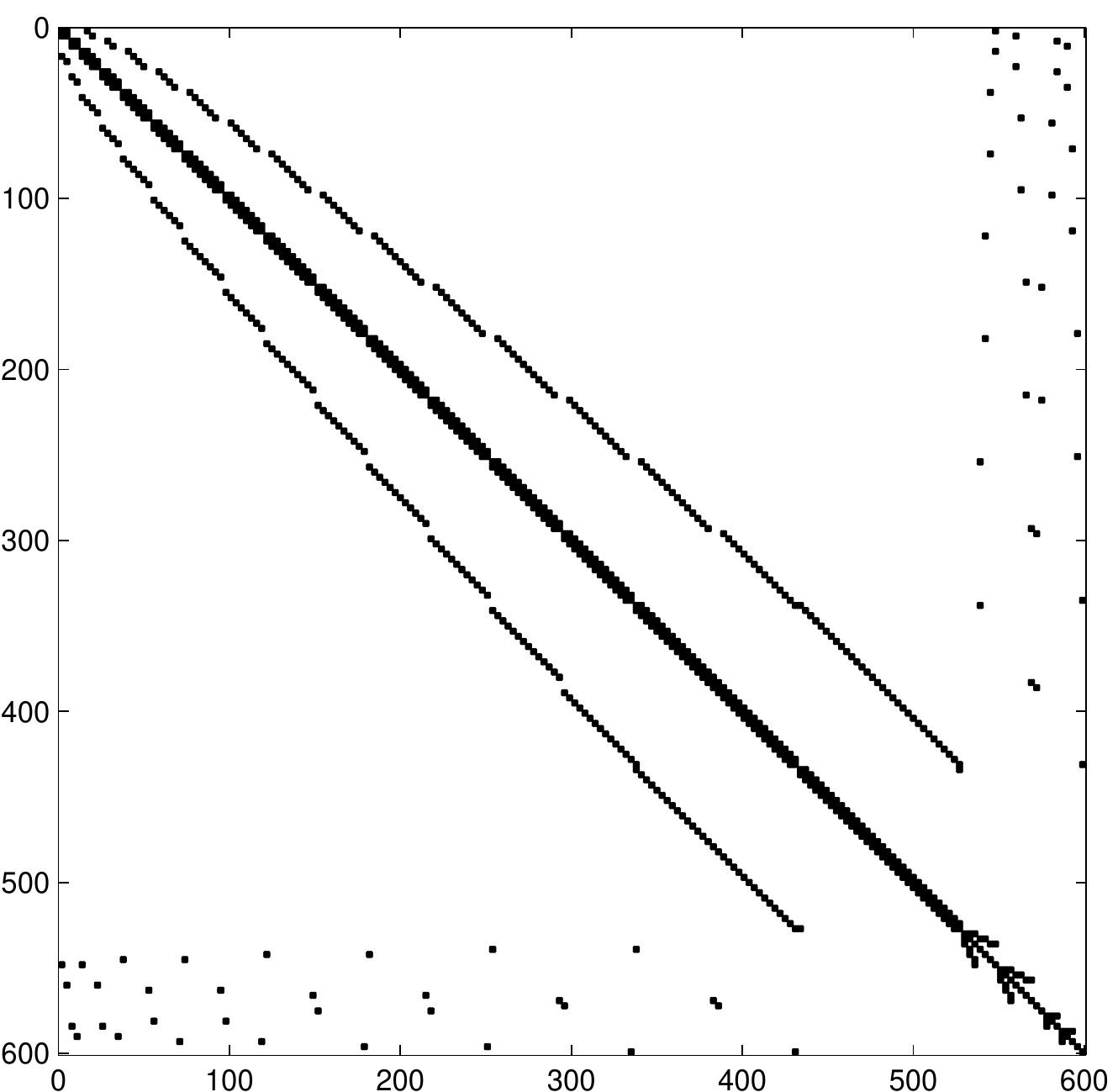}
\caption{Nonzero pattern of Jacobian. Low resolution, $N = 32$, is used to better show structure.}
\label{jacobian_nnz}
\end{figure}

This implies that we need the Jacobian of the entire nonlinear system of difference equations. 
Experiments involving approximations to the Jacobian and attempts to use various Jacobian-free optimization algorithms failed completely. 
Thus, we compute the form of the entire Jacobian analytically and implement a program to evaluate this analytic form for any given configuration of the model valve. 
The Jacobian has a large block in the upper left that corresponds to the leaflets. 
The lower right contains blocks internal to the various trees of chordae, and the two remaining blocks correspond to insertions of the tree into the leaflet. 
The nonzero pattern, shown in Figure \ref{jacobian_nnz}, of this matrix is symmetric, but the matrix is not, owing to blocks corresponding to the pressure that are skew-symmetric, and that are not equal to the transpose of their image block on the other side of the main diagonal of the Jacobian as a whole.

The overall Jacobian is built by taking Jacobians of individual 3-vector equations.  
All of the derivatives, gradients, and individual Jacobians that are needed are computed by hand.
Data structures determine which nodes are internal to the leaflets and which are on the boundaries, and manage constants and coefficients through all included structures. 
They specify which nodes are internal to the chordae, and how they connect to the leaflet. 
The Jacobian program reads these data to find which types of derivatives need to be computed for each term, calculates the associated three by three blocks, and places them into the appropriate location in the overall Jacobian.

Pressure blocks take the form of matrices which apply the cross product up to signs. 
Let 
\begin{align} 
C(\bb X) = 
\begin{bmatrix*}[c]   
 0 &  -X_{(3)} &  \phantom{-}X_{(2)} \\  
   \phantom{-}X_{(3)}  &   0 & -X_{(1)}  \\
  -X_{(2)}  & \phantom{-}X_{(1)}  &  0
 \end{bmatrix*}
\end{align} 
Note that $C(\bb X) \bb Y = \bb X \cross \bb Y$, so $C(\bb X)$ is the matrix representation of the linear operator whose action is taking the cross with a fixed vector $\bb X$. 
The pressure Jacobian with respect to $\bb X^{j+1,k}$ is 
\begin{align}
&J_{\bb X^{j+1,k}} ( (\bb X^{j+1,k} - \bb X^{j-1,k} )  \cross  ( \bb X^{j,k+1} - \bb X^{j,k-1} ) )  
= - C(\bb X^{j,k+1} - \bb X^{j,k-1})  
\end{align}
Other pressure terms take the same form up to signs. 

Tension blocks are computed as follows. 
Let $\bb X$ be a node of interest, and let $\bb X'$ be any one of its neighbors.
For functions
\begin{align}
f:\R^{3} \to \R , \quad \bb G : \R^{3} \to \R^{3} , 
\end{align}
we have the product rule 
\begin{align}
J_{\bb X}(  f(\bb X) \bb G (\bb X)  ) =    \grad_{\bb X} f  \bb G^{T} + f J_{\bb X}(\bb G) . 
\end{align}
Then
\begin{align}
J_{\bb X}\left(  \frac{\bb X' - \bb X}{|  \bb X' - \bb X | } \right) 
=  \frac{ (\bb X' - \bb X) (\bb X' - \bb X)^{T} }{ |  \bb X' - \bb X |^{3} }  -  \frac{1}{ |  \bb X' - \bb X |} I   , 
\end{align}
where $I$ is the three by three identity matrix. 
The Jacobian associated with an elastic link is then 
\begin{align} 
J_{X} &\left ( \left( 1 - \frac{1}{1 + \dfrac{ |\bb X' - \bb X|^{2} }{ a^{2} (\du)^{2} }} \right)  \frac{\bb X' - \bb X}{ |  \bb X' - \bb X |} \right) \\ 
= &-\left( \frac{   \dfrac{ 2 |\bb X' - \bb X| }{ a^{2} (\du)^{2} } }{ \left( 1 + \dfrac{ |\bb X' - \bb X|^{2} }{ a^{2} (\du)^{2} } \right)^{2} } \right) \frac{(\bb X' - \bb X)(\bb X' - \bb X)^{T}}{ |  \bb X' - \bb X |^{2}}   \nonumber \\ 
&+ \left( 1 - \frac{1}{1 + \dfrac{ |\bb X' - \bb X|^{2} }{ a^{2} (\du)^{2} }} \right) \quad \left(  \frac{ (\bb X' - \bb X) (\bb X' - \bb X)^{T} }{ |  \bb X' - \bb X |^{3} }  -  \frac{1}{ |  \bb X' - \bb X |} I  \right) . \nonumber 
\end{align}
This gives all the forms of Jacobians that we need to build the global Jacobian. 

\begin{figure}[ht]
\centering 
\includegraphics[width=.6\columnwidth]{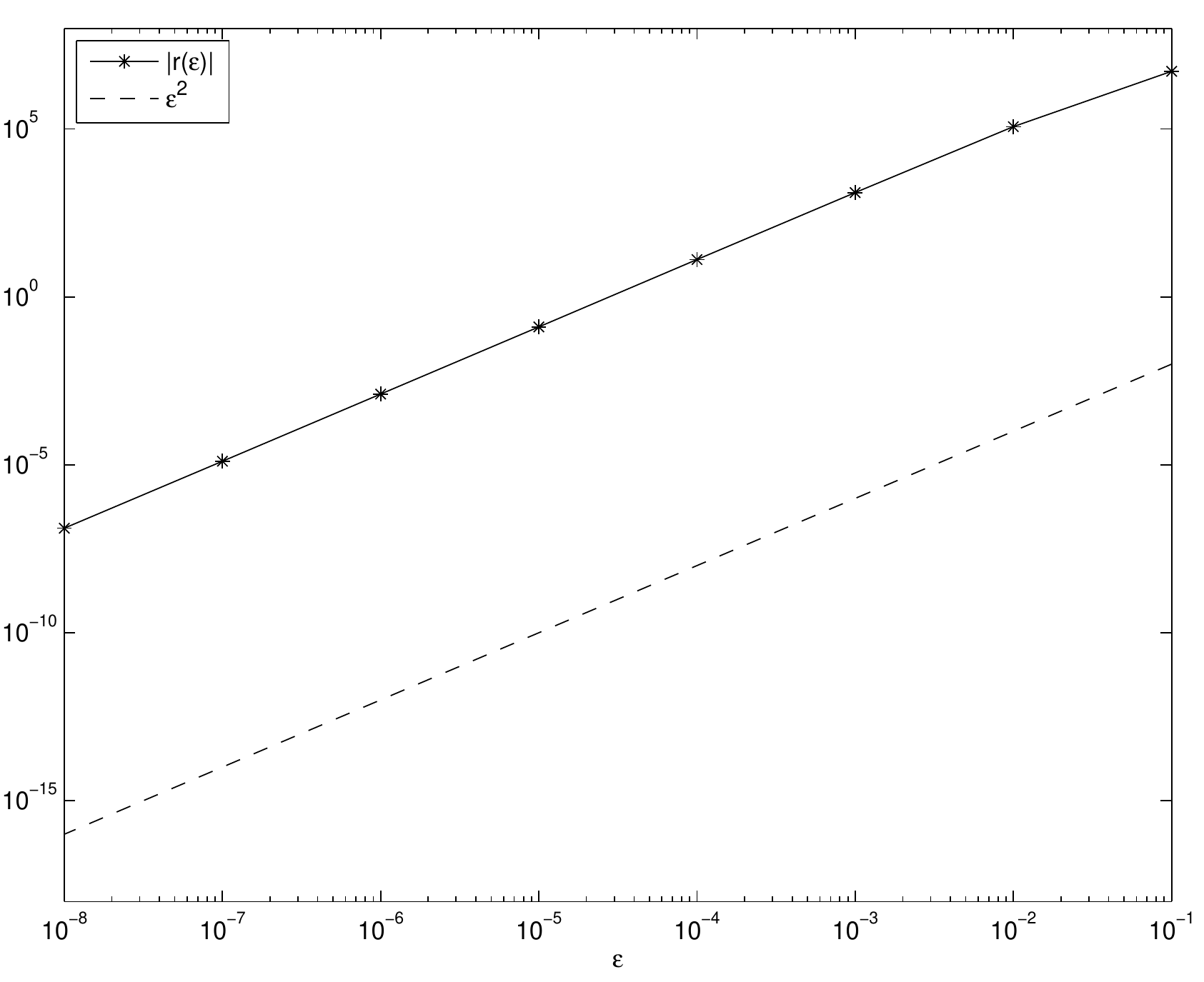}
\caption{Second order decrease in Jacobian comparison.} 
\label{jacobian_conv}
\end{figure}

To check the Jacobian, we use a test based on a Taylor expansion to show that the computed Jacobian is indeed the derivative of the function $F$. 
The Jacobian appears in the Taylor expansion of $F$ as 
\begin{align} 
F(\Phi + \eps \Psi) = F(\Phi) + \eps J(\Phi) \Psi  + O(\eps^{2}) 
\end{align} 
To check, let 
\begin{align} 
r(\eps) = F(\Phi + \eps \Psi) - F(\Phi) - \eps J(\Phi) \Psi 
\end{align} 
and compute $ | r(\eps) |$ for a variety of $\eps$. 
When evaluated, the exponent of $|r(\eps)|$ should decrease by $10^{-2}$ for every $10^{-1}$ decrease in $\eps$. 
Note that this tests the relationship between $F$ and $J$, that $J$ is the Jacobian of $F$, rather than anything about $F$ itself. 
Results are shown in Table \ref{jac_test} and Figure \ref{jacobian_conv}. 

\begin{table}[ht]
\centering 
\begin{tabular}{ c  |  c |  }
$\eps$	 & $| F(\Phi + \eps \Psi) - F(\Phi) - \eps J(\Phi) \Psi |$ \\ 
\hline 
1.0e-01	 & 5.194941e+06 \\ 
 \hline 
1.0e-02	 & 1.167860e+05 \\ 
 \hline 
1.0e-03	 & 1.283970e+03 \\ 
 \hline 
1.0e-04	 & 1.286958e+01 \\ 
 \hline 
1.0e-05	 & 1.287329e-01 \\ 
 \hline 
1.0e-06	 & 1.287367e-03 \\ 
 \hline 
1.0e-07	 & 1.287377e-05 \\ 
 \hline 
1.0e-08	 & 1.294217e-07 \\ 
 \hline 
\end{tabular}
\caption{Second order decrease in Jacobian comparison.}
\label{jac_test}
\end{table}

To solve for $J(\Phi_{i})^{-1} F(\Phi_{i})$, we use a direct method provided by Matlab's backslash operator. 
The matrix is stored in a sparse data structure, and the operator checks the matrix for sparsity and symmetry, and selects an algorithm automatically. 
Here, Matlab selects a solver provided by UMFPACK, which computes an LU factorization of the matrix using pivoting strategies to maintain sparsity and numerical stability \cite{UMFPACK}.

In the application of Newton's method to our problem, we make use of the method of continuation, using pressure as the parameter. 
Since Newton's method converges most robustly with a good initial guess, we use solutions at lower pressures as initial guesses, then restart the Newton's iteration accordingly. 
To accomplish this, first we try to solve at a specified target pressure.
With the decreasing tension model, lower pressures converge more robustly, so if that fails, try to solve at a specified easiest pressure, zero. 
Set the increment of pressure to one fourth the target pressure, add this to the pressure and try to solve. 
Every failure, reduce the size of the increment by four (which is empirically better than two) and continue the process. 
Repeat until we have a successful solve at the target pressure.

\subsection{Solutions to equations of equilibrium for various $N$}
\label{structure_convergence}

To verify that the solution to the equations of equilibrium does not change much as the mesh is refined, we check that the differences between solutions go to zero as $N$ is refined. 
We do not wish there to be a fiber centered on either leaflet, because each point on the free edge is assumed to receive exactly one branch from a chordal tree. 
A point in the center would have to attach to both trees (which is not anatomical) or no trees (which would make the tension in that fiber unsupported at the free edge) to maintain this symmetry. 
Thus, we select an even number of points on the ring, and corresponding to each leaflet. 
This implies that on each leaflet there are an even number of points at the valve ring, and thus an odd number of intervals. 
No point on the anterior leaflet is placed at $u=1/4$ (corresponding to $\theta = 0$ and the center of the leaflet) because there are an even number of equally spaced points centered on $\theta = 0$. 
Similarly, no point on the anterior leaflet is placed at $u=3/4$ (corresponding to $\theta = \pi$ and the center of the leaflet) because there are an even number of equally spaced points centered on $\theta = \pi$. 
Thus, on each leaflet there is no radial fiber in the center of the valve. 

Note that as $N$ is refined by factors of two, the set of unrefined values is not a subset of the set of refined values. 
To account for this when comparing valves of varying resolution, we use linear interpolation to evaluate the finer mesh at corresponding points on the coarse valve. 
Linear interpolation introduces second order errors, see theorem 4.3.3 in \cite{dahlquist2003numerical}. 
We expect less than second order accuracy here because of the hybrid nature of our model, which combines a continuum model of the leaflets with a discrete model of the chordae tendineae, so second order interpolation errors are unimportant. 

There is no such complication in $v$ when $N$ is doubled. 
One row is used for the boundary condition at the valve ring. 
The number of rings below the boundary condition is a power of two, as is the number of points on the leaflet below the rings. 
This implies that the total number of points in the $v$ direction is odd, so there are an even number of intervals. 
Thus in each refinement of $\Delta v$ by a factor of 2, the set of unrefined values of $v$ is a subset of the set of refined values.

Let $\bb X_{N}$ denote the solution for the valve with $N$ mesh points on the valve ring, and $\bb X_{N, interp}$ denote the solution interpolated onto the $N/2$ mesh. 
We compute 
\begin{align}
| \bb X_{N} - \bb X_{2N, interp} |  
\end{align}
in the discretized $L^{1}, L^{2}$ and supremum ($L^{\infty}$) norms. 
The results are shown in Table \ref{static_differences_check}. 
Absolute differences between leaflets at difference resolutions are less than .01 cm in the Euclidean ($L^{2}$) norm for comparisons of $N = 128$ and larger. 
These values are small in comparison to the circumference of the valve ring, which we take to be the length scale of this problem, which is of order 10 cm.
The differences decrease monotonically with $N$, except for the $L^{2}$ norm in the final step.

\begin{table}[ht]
\centering 
\begin{tabular}{ c  |  c |  c  |  c |  }
$N,2N$	 & $L^{1}$ & $L^{2}$ & $L^{\infty}$   \\ 
\hline  
64,128 & 5.65e-03 & 1.26e-02  &  1.06e-01    \\  
 \hline 
 128,256 & 4.38e-03 & 9.74e-03  &  7.83e-02    \\  
 \hline 
 256,512 & 3.51e-03 & 7.98e-03  &  6.34e-02    \\  
 \hline 
 512,1024 & 3.27e-03 & 8.07e-03  &  6.28e-02    \\  
 \hline 
 \end{tabular}
\caption{Norms of differences (cm) in solutions for various $N$. Differences between the model at the resolution used for simulations, $N = 512$, and models that are both more coarse and more fine are small relative to length scales involved.}
\label{static_differences_check}
\end{table}

These differences do not indicate convergence at any particular order of accuracy. 
The situation here is considerably more complicated than in the
 discretization of a partial differential equation.  The trees of our
 model have a fundamentally discrete character, and they are coupled
 to the continuum leaflets at the free edge of the valve, the
 configuration of which is not prescribed but is determined along with
 the configurations of the leaflets and chordae.  We have postulated a
 scaling of the various parameters of the model.  In particular, our
 scaling law for the trees of chordae is intended to produce models
 that strongly resemble each other despite being computed at different
 resolutions.  The results in Table \ref{static_differences_check} show that this is partly
 successful, since the differences from one level of resolution to the
 next are small in comparison to the dimensions of the valve.  On the
 other hand, it is clear that the convergence, if any, in Table \ref{static_differences_check} is
 slowing down as we move from one level of resolution to the next.
 Whether this can be overcome by changing the scaling law for the
 trees, or whether it is inherent in our modeling approach, is a
 subject for future research.

We also check the convergence of Newton's method to the desired solution. 
In the first few iterations, the line search parameter $s_{i}$ in equation \eqref{newtons_line_search}, is very small, but it systematically increases and finally becomes equal to one as the iteration proceeds. 
In this way, the algorithm achieves monotonic decrease of the Euclidean norm of the residual.
After a number of iterations, the solution appears to be in the region surrounding a zero of the difference equations. 
The algorithm finds that the norm of the error decreases with $s_{i} = 1$. 
Suddenly, the convergence takes on the squaring of error, that is, approximate doubling of the number of correct digits, or ``quadratic convergence'' that characterizes Newton's method in the neighborhood of a zero. 
This is shown in Table \ref{newton_converge_table}, which also shows $s_{i}$ and the time taken for each iteration. 
Iterations with $s_{i} = 1$ are generally faster, since the difference equations are evaluated fewer times. 
Over the course of fifteen iterations, the error improves eleven orders of magnitude, four orders of magnitude on the final iteration alone.

\begin{table}[ht]
\centering 
\begin{tabular}{ c  |  c |  c  |  c |  }
Iteration	 &  $ |F(\Phi_{m})| $ & $s_{m}$ & Time elapsed   \\ 
	 & (dynes)  &  &  (s)  \\ 
\hline 
initial &  2.40e+05 & - & - \\ 
\hline
1 & 2.32e+05 & 1/32 & 43.86 \\ 
 \hline 
2 & 2.14e+05 & 1/16 & 37.14 \\ 
 \hline 
3 & 1.79e+05 & 1/8 & 35.02 \\ 
 \hline 
4 & 1.29e+05 & 1/4 & 31.05 \\ 
 \hline 
5 & 7.02e+04 & 1/2 & 25.96 \\ 
 \hline 
6 & 4.93e+04 & 1/2 & 24.35 \\ 
 \hline 
7 & 4.62e+04 & 1/2 & 24.15 \\ 
 \hline 
8 & 4.18e+04 & 1/2 & 24.43 \\ 
 \hline 
9 & 3.33e+04 & 1/2 & 24.08 \\ 
 \hline 
10 & 2.34e+04 & 1/2 & 24.10 \\ 
 \hline 
11 & 1.47e+04 & 1 & 20.46 \\ 
 \hline 
12 & 4.28e+02 & 1 & 20.49 \\ 
 \hline 
13 & 3.50e+01 & 1 & 20.47 \\ 
 \hline 
14 & 1.91e-01 & 1 & 20.44 \\ 
 \hline 
15 & 9.47e-06 & 1 & 20.49 \\
 \hline 
\end{tabular}
\caption{Convergence of Newton's method. $N = 512$.}
\label{newton_converge_table}
\end{table}

\section{Fluid-Structure Interaction }

In this section, we describe technical details related to the fluid-structure interaction portion of the paper and present further results on alternative driving pressures. 

\subsection{Simulation setup}
\label{setup}

Details of the simulation setup are described here.

The fluid domain is a rectangular box of dimensions $[-L,L] \times [-L,L] \times [-3L,L]$, where $L = 3$ cm. 
This domain is discretized with $128 \times 128 \times 256$ points, and has an Eulerian mesh spacing of $\dx = 4.6875 \cdot 10^{-2}$ cm. 
The mitral ring radius is approximately 2.19 cm, so there are just over 90 fluid points across the valve ring at its largest diameter. 
On the walls, where $x$ or $y$ is equal to $\pm L$, we impose a no-slip boundary condition. 
Gravity is ignored throughout. 
The time step is set to $\dt = 1.5 \cdot 10^{-6}$ s in all simulations shown in this article, which is the largest time step that we found to be stable for this particular spatial discretization. 
Simulations were run on New York University's high-performance computing cluster on 84 cores across three nodes, each of which possesses two Intel Xeon E5-2690v4 2.6GHz CPUs (``Broadwell'') with 14 cores/socket or 28 cores/node and an EDR interconnect. 
Individual steps take approximately 0.4 seconds; the elapsed time required to complete each cardiac cycle of 0.8 s duration is therefore approximately four days.

The domain is taken to be periodic in z, but see below for the description of a numerical flow straightener that prevents the flow pattern downstream of the valve from influencing what happens upstream.  
We drive the flow by a spatially uniform body force (force per unit volume) that points in the $z$ direction.  
This is equivalent to specifying a pressure difference between the top and bottom of the domain, and we describe it that way, i.e., we report the applied pressure difference rather than the body force itself.
The force has value 
$\bb d = \left(0, 0, -P(t) / H \right)$, 
where $P(t)$ is the pressure difference and $H = 4L$ is the total domain height.

At the downstream end of the chamber, away from the valve, we add a mathematical flow straightener that applies force in the $x$ and $y$ directions to oppose those components of velocity, and an averaging force in the $z$ direction. 
The straightener serves to reduce the values of $u,v$, and thereby to prevent vortices that were shed from the valve leaflets from reappearing at the upstream boundary of our periodic domain.  
The averaging force similarly prevents localized regions of high or low velocity in the wake of the valve from reappearing upstream, and ensures a nearly uniform inflow profile. 
Thus, the flow straightener serves to hide the periodicity of the domain as much as possible. 
The body force of the flow straightener is of the form 
\begin{align}
\bb s =  -\eta (u, v, w - \overline w) , 
\label{straightener_force}
\end{align}
where 
\begin{align}
\overline w = \frac{1}{A} \int_{z = 0} w \; dx dy , 
\end{align}
where $A = (2L)^2 = $ is the cross-sectional area of the domain, so that $\overline{w}$ is the mean of the z-component of the velocity. 
Note that because of the zero normal velocity on the side walls of the domain, $\overline{w}$ is the same for any plane of constant $z$. 
We use the flow through the mitral ring here, because it is already computed for output purposes. 
The force is applied in a thin slab, $z \in (-3L + .5\text{cm} - \dx, -3L + .5\text{cm} + \dx)$, and is scaled by 
\begin{align}
\frac{1}{2} \left( \cos \left(  \frac{\pi ( z - (- 3L + 0.5\text{cm})) }{ \dx}   \right) + 1 \right)  
\end{align}
to slightly smooth the force. (With the staggered grid discretization, there are two planes in which this force is nonzero in the $x,y$ directions and one in the $z$ direction.)
The value of the coefficient  is selected as  $\eta = \rho / (4 \dt)$.

We tune the Lagrangian mesh to have local spacings on all discrete links of approximately $\ds_{goal} = \dx / 2$, and no larger than $\dx$. 
The fluid spacing is $\dx = $ 6cm / 128 $\approx  4.69 \cdot 10^{-2}$ cm. 
When building the model, we discretize the ring with $N$ points, so the spacing of Lagrangian points on the valve ring is then approximately $\ds = 2\pi r / N$. 
We use $N = 512$ for all simulations, and $r = 2.19$, which implies that the ring spacing is $\approx 2.69 \cdot 10^{-2}$ cm.
This is roughly half the fluid spacing, and we use this rough equivalence to determine what structure resolution corresponds with what fluid resolution.

After solving the equations of equilibrium, the links take lengths that are determined as the solution to the equations of equilibrium. 
Some may be above or below the desired Lagrangian spacing. 
If a link is longer than twice the desired Lagrangian spacing, it is split into two or more links. 
That is, if a link has length $L \geq 2\ds_{goal}$, we split it into $\lfloor  L / \ds_{goal} \rfloor$ links. 
This ensures that no link is longer than $2\ds_{goal} = \dx$, the fluid mesh spacing.  
When a link is broken up into two or more links connected in series, the tension/strain relation is the same for all of the sub-links as it was for the original link. 
This re-meshing step takes place immediately before placing the model valve in fluid. 
This has the effect that the leaflet mesh is no longer Cartesian, but this is of no consequence, since IBAMR is able to cope with an arbitrary network of elastic links. 
If a link in the Lagrangian mesh is shorter than the estimated spacing, we view that as acceptable and do nothing. 
Points on the partition between the upstream (atrial) and downstream (ventricular) chambers of the model are not the solution to the equations of equilibrium and thus can be spaced to our preferences.
They are taken with spacing $\ds_{goal} = \dx / 2$.

\begin{figure}[ht]
\centering
\includegraphics[width=.4\columnwidth]{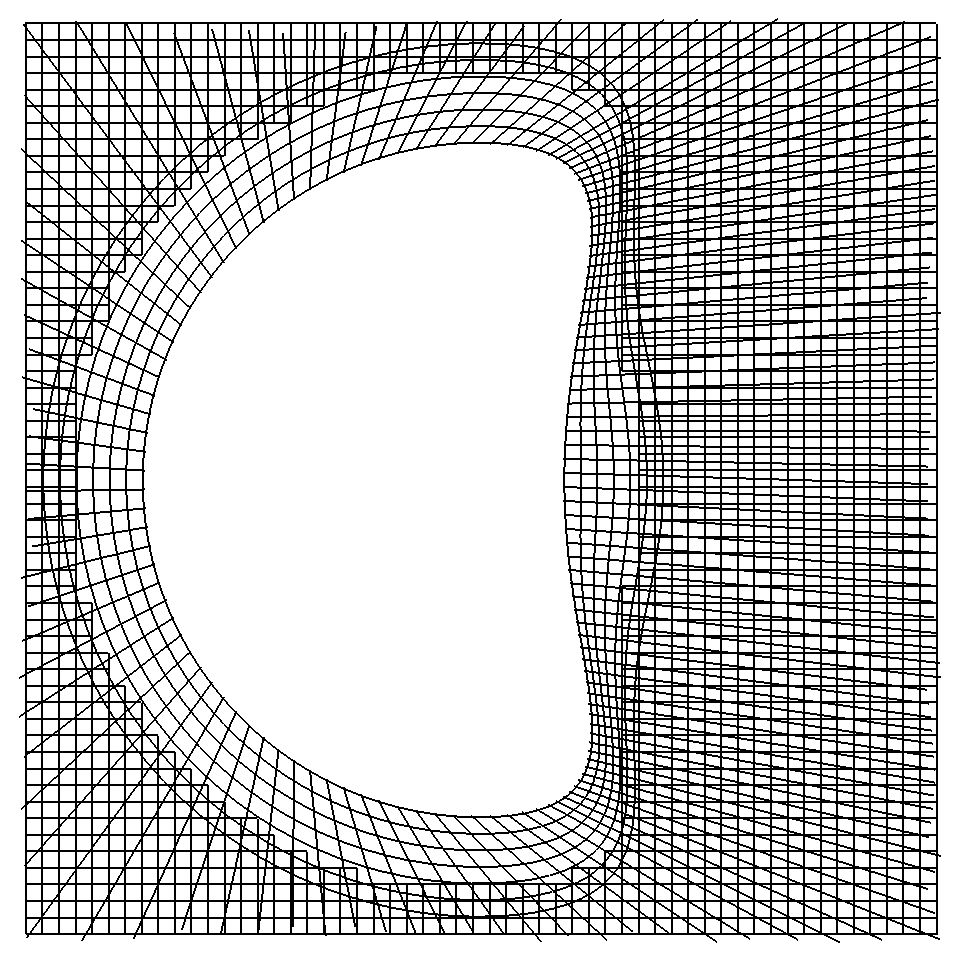}
\caption{Partition without valve, viewed from above (left atrial side). $N=128$.}
\label{top_view_mesh_only0000}
\end{figure}

The partition is shown in Figure \ref{top_view_mesh_only0000}, viewed from the left-atrial side and without the valve for clarity. 
The partition also uses a fiber based elastic model. 
There are three distinct types of fibers included. 
The first family is a system of rings, the innermost of which is the mitral valve ring. 
Starting with the mitral ring, successive rings are generated by propagating each point a fixed distance $\ds$ radially outward from the origin of our coordinate system. 
(The origin lies within the plane of the partition and is interior to the mitral ring; it is the center of the semicircle that forms the posterior part of the mitral ring.) 
Rings continue to be added in this manner as long as they fit within the domain.

The second is a system of rays. 
These are arranged to be continuous with the radial fibers in the valve model. 
These are placed so that the radial fibers in the leaflet do not ``end'' and continue into the partition.  
This allows tension in the radial fibers to be supported by tension and target point forces (described in Section \ref{papillary_target}) in the partition. 
To compute the angle at which the ray should propagate, we rotate the tangent to the radial fiber at its intersection with the mitral ring around the tangent to the mitral ring at the same point until the new tangent lies in the plane $z = 0$. 
Points are placed with spacing $\ds = \dx/2$ along this ray so long as they remain in the domain.

Finally, a certain distance from the valve ring, we place a two-dimensional Cartesian mesh.
This serves to prevent leaks in the corners of the box, where the previous fiber families have inadequate coverage. 
The mesh begins $\ds = \dx/2$ from the no-slip walls on the side of the chamber. 
Links in this mesh are included if they are outside of the fourth ring in the first family of fibers described above. 

All points in the partition are held approximately still with forces due to \emph{target points}. 
The details of how such forces are applied are described in Section \ref{papillary_target}.

\subsection{Papillary muscles and target points}
\label{papillary_target}

Both the partition and the model papillary tips are held in place with \emph{target points}. 
Target points provide a method to specify an approximate location for an immersed boundary point. 
An immersed boundary point is connected to the target point, and a force is applied to keep them close together. 
The location of the target points is prescribed, and for the papillary muscles, time dependent. 
This is an example of a penalty method, which approximately enforces a boundary condition. 
Note that enforcing these locations exactly as boundary condition would require phrasing the conditions as a constraint on the fluid velocity field, and thus complicate the equations dramatically. 

Suppose that $\bb X_{node}$ denotes the position of an immersed boundary point, the dynamics of which are specified by the interaction equations. 
Let $\bb X_{target}$ denote the target location.
The target point does not interact with the fluid directly. 
The location of this point is prescribed as a boundary condition, as is its derivative in the time dependent case. 
It appears in the equation for computing Lagrangian force, but not elsewhere. 

Target points exert force as follows. 
Each target point is connected to an immersed boundary point with a Kelvin-Voigt viscoelastic element \cite{howell}.
This exerts tension as a linear spring, chosen to have zero rest length, in parallel with a dashpot. 
The dashpot term was added to reduce a ``ringing'' effect that we observed in preliminary simulations, in which the oscillations in flow through the mitral ring during at valve closure persisted for an unphysically long time with little decay in amplitude.
That is, the tension is the sum of a term proportional to length $L$ of the element, plus a term proportional to the derivative of length,
\begin{align}
T = k L + \eta \frac{dL}{dt}.  
\label{target_force}
\end{align}
In the immersed boundary framework, the relevant length for use in equation \eqref{target_force} is given by 
\begin{align}
 L = | \bb X_{target} - \bb X_{node} |  . 
\end{align} 
The force associated with a target point is the tension multiplied by a unit vector pointing from the current node to the target point 
\begin{align}
\bb F = T \left( \frac{\bb X_{target} - \bb X_{node} }{L} \right) . 
\end{align}
Take the time derivative of the expression 
\begin{align}
 L^{2} = | \bb X_{target} - \bb X_{node} |^{2}  
\end{align}  
to obtain 
\begin{align}
L \frac{dL}{dt} =  \left( \bb X_{target} - \bb X_{node} \right) \cdot \left(  \frac{d \bb X_{target} }{dt} - \frac{d \bb X_{node} }{dt} \right),   
\end{align} 
which we use to calculate the time derivative of the spring length. 
Since we prescribe the position of $\bb X_{target}$, we know both its position and derivative. 
The time derivative of the Lagrangian variable $\bb X_{node}$ is available from interpolation of the fluid velocity field; we denote it $\bb U_{node}$. 
Because this force is computed with known Lagrangian velocity, it is explicit in time in the discretized version of the model.

The scalar tension associated with the dashpot is thus computed as 
\begin{align}
\eta \left ( \frac{  \bb X_{target} - \bb X_{node}  }{L} \right)  \cdot  \left(  \frac{d \bb X_{target} }{dt} -  \bb U_{node} \right) 
\end{align} 
The full force is then given as
\begin{align}
F = \Big( k  L    
              + &\eta \Big( \frac{  \bb X_{target} - \bb X_{node}  }{L}  \cdot  \Big( \frac{d \bb X_{target} }{dt} -  \bb U_{node} \Big) \Big)  \Big)   
     \frac{ \bb X_{target} - \bb X_{node}}{L} .   
\end{align}
Note that the dashpot force depends on the difference in the prescribed target velocity and the fluid velocity at the node. After scaling by the damping coefficient $\eta$, this quantity is then orthogonally projected onto a vector between the target point and node. 

The partition is not designed to mimic any particular material. 
Similarly, we are not modeling any details of the papillary muscle mechanics. 
Thus, we set both the spring and damping coefficients empirically in both the partition and the papillary muscle target points. 
In the partition, the coefficients are set to $k = 1.89 \cdot 10^{4}$ dynes/cm and $\eta = 3.79$ (dynes s)/cm. 
At the papillary tips, the coefficients are set to $k = 7.58 \cdot 10^{5}$ dynes/cm and $\eta = 1.52 \cdot 10^{3}$ (dynes s)/cm.

In echocardiographic studies, ``Distance from the annulus to the papillary muscle tip was measured both in early and at peak ventricular systole. In normal subjects, this distance did not change significantly through systole'' \cite{sanfilippo1992papillary}. 
Thus, we keep the position of the chordae attachment to the model papillary tips constant through systole. 
The location in diastole was chosen by trial and error such that tension is sufficiently reduced in diastole and the leaflets can move freely, as observed in echocardiograms. 
When moving from systolic to diastolic positions and back, all papillary target points move with the same velocity. 
The motion between the positions is piecewise linear, so the prescribed velocity is piecewise constant. 
The papillary tips move 1.25 cm towards the anterior leaflet during diastole, and 0.25 cm towards the valve ring. 
The papillary tips are in their systolic position for $t \in [.5, .8]$, and their diastolic position for $t \in [.1, .46]$. 
These motions are repeated in time every cardiac cycle.

\subsection{Thickening the structure}
\label{layers}

\begin{figure}[ht]
\centering 
\includegraphics[width=.5\columnwidth]{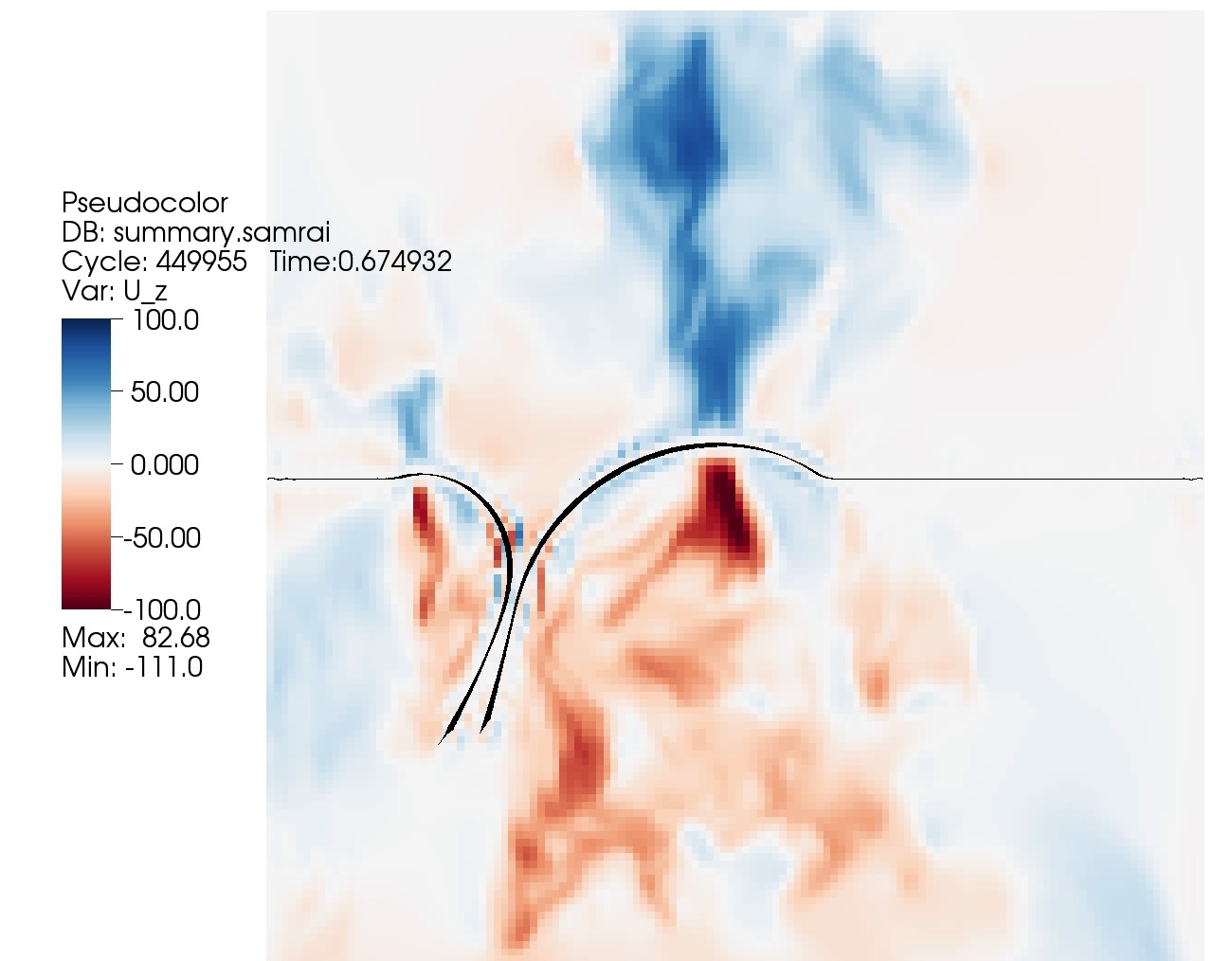} 
\caption{Grid aligned artifacts on a mitral valve simulation with an infinitely thin (two-dimensional surface) valve model. This slice view shows the $z$ component of velocity and the model valve.}
\label{artifact_valve}
\end{figure}

In many simulations, we noticed a strange spurious flow.  
At locations at which the valve is aligned with the coordinate direction, a jet appears going away from the valve in both directions. 
After examining flow fields, we could find no physical explanation for such flows, which suggests that the problem may be numerical. 
An example is shown in Figure \ref{artifact_valve}, which shows a slice view of the $z$-component of velocity during systole. 
Approximately centered in the frame, the model valve has a tangent plane that is approximately flat; that is, it is aligned with the $x,y$ plane. 
Near this point, a jet shoots away from the membrane. 
(One may consider the possibility that this is an artifact of the slice view, and that a physically relevant flow may be in a plane that is not shown here. 
The experiments described in this section strongly suggest that this is not the case, and that this flow is indeed a numerical artifact.) 

\begin{figure}[ht]
\centering 
\includegraphics[width=.475\columnwidth]{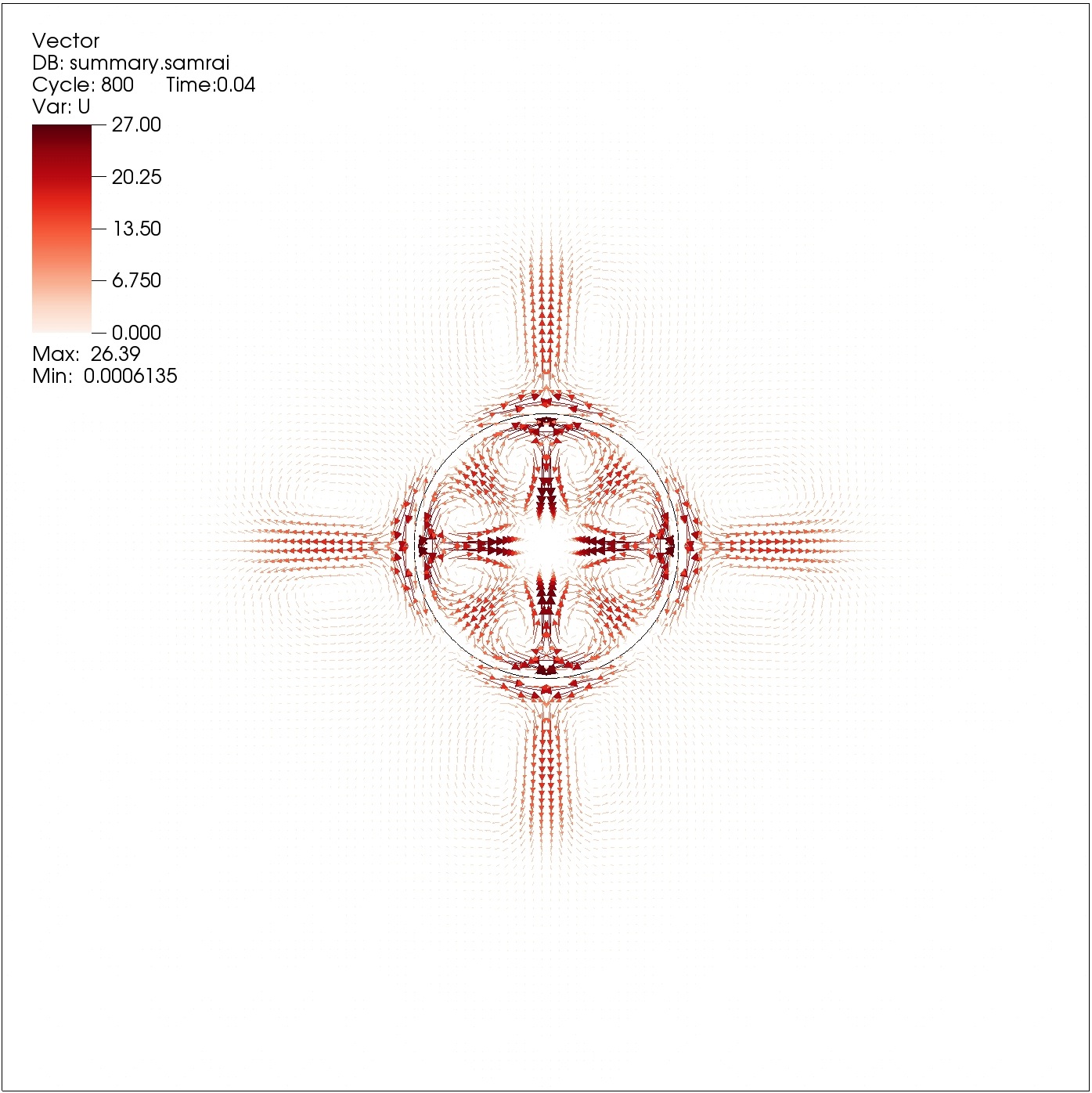} 
\caption{A thin membrane under tension. Large grid aligned artifacts are clearly visible. Note that all flow is numerical error, since the exact solution to this problem is $\bb u = 0$.}
\label{artifact_thin}
\end{figure}

This phenomenon is not particular to our valve model; we found out that it is easy to create such flows. 
Consider a one-dimensional circular membrane of radius $1/4$ cm under tension in a two-dimensional flow. 
The fluid domain is a 1cm by 1cm box with periodic boundary conditions. 
This has an exact solution with $\bb u = 0$ and piecewise constant pressure.
The membrane is discretized as springs with zero rest length and initial spacing $\ds = \dx / 2$. 
Springs with zero rest length exert tension as 
\begin{align}
T = k L, 
\end{align}
where $L$ is the length of the spring in question. 
Spring constants are set to the arbitrary value of $k = 10^{6}$ dynes/cm, which gives a pressure difference of approximately 23.4 mmHg. 
Precisely the same type of artifact occurs, shown in Figure \ref{artifact_thin}. 
It is repeated at locations in which the membrane aligns with the grid, here forming an approximate four-fold symmetry of artifacts. 
Further, we observed similar effects in unrelated physical simulations involving flows in arteries (unpublished coursework).  
In the artery simulations, we moved the structure by various angles, yet the artifact always remains aligned with the grid. 
It seems that any situation involving a discontinuous pressures across an infinitely thin membrane is prone to these issues. 

This effect is severe and needs to be mitigated. 
It persists with changes in delta functions, discretization of the advective term and time step (though the precise velocity field may change slightly).  
One might suppose that the problem is caused by the Cartesian product nature of the discrete delta function, because this is a phenomenon that is aligned with Cartesian grids. 
The effect persists, however, with a radially symmetric delta function. 
Increasing the resolution decreases the magnitude of the artifact and makes it more localized, but it remains at all reasonable fluid resolutions for use in valve simulations, since these simulations always involve a large pressure difference across the valve when it is closed. 
Note that 100 mmHg = $1.3332 \cdot 10^{5}$ dynes/cm$^{2}$, and that all other relevant quantities are moderate in cgs units; thus, the pressure is large in a dimensionless sense. 

\begin{figure}[ht]
\centering 
$ \begin{array}{cc}
\includegraphics[width=.475\columnwidth]{1_shell_bad.jpeg} 
\includegraphics[width=.475\columnwidth]{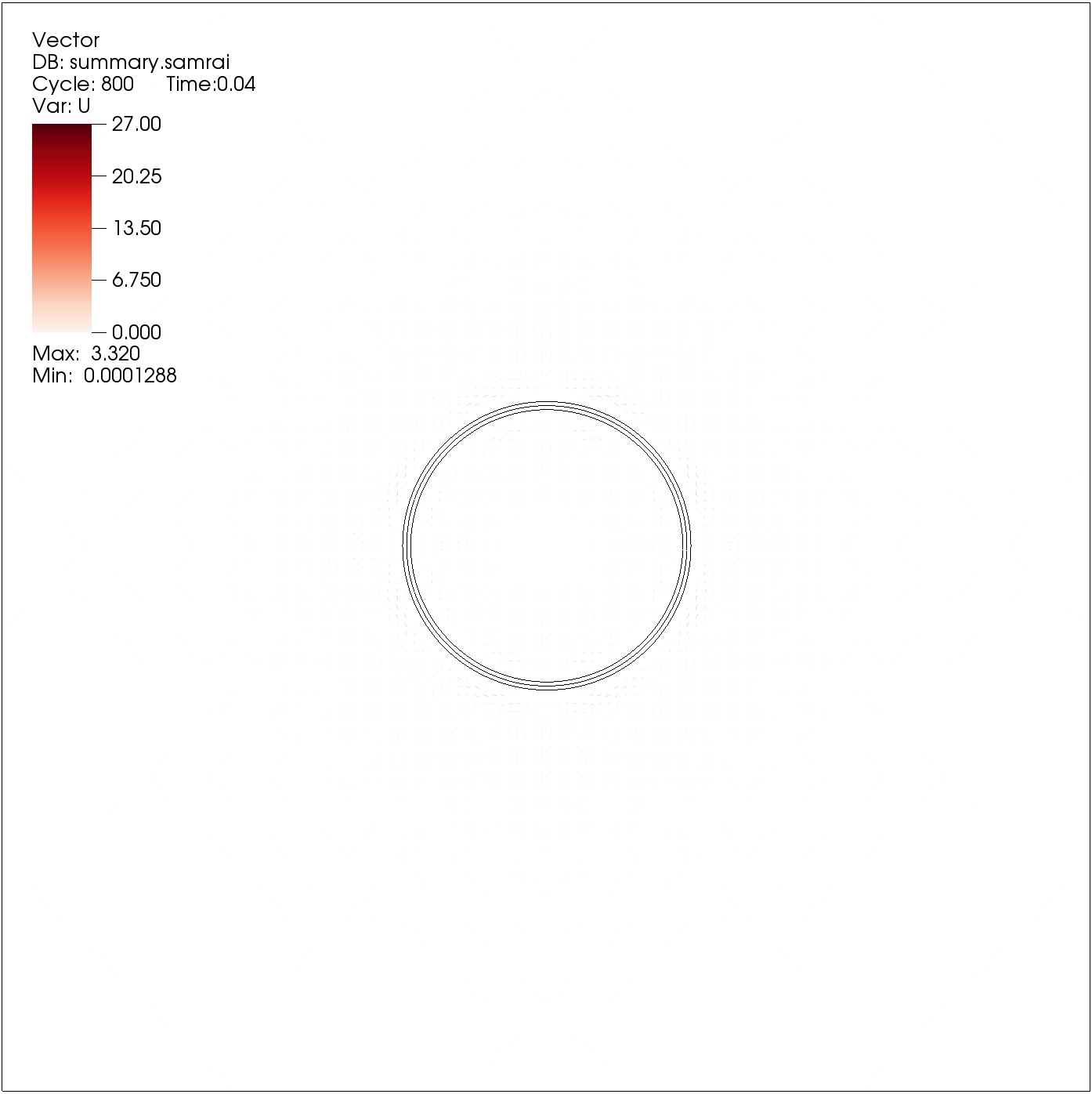} 
\end{array} $
\caption{Comparison of a simulation with a thin structure (left) and a thickened structure (right) under tension. Artifacts with the thickened structure are an order of magnitude smaller and are invisible at this scale.}
\label{artifact_shell}
\end{figure}

We have found, however, that thickening the structure reliably decreases the size of the artifact by an order of magnitude or more. 
Figure \ref{artifact_shell} shows the result shown in Figure \ref{artifact_thin} again, along with the same experiment with a thickened shell. 
Here we replace the single ring under tension with three rings with one third the stiffness. 
With the same scales in both panel no visible artifact appears in the thickened model. 
	
This suggests that all immersed structures that support a pressure should be thickened to avoid this type of problem. 
To quantify this, we run a number of simulations with various numbers of layers and spacings in two-dimensional flows. 
This will advise us on how to avoid such numerical artifacts on a real flow. 
Table \ref{shell_test} shows the numerical error in the maximum norm for these experiments. 
The minimum spacing we investigate is one half the Eulerian mesh width, or $\dx/2$, because that is the target Lagrangian mesh width in general, and we do not wish to have this dimension be much thinner than any other mesh spacings. 
This test reveals that three layers spaced $\dx/2$ apart reduces the error the most. 

\begin{table}[ht]
\centering 
\begin{tabular}{ c | c | c | c | c | c | c | }
	\multicolumn{2}{c}{ }  & \multicolumn{5}{c}{Number of layers}  \\ 
\cline{3-7}
\multicolumn{2}{c|}{ } & 1 & 2 & 3 & 4 & 5 \\ 
\cline{2-7}
\multirow{3}{*}{Spacing} 
& $ \dx / 2$ & 26.10 & 15.47 & 3.38 & 4.22 &  4.72 \\ 
\cline{2-7}
&  $ \dx $ & 26.10 & 14.80 & 9.54 & 6.79  &  5.83 \\ 
\cline{2-7}
&  $ 2 \dx $ & 26.10  & 13.81  &  9.87 & 7.85 & 6.93 \\ 
\cline{2-7}
\end{tabular}
\caption{Maximum modulus of $\bb u$, which is equal to the maximum norm of the numerical error, for various spacings and number of layers}
\label{shell_test}
\end{table}

Thus, we thicken the entire Lagrangian structure in all mitral valve simulations. 
We add additional copies of the entire valve structure and partition, simply placing the additional copies the desired spacing uniformly offset in the $z$ direction. 
We select three layers spaced $\ds = \dx/2$ apart. 
Conveniently, this corresponds to the target Lagrangian mesh width through the entire valve structure. 
Spring constants are uniformly reduced by the number of copies, three, throughout.  
At full resolution, this models the Lagrangian structure as approximately $0.07$ cm thick.
Also conveniently, this is approximately the thickness of a mitral valve leaflet. 

\begin{figure}[ht]
\centering 
\includegraphics[width=.5\columnwidth]{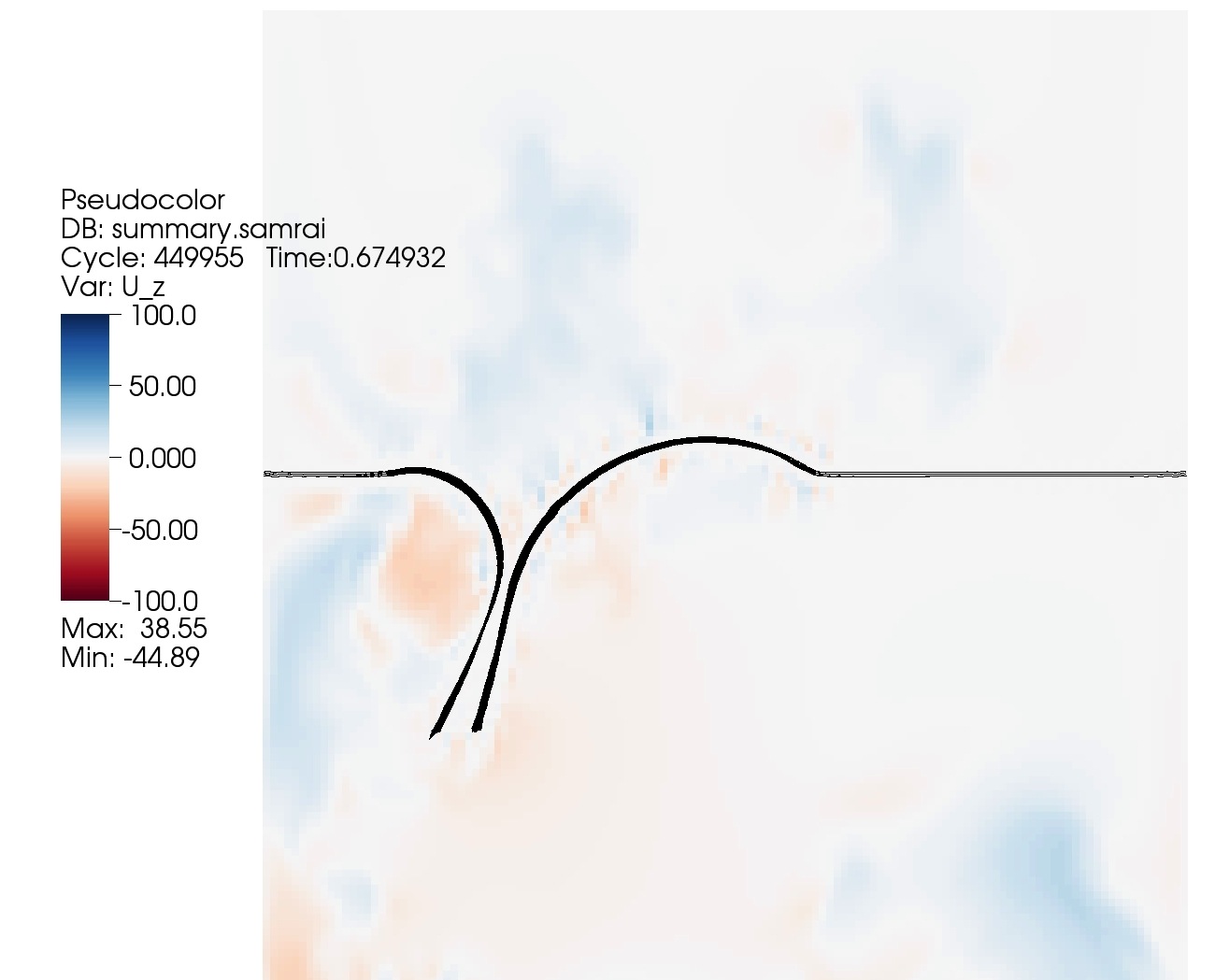} 
\caption{Reduction of artifacts with thickened model valve. 
Note that this simulation is otherwise identical to that of Figure \ref{artifact_valve}, and both figures show the same time step.}
\label{artifact_gone_thickened}
\end{figure}

This dramatically reduces the appearance of the artifact, as shown in Figure \ref{artifact_gone_thickened}. 
The field is qualitatively different, with much lower magnitude throughout. 

The layers are placed close together, and because they move in a continuous velocity field, they may be expected to stay close together. 
However, we observed that oscillations during closure grew as the simulation progressed, and in some locations on the valve mesh it appeared that the layers were not well aligned. 
Thus, we tie the layers together with linear springs of rest length $\ds = \dx/2$. 
We determine the constants, which are uniform, empirically by picking the largest value that does not contribute to time-step restrictions. 
This only serves to keep the layers together, is not intended to represent specific mechanical properties of the anatomy.  
We found it most effective to keep relative spring constants constant across resolutions.
Layers in the chordae are connected in the same manner. 
Layers on the partition are not connected, as they are held in place by target points.  

Despite its persistence under the changes in methodology mentioned above, two-dimensional tests suggest that this artifact can be removed by use of a new immersed boundary method with a divergence free velocity interpolation scheme \cite{div_free_ib}. 
Development of a production-scale implementation of this method and validation of its performance on valve applications remain for future work.

\subsection{Driving pressures}
\label{driving_pressures_appendix}

To represent each of the two pressure waveforms, one for the atrial and one for the ventricular pressure, we use a finite Fourier series with 600 frequencies. 
This ensures that each pressure is periodic in time and smooth.  
We select relevant points by eye from the experimental records, then construct a piecewise linear interpolant to these points. 
To ensure the function is smooth, we then take the convolution of the piecewise linear interpolant with a smooth compactly supported function $\phi$. 
The function $\phi$ is supported on $|t| < \tau$ (not to be confused with $\tau$ used for tensions earlier) and is defined as 
\begin{align}
\phi(t) = 
\begin{cases}
\frac{1}{\tau} \cos^{2} \left( \frac{\pi}{2 \tau } t  \right)  & : \; |t| < \tau \\ 
0                         & : \; |t| \geq \tau . 
\end{cases}
\end{align}
Note that this function integrates to one for all $\tau$. 
We exclusively use $\tau = 0.05$ s here. 
The resulting pressure waveform for physiological pressures is shown in Figure 11 of the main text. 
The maximum systolic pressure difference across the mitral valve is approximately 116 mmHg for this waveform. 
Alternative driving waveforms are shown in Figures \ref{pressure_flux_low},  \ref{pressure_high} and \ref{pressure_flux_nokick}. 

\subsection{Convergence}
\label{fluid_convergence}

Here we investigate the convergence of the fluid-structure interaction simulations. 
We use a fluid mesh spacing of $\dx = 4.6875 \cdot 10^{-2}$ cm, which we refer to as the fine fluid spacing, everywhere in the paper besides this section. 
We compare to a coarse fluid mesh spacing of $\dx = 9.375 \cdot 10^{-2}$ cm. 
The structure spacing is targeted to $\ds = \dx/2 = 2.34375 \cdot 10^{-2}$ cm as described in Section \ref{setup}, which we refer to as the fine structure spacing. 
We compare to the coarse structure spacing, which is targeted to $\ds = \dx/2 = 4.6875 \cdot 10^{-2}$. 
Note that the maximum Reynolds number is estimated to be approximately 2700 (see Section \ref{Reynolds_number} for computation), and flow fields have associated physical instability. 
Additionally, the structure possesses small differences that do not show convergence at any particular order (See Section \ref{structure_convergence}). 
This means that precise convergence of the methods at a particular order is not available. 
Further, the width of the delta function in the interaction equations increases with the larger mesh spacing.
This means that coarsening the resolution decreases the effective orifice area of the open valve, and thus the forward flow.

We compare four simulations, and these comparisons suggest that the simulations are well-resolved at the fine fluid and fine structure resolution. 
The first is with fine fluid and fine structure, as described throughout Section \ref{Fluid-Structure_Interaction}. 
The second is coarse fluid and coarse structure, where we adjust values of coefficients slightly to create a larger orifice area, and thus comparable, physiological forward flow rates. 
We place 1 edge connector (rather than four, which would be half the number placed at fine structure resolution) and set its maximum tension to be $\alpha = 0.25 \tau = 0.61 \cdot 10^{6}$ dynes, or one fourth that of the edge connectors at fine structure resolution. 
The third is coarse fluid and fine structure, and fourth is coarse fluid and coarse structure, without any adjustments to constants at the edge connectors. 

\begin{figure}[ht]
\centering 
\includegraphics[width=\columnwidth]{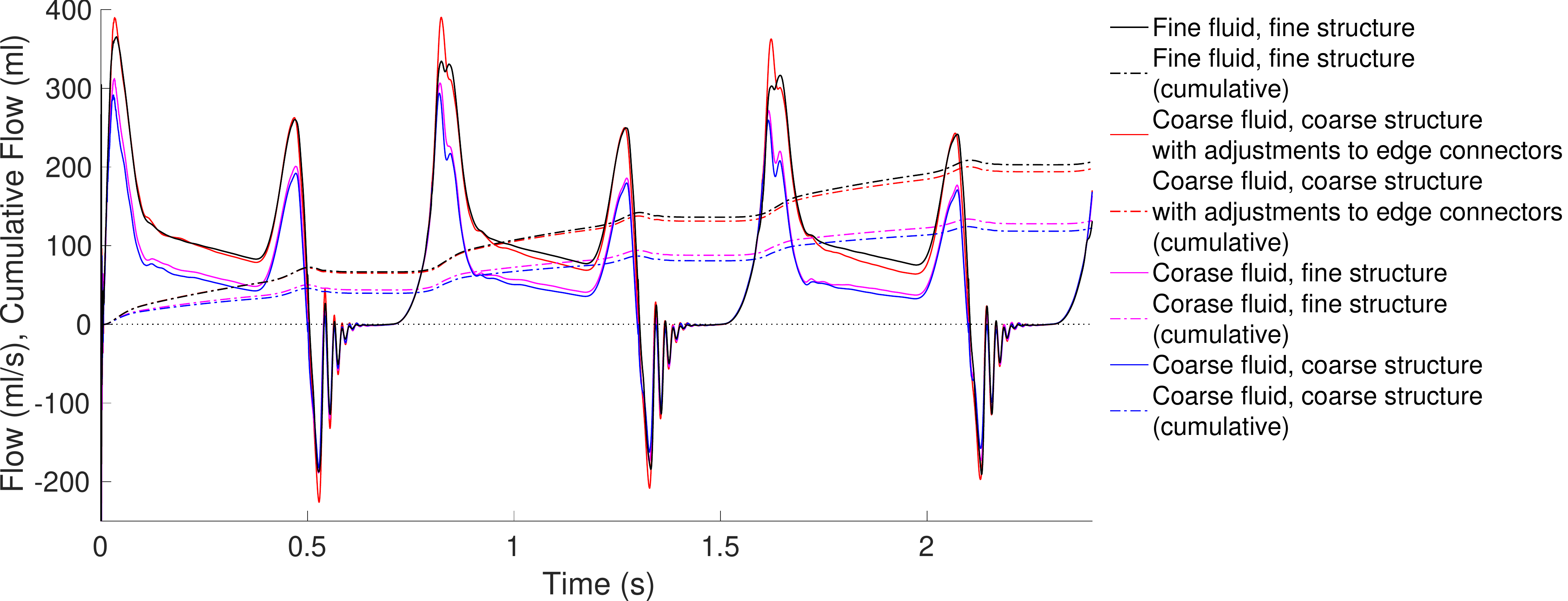}
\caption{ Emergent flows at fine resolution, coarse resolution of fluid and structure, but with adjustments to the edge-connectors of the structure, coarse fluid with fine structure, and coarse fluid and structure. 
The closing transients and flows at all times in which the structure is loaded are consistent across all four simulations. 
Comparing coarse fluid and fine structure resolution with coarse fluid and coarse structure resolution, forward flows are very similar indicating that changing the structure resolution has little effect. 
Forward flows are reduced with coarse fluid and coarse structure resolution compared to forward flows with fine fluid and fine structure resolution because of the width of the delta function in the immersed boundary method. 
The simulation with fine fluid and fine structure resolution has similar forward flow to the simulation flow with coarse fluid and coarse structure resolution with adjustments to the edge connectors of the structure. 
}
\label{flux_comparisons.eps}
\end{figure}

Flow rates and cumulative flows are shown in Figure \ref{flux_comparisons.eps}. 
The closing transient is nearly identical across all simulations. 
The frequency of the associated vibration appears to be consistent. 
The unloading transient after closure looks nearly identical until a true forward flow is established. 
Due to differences in effective orifice area because of the width of the delta function in the immersed boundary method, both coarse fluid with fine structure, and coarse fluid and coarse structure have reduced forward flow rates. 
Reducing the number of edge connector links at the commissures and decreasing their maximum tension can allow a slightly larger opening at coarse resolution, which then produces similar flows to fine resolutions at all phases in the cardiac cycle. 

\begin{figure}[!t]
\setlength{\tabcolsep}{0.5pt}
\centering 
\begin{tabular}{ p{.11\columnwidth}  cccc}
\hspace{0pt} & fine fluid        & coarse fluid, coarse structure & coarse fluid & coarse fluid \\
\hspace{0pt} & fine structure & adjustments to the edge- & fine structure & coarse structure \\
                     &                       & connectors of the structure  & & \\ 
	\vspace{-200pt}
	\begin{tabular}{l}
	velocity \\ 
	cm/s \\ 
	\includegraphics[width=.07\columnwidth]{colorbar_velocity_kick_compare.jpeg} \\ 	
	\vspace{5pt} \\ 
	t = 1.674 s \\ 
	early filling 
	\end{tabular}  

&
\includegraphics[width=.22\columnwidth]{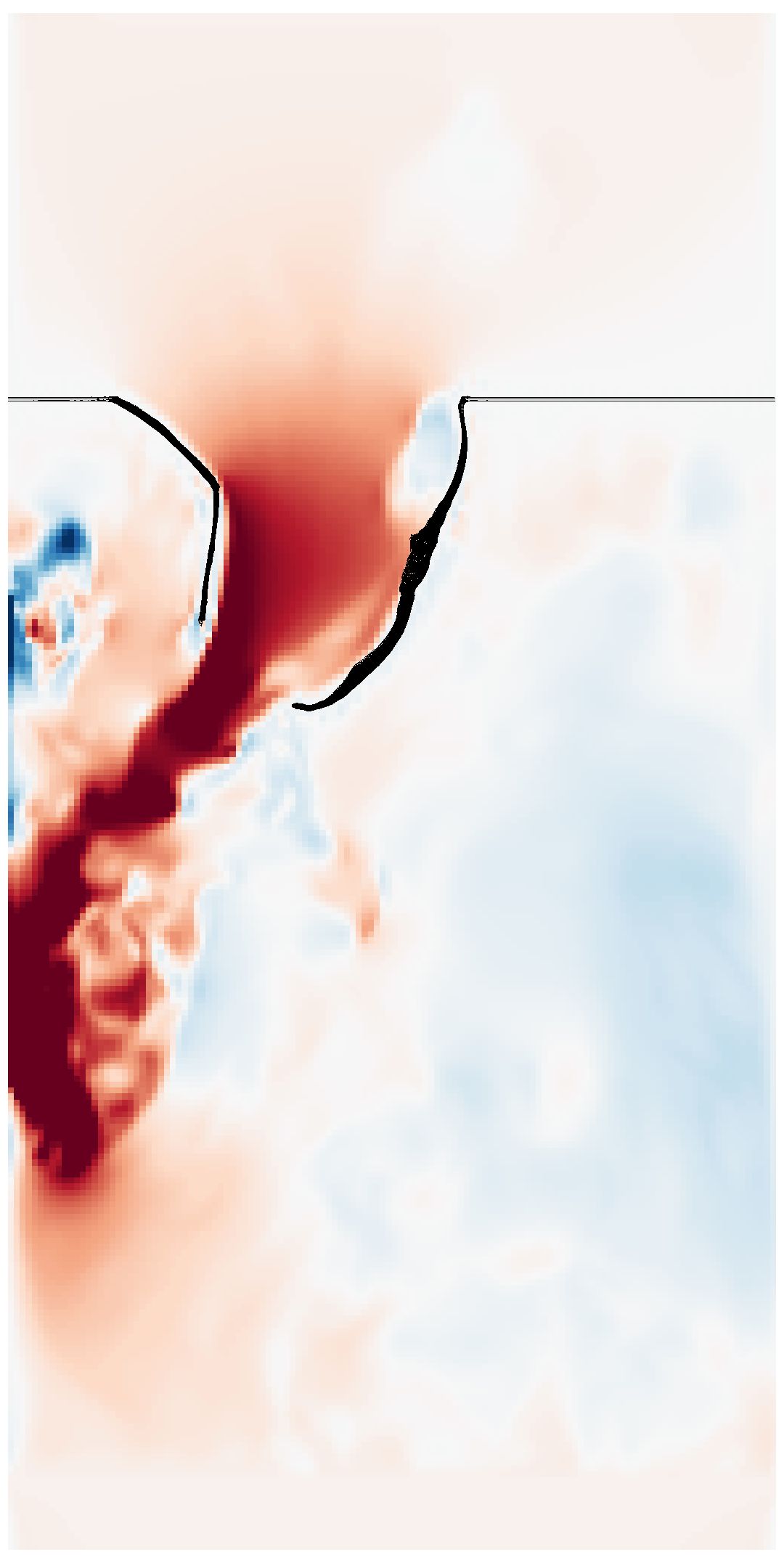} & 
\includegraphics[width=.22\columnwidth]{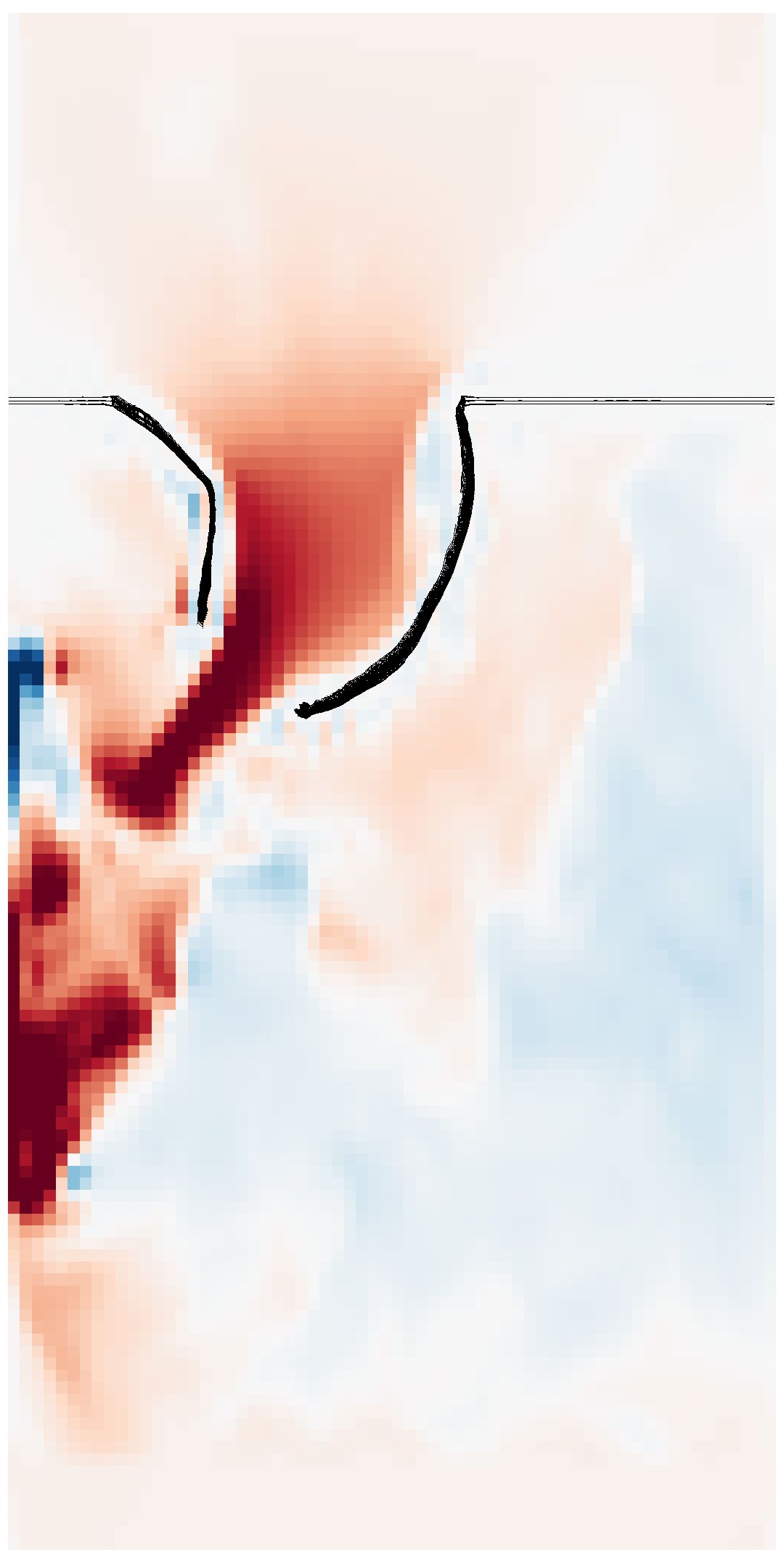} &
\includegraphics[width=.22\columnwidth]{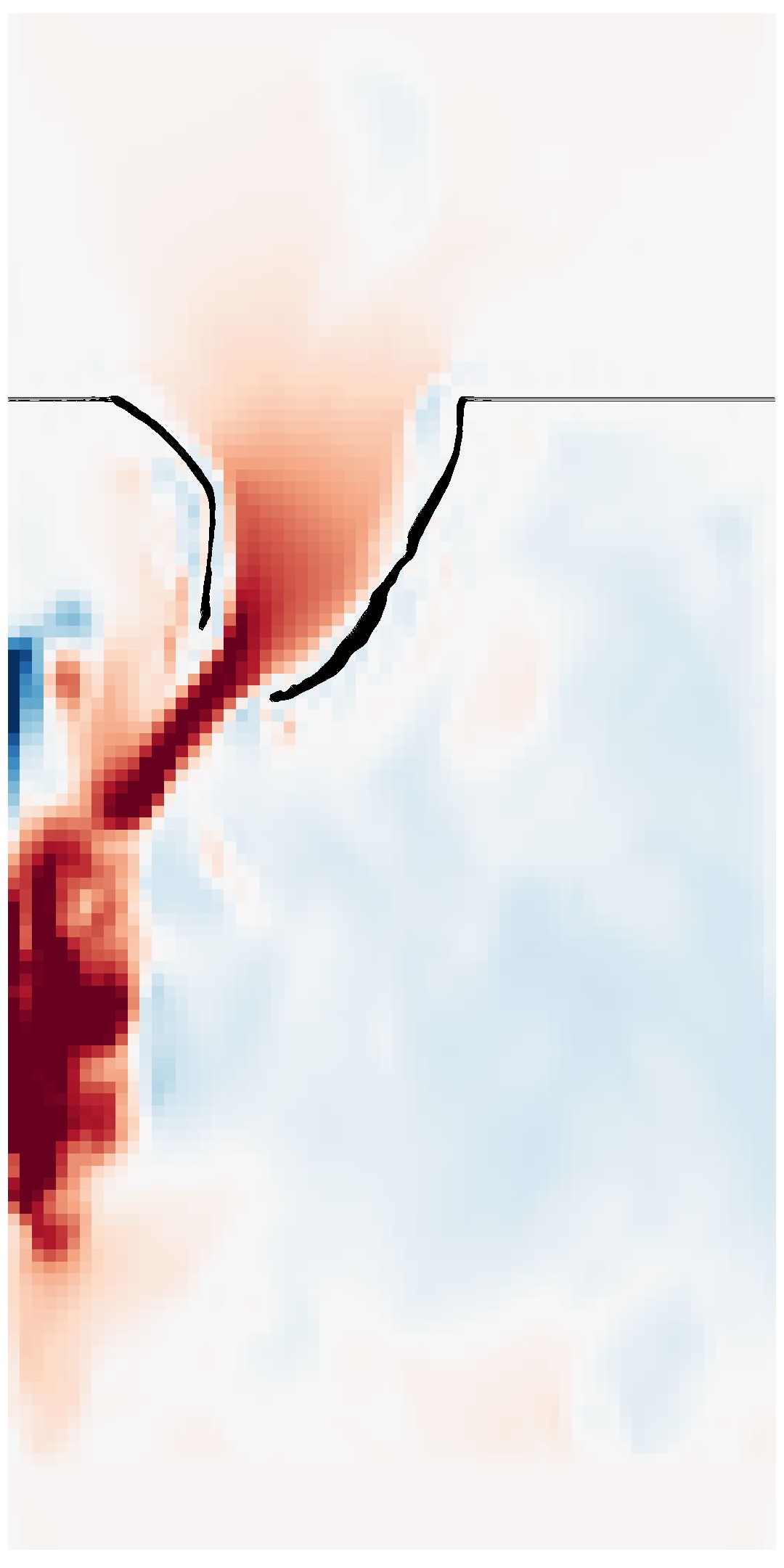} & 
\includegraphics[width=.22\columnwidth]{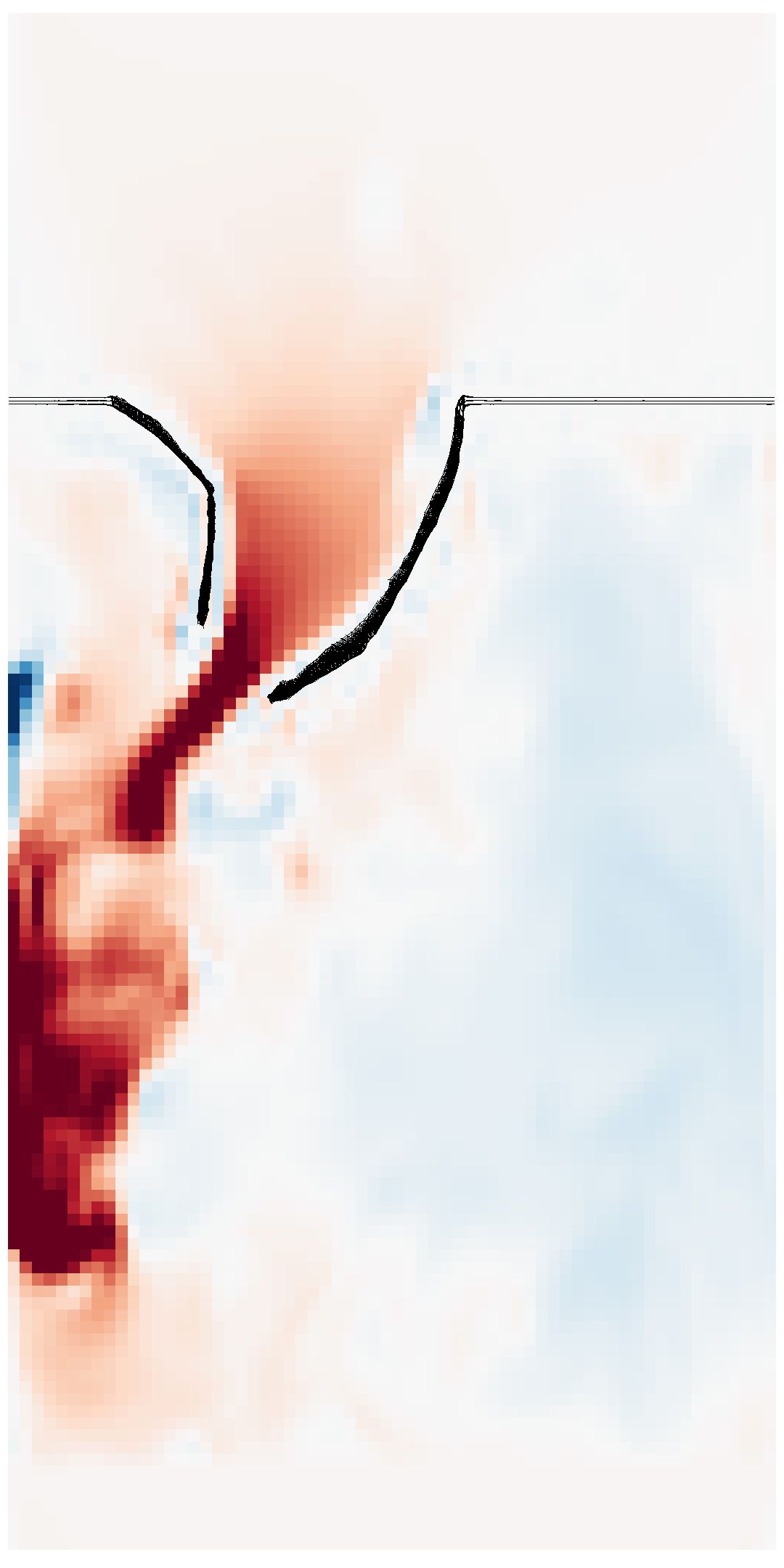} \\  

	\vspace{-200pt}
	\begin{tabular}{l}
	velocity \\ 
	cm/s \\ 
	\includegraphics[width=.07\columnwidth]{colorbar_velocity_low_med_high_comparison.jpeg} \\
	\vspace{5pt} \\ 
	t = 2.303 s \\ 
	mid-systole 
	\end{tabular}  
	
&
\includegraphics[width=.22\columnwidth]{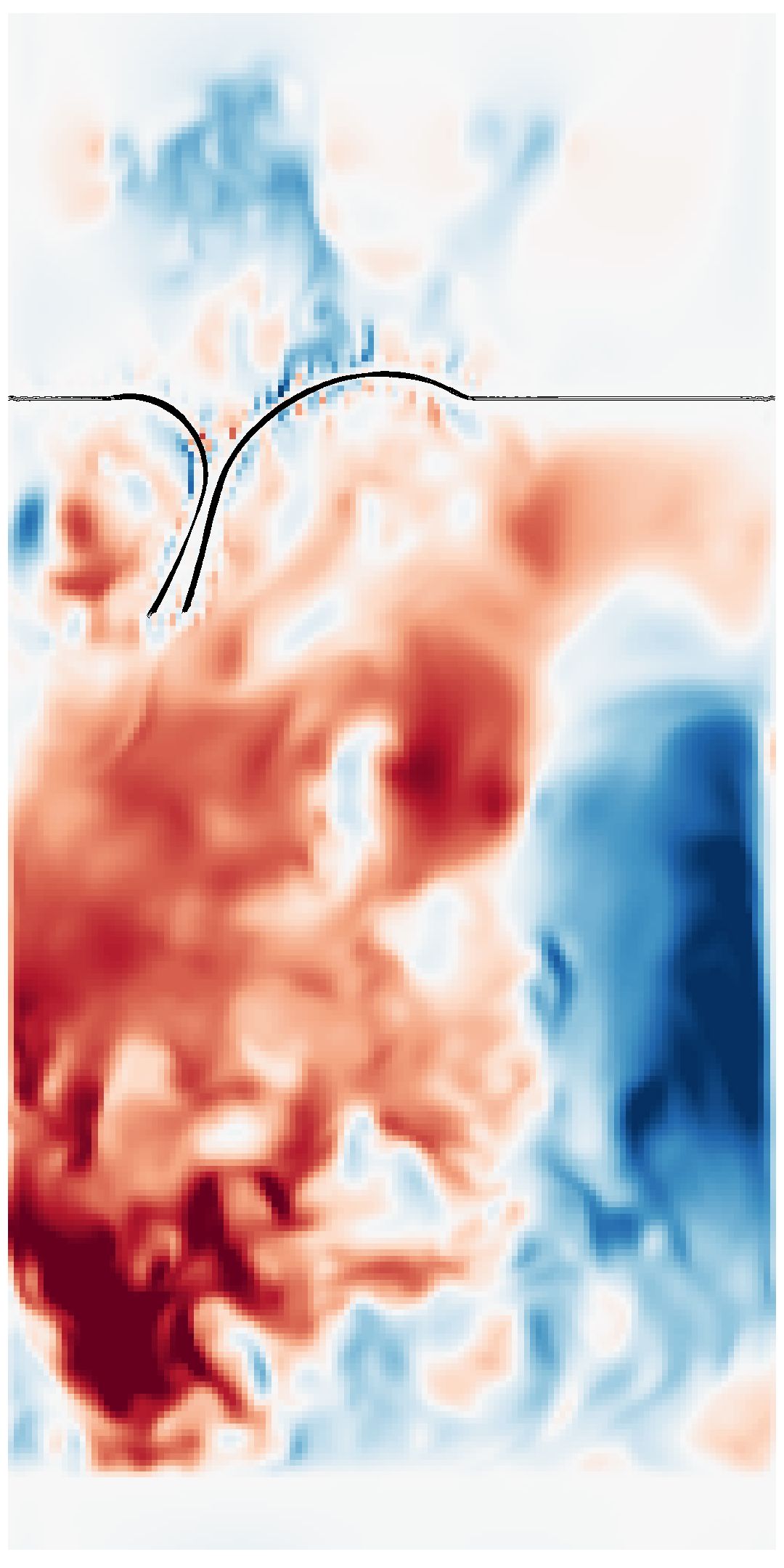} & 
\includegraphics[width=.22\columnwidth]{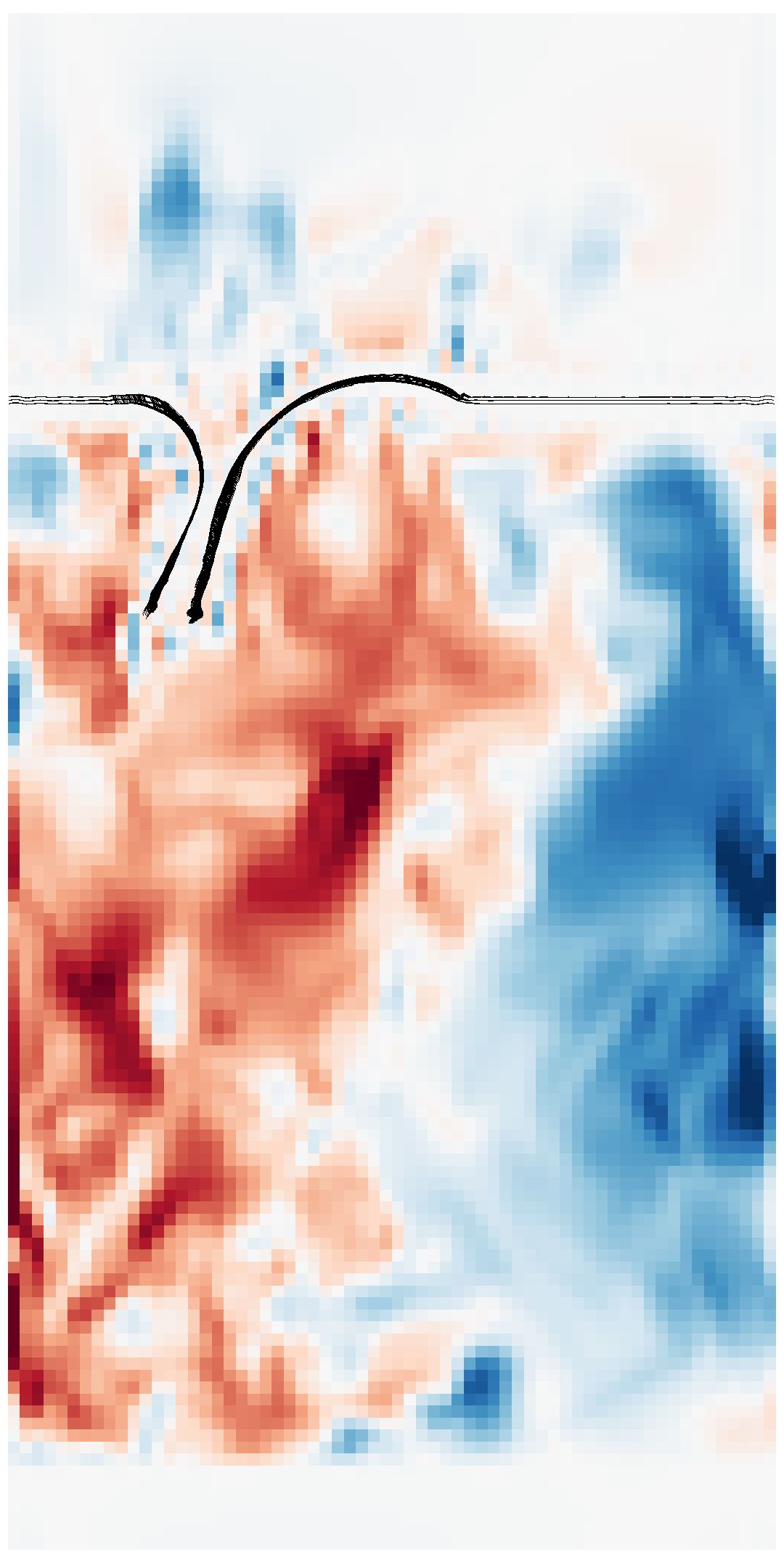} & 
\includegraphics[width=.22\columnwidth]{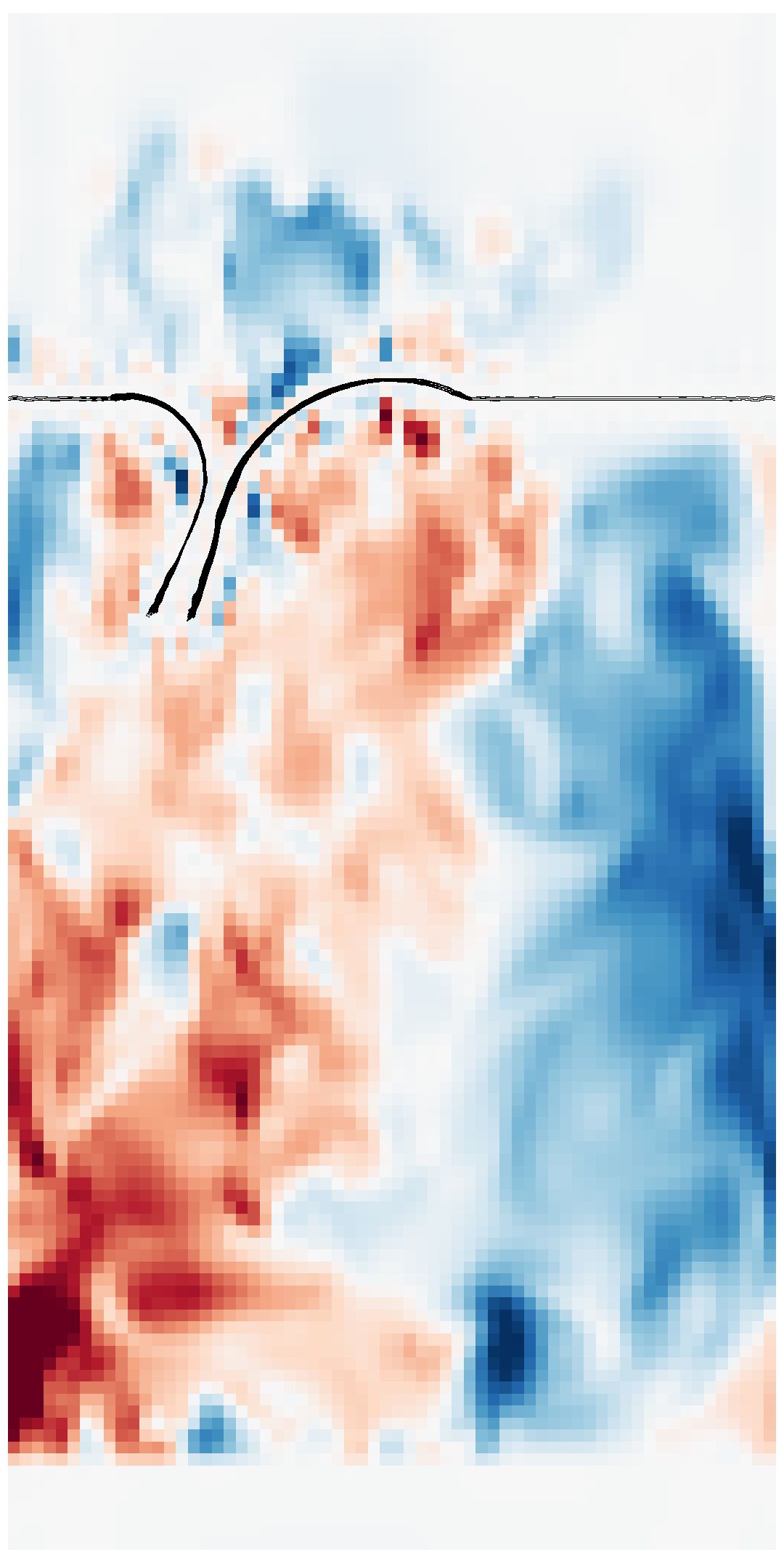} &
\includegraphics[width=.22\columnwidth]{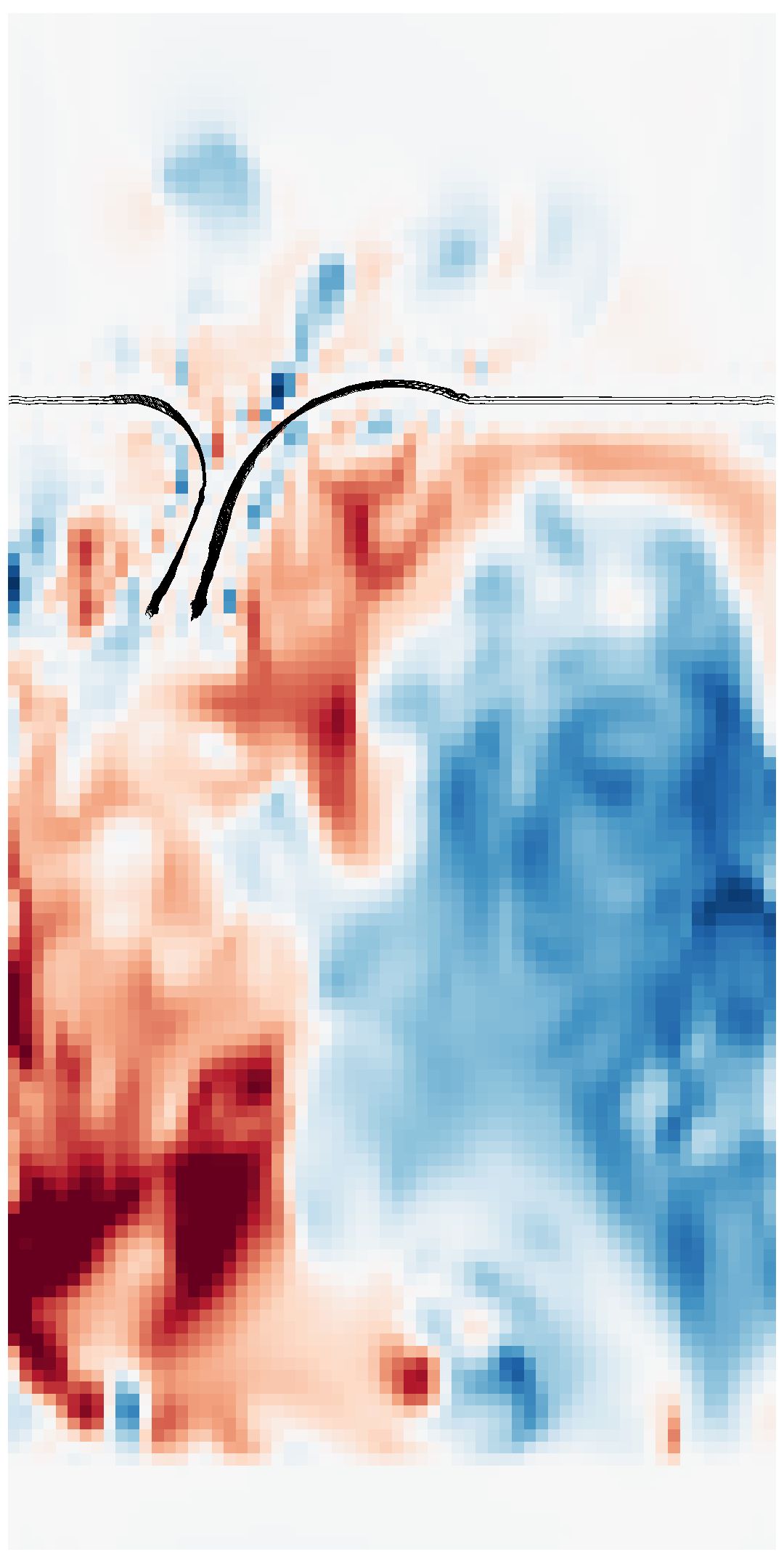} \\  

\end{tabular}
\caption{
Slice view of the velocity fields at (left to right) fine fluid and fine structure resolution, coarse fluid and coarse structure resolution with adjustments to the edge-connectors of the structure, coarse fluid with fine structure, and coarse fluid and coarse structure showing early filling (top row) and closure during mid-systole (bottom row).  
Note that the color scale is 4 times as fine during closure to better illustrate flow features when magnitudes are much lower. 
}
\label{fields_convergence}
\end{figure}

Velocity fields of each of the four simulations are shown in Figure \ref{fields_convergence}, which depicts flow during filling and closure. 
Larger-scale flow features are consistent with changing resolution of the fluid or structure. 
During filling, the forward jet appears in a similar location and begins to break up a similar distance away from the free edges of the leaflets. 
The effects of the width of the delta function are visible as a thin layer of reduced flow near the leaflets on all simulations, but this width is half as large with fine fluid resolution.  
The width of the jet is wider at fine fluid resolution as a result. 
With coarse fluid resolution and comparing fine structure and coarse structure resolution, the jets look very similar. 
During closure, flows are much lower in magnitude in all cases; all valves close effectively.
A circulation appears in the test chamber in all simulations. 

Thus, when the valve is loaded, changes in resolution make very little difference in flow rates. 
Changes in resolution that change the delta function and width and thus the effective orifice area affect the quantitative values of forward flow. 
Adjustments to the edge connector region can alter the effective orifice area during filling, while maintaining consistent behavior during closure. 
We conclude that the results presented at fine resolution are well resolved. 

Similar studies on modeling valves showed similar results. 
One study using IBAMR to study aortic valve mechanics compared simulations with fine-grid resolutions of $\dx = 7.8  \cdot 10^{-2}$ cm and $3.9  \cdot 10^{-2}$ cm \cite{Flamini2016}, and another compared simulations with fine-grid resolutions of $\dx = 8.6  \cdot 10^{-2}$ cm and $4.3  \cdot 10^{-2}$ cm \cite{Hasan2017}.
Both found only minor differences between flow fields and cumulative flow through the valve at these resolutions when controlling for delta function widths. 
Further, convergence of the IBAMR solver library has been extensively verified. 
For formal checks of order of accuracy, see \cite{GRIFFITH200710, Griffith_immersed_fe}.

\subsection{Reynolds number}
\label{Reynolds_number}

We estimate the Reynolds number of this flow under standard driving pressures as follows. 
The density of blood is taken to be $\rho = $ 1 g/cm$^{3}$ and viscosity $\mu = 0.04$ Poise. 
The area of the mitral ring is 
\begin{align}
A = \pi r^{2} (1 - \rho_{dip}/2 + 3\rho_{dip}/16) \approx 10.99 \text{ cm}^{2}
\end{align}
The average distance of the ring to the origin is 
\begin{align}
\bar r = r (1 - \rho_{dip}/4)  \approx 1.78 \text{ cm}
\end{align}
(This is not the average radius from the centroid of the ring, but we use it anyway for simplicity.)
For computing the maximum Reynolds number, we do not include the first .1 s of the simulation, as the initial opening has a slightly higher peak than we see at other times in the simulation. 
The maximum flow on cycles after the first one is $Q_{max} = 334.5 $ cm$^{3}$/s. 
This gives the maximum Reynolds number as 
\begin{align}
Re_{max} = \frac{\rho U_{max} D}{\mu} = \frac{\rho (Q_{max}/A) 2 \bar r}{\mu} = 2706
\end{align}
The average flow during forward flow on cycles after the first one is $Q_{avg} = 122.2$ cm$^{3}$/s. 
This gives the average Reynolds number as 
\begin{align}
Re_{avg} = \frac{\rho U_{max} D}{\mu} = \frac{\rho (Q_{avg}/A) 2 \bar r}{\mu} = 988
\end{align}
We thus estimate the maximum Reynolds number as 2700 and the average Reynolds number as 1000. 
The experimental trace shown in Figure \ref{yellin_dog_flow_stats} and taken from \cite{yellin_book} shows an approximate maximum flow of $Q_{max} = 250$ ml/s, which suggests $Re_{max} = 2022$.
One in vitro study computed a peak Reynolds number of 3370 \cite{Rabbah2013}. 
Thus. we believe our estimates of the Reynolds are in the expected range. 

These Reynolds numbers are much greater than one and the flow is inertially dominated. 
Note that we use flow to compute the characteristic velocity. 
If we used the maximum velocity as the characteristic velocity, rather than flow divided by area, these estimates would increase.  
Thus, these are conservative estimates on the Reynolds number of this flow.

\section{Movies}

We include the following movies for visualization purposes. 
The reader is encouraged to view them, as they convey the dynamics and competency of the model valve in ways that still images cannot. 
\\

\noindent {\bf M1\_physiological\_pressure\_with\_sound. }
Motion of the model valve when driven by physiological pressures as described in Section \ref{control_results}.  
This animation shows pathlines of flow colored by the local magnitude of velocity. 
The simulation is shown in real time. 
Every fourth fiber in the leaflets is shown for visual clarity.
The audio plays the S$_{1}$ heart sound, as synthesized from the flow during systole. 
\\ 

\noindent {\bf M2\_physiological\_pressure\_10x\_slow\_motion} 
Motion of the model valve when driven by physiological pressures as described in Section \ref{control_results}.
This animation shows pathlines of flow colored by the local magnitude of velocity. 
The simulation is shown in 10x slow motion, so that the viewer can see more details of this fast-moving flow. 
Every fourth fiber in the leaflets is shown for visual clarity.
\\ 

\noindent {\bf M3\_physiological\_pressure\_slice\_view\_with\_sound} 
Motion of the model valve when driven by physiological pressures as described in Section \ref{control_results}.
This shows planar slice views of the vertical component of velocity. 
The simulation is shown real time. 
Audio includes the S$_{1}$ heart sound, synthesized from the flow during systole. 
 All fibers in the leaflets are shown, including the links between layers as described in Section \ref{layers}, since we lack a straightforward way to remove a subset of fibers with this plotting software.
\\ 

\noindent {\bf M3\_physiological\_pressure\_slice\_view\_10x\_slow\_motion} 
Motion of the model valve when driven by physiological pressures as described in Section \ref{control_results}.
This shows planar slice views of the vertical component of velocity. 
The simulation is shown in 10x slow motion, so that the viewer can see more details of this fast-moving flow. 
 All fibers in the leaflets are shown, including the links between layers as described in Section \ref{layers}, since we lack a straightforward way to remove a subset of fibers with this plotting software.
\\ 

\noindent {\bf M5\_low\_pressure\_10x\_slow\_motion} 
Motion of the model valve when driven by low pressures as described in Section \ref{further_results}.  
This animation shows pathlines of flow colored by the local magnitude of velocity. 
The simulation is shown in 10x slow motion, so that the viewer can see more details of this fast-moving flow. 
 Every fourth fiber in the leaflets is shown for visual clarity.
\\ 

\noindent {\bf M6\_high\_pressure\_10x\_slow\_motion  }      
Motion of the model valve when driven by high pressures as described in Section \ref{further_results}.  
This animation shows pathlines of flow colored by the local magnitude of velocity. 
The simulation is shown in 10x slow motion, so that the viewer can see more details of this fast-moving flow. 
 Every fourth fiber in the leaflets is shown for visual clarity.
\\ 

\noindent {\bf M7\_no\_atrial\_systole\_10x\_slow\_motion}  
Motion of the model valve when driven by pressures without atrial systole as described in Section \ref{further_results}.  
This animation shows pathlines of flow colored by the local magnitude of velocity. 
The simulation is shown in 10x slow motion, so that the viewer can see more details of this fast-moving flow. 
 Every fourth fiber in the leaflets is shown for visual clarity.

\end{document}